\documentclass[12pt]{article}
\usepackage{authblk}
\usepackage{geometry}
\usepackage{amsthm}
\usepackage{mathrsfs}
\usepackage{subfigure}
\usepackage{customcommands}
\usepackage{bm}
\usepackage{Chrisstyle}

\usepackage{amsmath}
\usepackage{amssymb}
\usepackage{amsfonts}
\usepackage{braket}
\usepackage[normalem]{ulem}
\usepackage{color}
\usepackage{bbold}
\usepackage{ulem}
\usepackage{mathtools}
\usepackage{tensor} 

\newcommand{\eqa}{\stackrel{(a)}{=}}

\DeclareMathOperator{\tr}{tr}

\title{Leading order corrections to the quantum extremal surface prescription}

\author[1]{Chris Akers,}
\author[2,3]{Geoff Penington}

\affiliation[1]{Center for Theoretical Physics,\\
Massachusetts Institute of Technology, Cambridge, MA 02139, USA}
\affiliation[2]{Center for Theoretical Physics,\\ University of California, Berkeley, CA 94720 USA}
\affiliation[3]{Google, Mountain View, CA 94303 USA}

\emailAdd{cakers@mit.edu}
\emailAdd{geoffp@stanford.edu}

\abstract{We show that a na\"{i}ve application of the quantum extremal surface (QES) prescription can lead to paradoxical results and must be corrected at leading order. The corrections arise when there is a second QES (with strictly larger generalized entropy at leading order than the minimal QES), together with a large amount of highly incompressible bulk entropy between the two surfaces. We trace the source of the corrections to a failure of the assumptions used in the replica trick derivation of the QES prescription, and show that a more careful derivation correctly computes the corrections. Using tools from one-shot quantum Shannon theory (smooth min- and max-entropies), we generalize these results to a set of refined conditions that determine whether the QES prescription holds. We find similar refinements to the conditions needed for entanglement wedge reconstruction (EWR), and show how EWR can be reinterpreted as the task of one-shot quantum state merging (using zero-bits rather than classical bits), a task gravity is able to achieve optimally efficiently.}

\begin{document}
\maketitle

\section{Introduction}\label{sec:intro}

 The quantum extremal surface (QES) prescription \cite{Engelhardt:2014gca} says that the entropy of a boundary region $B$ in AdS/CFT is given by
\begin{align}
  S(B)  = \min \text{ext}_\gamma \left[ \frac{A(\gamma)}{4 G} + S_\text{bulk}(\gamma) \right].
\end{align}
Here we are extremizing over bulk surfaces $\gamma$ that are homologous to $B$, $A(\gamma)$ is the area of the surface $\gamma$, $G$ is Newton's constant and $S_\text{bulk} (\gamma) = -\tr(\rho \ln(\rho))$ is the von Neumann entropy of the state $\rho$ of the  fields in the bulk region (known as the entanglement wedge) bounded by the surface $\gamma$ and the boundary region $B$.\footnote{In this paper, we will maintain the traditional fiction that bulk subregions are associated with subsystems of the bulk Hilbert space, and hence that we can construct the reduced state on a subregion by taking a partial trace. In reality, bulk subregions should instead be associated with von Neumann subalgebras of bulk operators.} The combination of the two terms is known as the generalized entropy.

The perceived role of the $S_\mathrm{bulk}$ term has shifted over time.
Because of the explicit factor of $1/G$, the area term becomes very large in the semiclassical limit $G \to 0$. The bulk entropy term, which has no such factor, was therefore initially regarded as a small, perturbative correction.
In the last year or two, this view has changed.
It has become clear that the QES prescription is still valid -- and indeed plays a crucial role -- even in situations where $S_\text{bulk}$ is very large, and so competes with the area term.
In particular, it was shown in \cite{Penington:2019npb, Almheiri:2019psf} that the QES prescription gives a unitary Page curve for the entanglement entropy of an evaporating black hole. This Page transition happens when the bulk entropy of the trivial `empty' QES becomes larger than the area term for a non-trivial QES that lies near the horizon.

However, as we shall see, in such situations considerable care is needed when applying the QES prescription. For many states, a na\"{i}ve application of the QES prescription gives contradictory answers, which are incompatible with basic properties of von Neumann entropies, even at leading order in $1/G$.

The primary aim of this paper is to (1) show that such contradictions exist, (2) show how the contradictions are resolved by more careful calculations, producing leading order corrections to the QES prescription, and (3) give general conditions for when the na\"{i}ve QES prescription is valid, and when it needs to be replaced by a more refined version.

The contradictions can arise whenever there are two extremal surfaces, with $\mathcal{O}(1/G)$ bulk entropy in the intermediate region between the two. 
While common enough in AdS/CFT, this situation is also central to the phase transition that provides the Page curve of the evaporating black hole. Indeed, in Section \ref{sec:contradictions} we show that black hole evaporation can still lead to entropies inconsistent with unitarity, when using the  na\"{i}ve QES prescription.

\begin{figure}
\centering
\includegraphics[width = 0.5\textwidth]{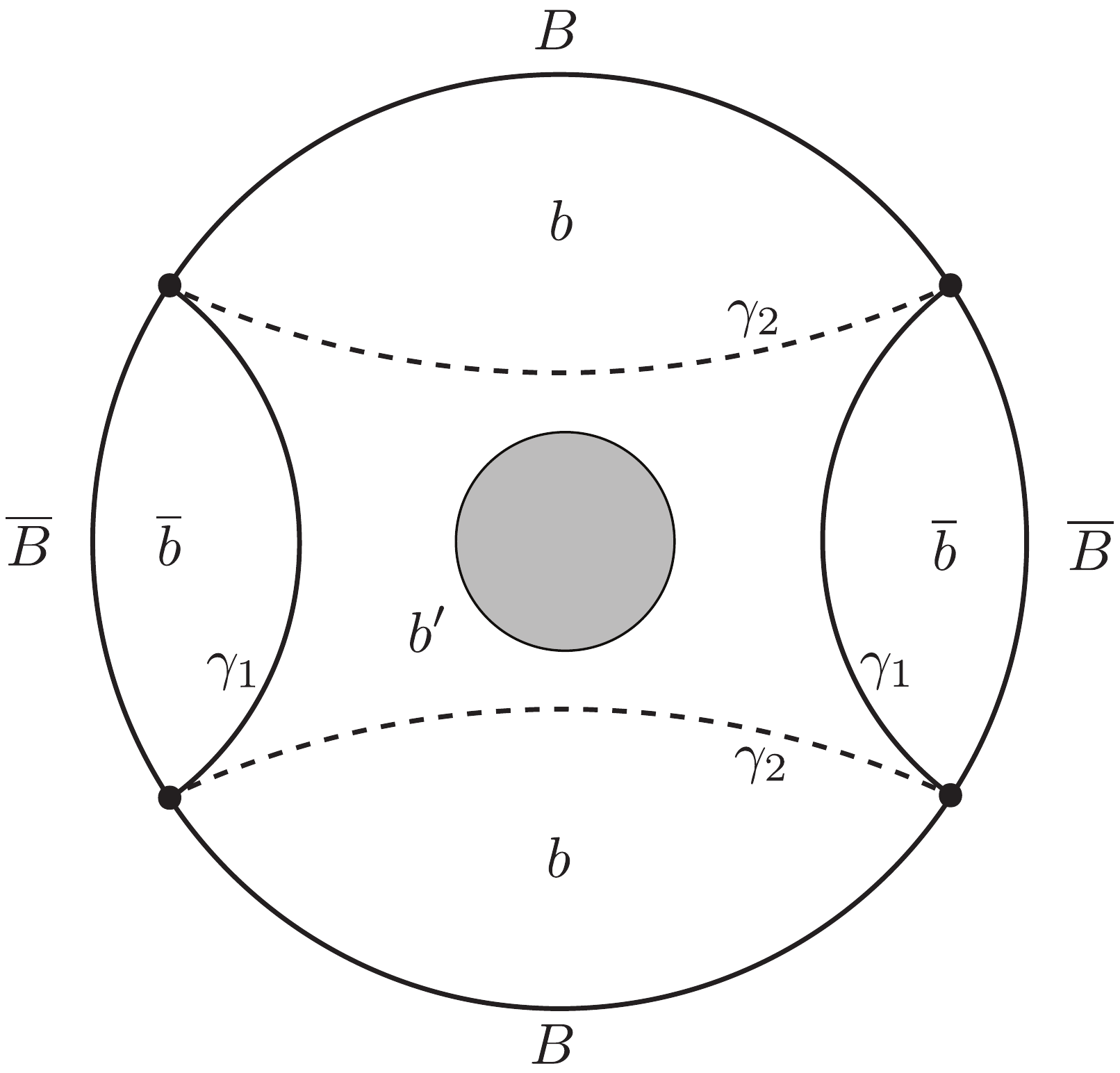}
\caption{Setup in which we derive a contradiction from a na\"{i}ve application of the QES prescription. The boundary is divided into two subregions, $B$ and $\overline{B}$. For both, there are two competing quantum extremal surfaces, $\gamma_1$ and $\gamma_2$, with $\gamma_1$ homotopic to $\overline{B}$ and $\gamma_2$ to $B$. We take $B$ to be larger, such that the area of $\gamma_2$ is bigger than that of $\gamma_1$ at $\mathcal{O}(1)$. Between these surfaces is a large amount of matter (the ``dustball''), such that some states of the matter have entropy much larger than the difference in areas of the two surfaces.}
\label{fig:dustball}
\end{figure}

Returning to AdS/CFT, a useful example setup was given in \cite{Akers:2019wxj}. Consider 2+1d AdS, with the boundary divided into four regions as shown in Figure \ref{fig:dustball}. 
Let two diametrically opposed regions be slightly larger than the other two, such that the union of those two, named $B$, has a connected entanglement wedge in the absence of bulk entropy. The complement of the boundary region $B$ shall be labelled $\overline{B}$. There are two extremal surfaces homologous to $B$: one homotopic to $\overline{B}$ and labelled $\gamma_1$, and one homotopic to $B$ and labelled $\gamma_2$. These surfaces divide the bulk into three regions: one named $b$ that neighbours $B$, one named $\overline b$ that neighbours $\overline B$, and a central region labelled by $b'$ that is bounded by the two extremal surfaces.
Let there be matter in $b'$ with energy $\mathcal{O}(\varepsilon/G)$, for some $\varepsilon  \ll 1$. 

The backreaction is under control for small enough $\varepsilon$, and the bulk matter can have entropy roughly equal to its energy. 
We can therefore easily dial the size of $B$ such that some bulk states have a bulk entropy larger than the area difference, while all states have the same approximate classical geometry.

Consider two states: In the first, the bulk matter is in a pure energy eigenstate. The matter therefore does not contribute to $S_\text{bulk}$, and the entanglement wedge of $B$ is connected. In the second, the matter is in a thermal state with the same average energy.
We tune the region $B$ such that the large entropy of the thermal state causes its entanglement wedge to be disconnected. Hence, the von Neumann entropies are 
\begin{alignat}{2}
    &\text{Matter pure: }&&S(B) = A_1/4G~, \\
    &\text{Matter thermal: }&&S(B) = A_2/4G~. 
\end{alignat}
Here $A_1$ and $A_2$ are the areas of $\gamma_1$ and $\gamma_2$ respectively. 

Now we can formulate the contradiction. What is $S(B)$ for a state that is a mixture of the pure state and the thermal state? In other words,
\begin{equation}\label{eq:mixture_state}
\rho_\mathrm{matter} = p \ket{\psi}\bra{\psi} + (1-p) \rho_\mathrm{thermal}~.
\end{equation}
A na\"{i}ve application of the QES prescription tells us that, at leading order, the answer is
\begin{equation}
\text{Mixture: } S(B)_\text{na\"{i}ve} = \min\bigg(A_1/4G + (1-p) S_\mathrm{thermal},~ A_2/4G\bigg)~.
\end{equation}
However, this can’t be correct. The AdS/CFT bulk-to-boundary map is linear, so the global boundary state must also be a mixture of the two boundary states. And, if the global state is a mixture of the two states, the reduced state will also be a mixture of the two reduced states. 
In general, the von Neumann entropy $S(\rho)$ of a mixture of quantum states 
\begin{align}
    \rho = \sum_i p_i \rho_i
\end{align}
of density matrices $\rho_i$ is bounded from above and below by 
\begin{align} \label{eq:upperlowerboundmixture}
    \sum_i p_i S(\rho_i)\leq S\left(\rho\right) \leq \sum_i p_i S(\rho_i) - p_i \ln p_i~,
\end{align}
see e.g. \cite{nielsen&chuang}.\footnote{The lower bound formalizes an intuitive fact: the uncertainty of a mixture of states must be at least as large as the average uncertainty of each of those states. The upper bound is true because $\rho$ must have less entropy than a state that includes a correlated reference system with orthonormal basis $\ket{i}$, $\sum p_i \rho_i \otimes \ket{i}\bra{i}$. The lower (upper) bound is saturated if and only if the $\rho_i$ are all identical (all orthogonal).}
Together, the bounds \eqref{eq:upperlowerboundmixture} are quite restrictive, forcing a mixture of $k$ states to have entropy at most $\mathcal{O}(\ln k)$ different than the average entropy of those states.
In particular, the entropy of a mixture of $\mathcal{O}(1)$ states is within $\mathcal{O}(1)$ of the average entropy within the mixture.

\begin{figure}
    \centering
    \includegraphics[width = \textwidth]{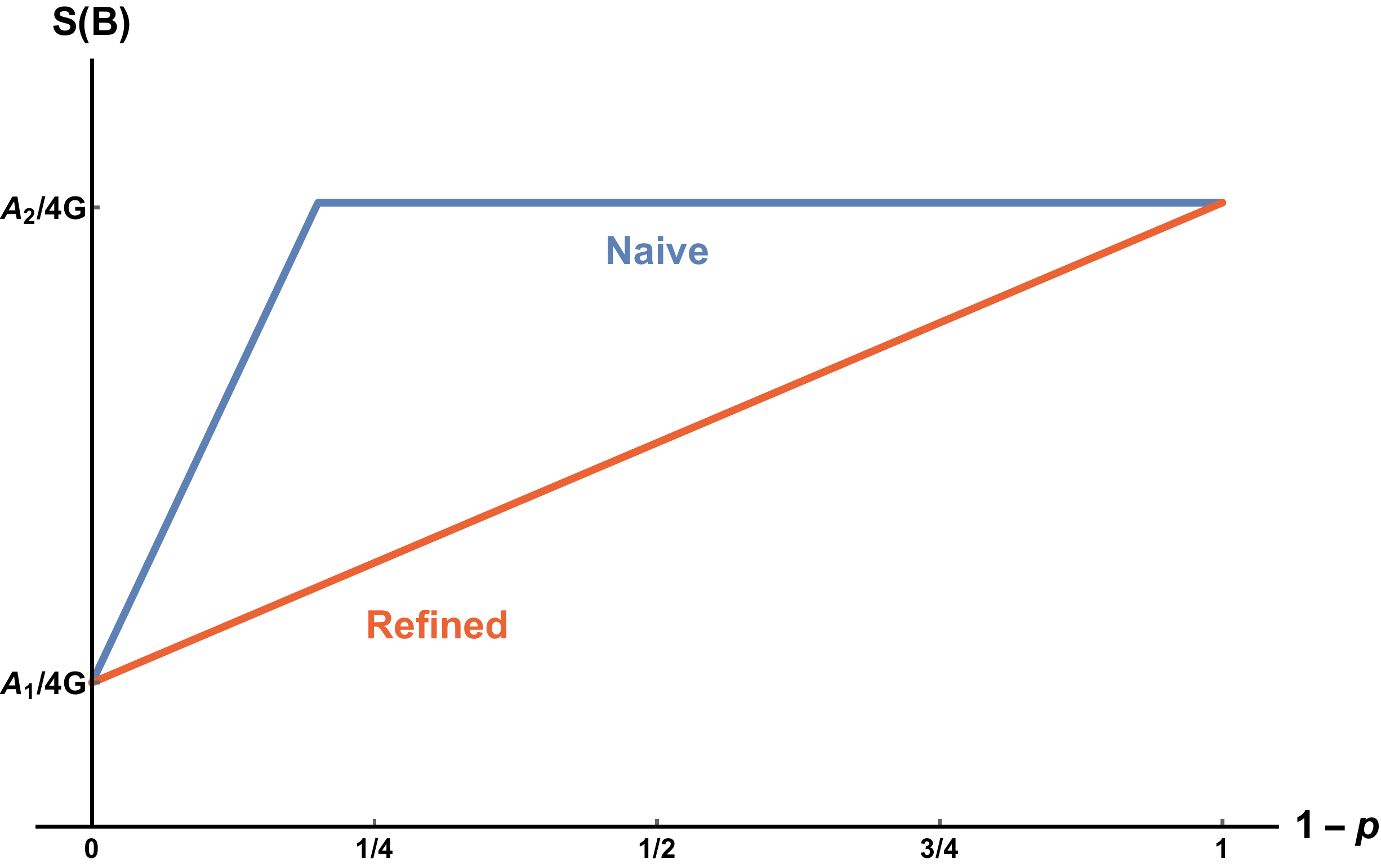}
    \caption{Comparison of the na\"{i}ve QES entropy to the correct, ``refined'' answer, for the state \eqref{eq:mixture_state} in the setup of Figure \ref{fig:dustball}. While the slope of the refined answer is controlled by $(A_2 - A_1)/4G$, the slope of the na\"{i}ve answer is controlled by $S_\mathrm{thermal} > (A_2-A_1)/4G$. 
    The na\"{i}ve answer is in general larger than the refined one by an $\mathcal{O}(1/G)$ amount.}
    \label{fig:naive_vs_refined}
\end{figure}

So in our example, it must be the case that the correct (or ``refined'') answer is
\begin{equation}
    \text{Mixture: }S(B)_\mathrm{refined} = p A_1/4G + (1-p) A_2/4G + \mathcal{O}(1)~. 
\end{equation}
This is generally different from the na\"{i}ve QES at leading order, $\mathcal{O}(1/G)$. See Figure \ref{fig:naive_vs_refined}.

Why does the na\"{i}ve QES fail for the mixture when (we claim) it gives the correct answer for both the pure state and the thermal state individually? Intuitively, this is because it is a mixture of states that are on different sides of the phase transition. But this notion is not very precise: the thermal state can itself be written as a mixture of states that are on either side of the transition (admittedly in this case one either needs a large number of states or some probabilities in the mixture to be very small), and yet it doesn't receive large corrections.

A more precise answer is that, unlike the pure state and the thermal state, the mixture of the two is not {\it perfectly compressible}. 
We say a state $\rho$ is perfectly compressible if we can throw away all but $e^{S(\rho)}$ of the states in its support without changing the state very much.
More precisely, there must exist another state $\sigma$ close to $\rho$  such that $\ln \mathrm{rank}(\sigma) = S(\rho)~+~\text{subleading}$. In the thermodynamic limit, thermal states are dominated by energies close to the saddle point energy, and are therefore perfectly compressible. A pure state has rank one and hence is trivially perfectly compressible. 

A general mixture of the two is not: any state $\sigma$ close to $\rho_\text{matter}$ will have almost the same rank as an approximation to the thermal state itself, because
\begin{align}
    \frac{1}{1-p} \sigma -\frac{p}{1-p} \ket{\psi}\bra{\psi} \approx \rho_\text{thermal}~.
\end{align}
In general the compressibility of a quantum state is characterized not by its von Neumann entropy, but by a quantity known as the smooth max-entropy $H_\text{max}^\varepsilon(\rho)$ \cite{renner2004universally, renner2005security}.
This is defined by the fact that you can throw away at most all but $e^{H^\varepsilon_\mathrm{max}(\rho)}$ of the states in the support of $\rho$, without changing $\rho$ very much, for small $\varepsilon$.
For thermal and pure states, the smooth max-entropy is approximately equal to the von Neumann entropy -- implying those states are perfectly compressible -- but in general it can be much larger. For example, in the mixture of a thermal and pure state, we have\footnote{In general, we take $\varepsilon$ to be polynomially small in $G$, though its exact size does not matter much. What's important is that it's not exponentially small, because we will need to assume $\ln \varepsilon \ll \mathcal{O}(1/G)$. Physically, this is closely related to the fact that bulk reconstruction necessarily has exponentially small errors \cite{Hayden:2018khn,Penington:2019kki}.}
\begin{align}
    H_\text{max}^\varepsilon(\rho) \approx \frac{S(\rho)}{1-p} ~.
\end{align}

To understand why this should be relevant to the QES prescription, we need to introduce the concept of entanglement wedge reconstruction (EWR) \cite{Czech:2012bh, Almheiri:2014lwa, Dong:2016eik}. This says that the bulk matter in the entanglement wedge is encoded in the boundary state on the boundary subregion $B$. 
``Encoded,'' here, means that the set of bulk operators local to the entanglement wedge has a representation on $B$ that acts faithfully on a ``code'' subspace of $B\overline{B}$.
It turns out EWR is implied by the QES prescription \cite{Dong:2016eik, Hayden:2018khn}.\footnote{More carefully, EWR (as usually defined) is possible if and only if the QES prescription holds for every state (pure or mixed) in the \emph{code subspace} of states for which the reconstruction is supposed to be valid \cite{Hayden:2018khn}.}

EWR (and hence QES) is deeply connected to compressibility. 
The intuition is that the number of degrees of freedom available in $B$ to describe the bulk state in region $b'$ is given by the difference in areas $(A_2 - A_1)/4G$ between the two extremal surfaces. If the bulk state in region $b'$ cannot be compressed into these degrees of freedom, EWR for region $b'$ cannot be possible, and hence the QES prescription, with $\gamma_1$ the minimal QES, cannot be valid, even if $\gamma_1$ is the surface with the smallest generalized entropy.
One of the main goals of this paper will be to formalize this intuition by showing that EWR can be reinterpreted as a particular information-theoretic task, called \emph{one-shot quantum state merging}, where Alice has to communicate a compressed version of a quantum state to Bob.

To make a precise statement, detailing how the QES prescription needs to be modified given the discussion above, it is helpful to write the na\"{i}ve QES prescription in the following form:
\begin{equation}\label{eq:naive_QES}
    S(B)_\text{na\"{i}ve} = 
    \begin{cases}
        A_1/4G + S(bb') ,& S(b'|b) \le \frac{A_2 - A_1}{4G} \\
        A_2/4G + S(b) ,& S(b'|b) \ge \frac{A_2 - A_1}{4G}~.
    \end{cases}
\end{equation}
The quantity $S(b'|b) = S(bb') - S(b)$ is the conditional von Neumann entropy. 

We will argue that this na\"{i}ve prescription only works when bulk states are perfectly compressible, because it implies the inclusion (or not) of $b'$ in the bulk entropy term only depends on the von Neumann entropy $S(b'|b)$. In reality, the information from $b'$ is only accessible in $B$ (and hence its entropy is only included in $S(B)$) when the quantum information in $b'$ can be compressed into $(A_2 - A_1)/4 \ln(2) G$ qubits. The relevant bulk entropy is therefore not the conditional von Neumann entropy $S(b'|b)$, but the conditional smooth max-entropy $H^\varepsilon_\mathrm{max}(b'|b)$. We'll explain $H^\varepsilon_\mathrm{max}(b'|b)$ in detail in Section \ref{sec:min_max_entropy} (along with the smooth conditional min-entropy $H^\varepsilon_\mathrm{min}(b'|b)$), but, roughly speaking, $H^\varepsilon_\mathrm{max}(b'|b)$ characterizes the compressibility of $b'$ when there is entanglement between $b'$ and $b$ (and $H^\varepsilon_\mathrm{min}(b'|b)$ is complementary to $H^\varepsilon_\mathrm{max}(b'|b)$). For all states, we have $H^\varepsilon_\mathrm{max}(b'|b) \geq S(b'|b) \geq H^\varepsilon_\mathrm{min}(b'|b)$.

A central result of this paper will be to refine the conditions for the QES prescription \eqref{eq:naive_QES}, replacing it by\footnote{We defer to Section \ref{sec:refined_prescription} for more general conditions, applicable to setups with more than two competing QES. }
\begin{equation}\label{eq:refined_QES}
    S(B)_\mathrm{refined} = 
    \begin{cases}
        A_1/4G + S(bb') ,& H^\varepsilon_\mathrm{max}(b'|b) \le \frac{A_2 - A_1}{4G} \\
        \text{(depends on details)} ,& H^\varepsilon_\mathrm{min}(b'|b) \le \frac{A_2 - A_1}{4G} \le H^\varepsilon_\mathrm{max}(b'|b) \\
        A_2/4G + S(b) ,& H^\varepsilon_\mathrm{min}(b'|b) \ge \frac{A_2 - A_1}{4G}~.
    \end{cases}
\end{equation}
We will not give a ``one answer fits all'' description of the middle regime; it does not admit one as convenient as the na\"{i}ve QES prescription.
The entropy there depends on the details of the bulk entanglement.
(That said, one can often estimate the answer by finding the average entropy of a set of constituent states, up to a Shannon term.) This refinement can be derived using replica trick calculations, and resolves the contradictions discussed above.

A heuristic way to understand the difference between these two prescriptions is that our refinement of the QES prescription recognizes that different parts of the wavefunction might be on different sides of the phase transition, whereas the na\"{i}ve prescription assumes that the entire state has to be on one side or the other. 
The min-/max-entropies appear because they describe the largest/smallest parts of the wavefunction respectively. 
If the smooth max-entropy is less than $(A_2 - A_1)/4G$, we can be sure that no significant part of the wavefunction has undergone the transition.
Similarly, if the smooth min-entropy is greater than $(A_2 - A_1)/4G$, we know that almost the entire wavefunction has undergone the transition. 
If they straddle $(A_2 - A_1)/4G$, then the entropy will depend on which parts of the wavefunction have crossed the transition.

\subsection*{Overview of paper}
The paper is organized as follows. 

In {\bf Section \ref{sec:contradictions}}, we illustrate the problem with a na\"{i}ve application of the QES prescription in more detail. We give several closely related examples of the na\"{i}ve QES prescription violating the bounds on the von Neumann entropy of mixtures of states. 

In {\bf Section \ref{sec:min_max_entropy}}, we review two quantities that are crucial for understanding the refined QES prescription: the smooth conditional min-entropy $H_\mathrm{min}^\varepsilon(A|B)$ and max-entropy $H_\mathrm{max}^\varepsilon(A|B)$. 

In {\bf Section \ref{sec:examples}}, we return to the simple examples from Sections \ref{sec:intro} and \ref{sec:contradictions} and carefully calculate their entropies using the replica trick. By avoiding using the Lewkowycz-Maldacena assumption, we find an answer that disagrees with the na\"{i}ve QES prescription but is consistent with the bounds on entropy of mixtures. This answer depends on the relative sizes of three quantities: the smooth conditional min- and max-entropy, and the difference in area of the two competing quantum extremal surfaces.

In {\bf Section \ref{sec:when_corrections}}, we present \emph{general} arguments that justify the conditions given in \eqref{eq:refined_QES} for the existence of large corrections to the na\"{i}ve QES prescription. We start by arguing this for so-called fixed-area states, and then argue that this extends to general holographic states, up to subleading corrections. A key tool is the connection between gravity calculations in fixed-area states and calculations in random tensor networks.

In {\bf Section \ref{sec:EWR}}, we update the conditions for entanglement wedge reconstruction (EWR), explaining how to generalize the results of Dong, Harlow, Wall \cite{Dong:2016eik} and Hayden, Penington \cite{Hayden:2018khn}, given this refinement of the QES prescription. These updated conditions clarify the relationship between EWR and a well-known quantum information task, one-shot quantum state merging. Our results demonstrate that EWR can be a maximally efficient form of one-shot quantum state merging, using zero-bits instead of the usual classical bits.

In {\bf Section \ref{sec:refined_prescription}}, we present a more general refinement of the QES prescription conditions, applying in situations where there are more than two competing extremal surfaces. To do so, we first introduce two interesting new physically relevant subregions of the bulk: the min-entanglement wedge (min-EW) and max-entanglement wedge (max-EW). The na\"{i}ve QES prescription applies if and only if the min-EW and max-EW are the same.

In {\bf Section \ref{sec:discussion}}, we mention some further implications of these results. In particular, we discuss how the smooth min- and max-entropies should be renormalized to get a UV-finite quantity.

\subsection*{Related work}
This paper has some technical overlap with the recent papers \cite{Dong:2020iod, Marolf:2020vsi}. 
They too find corrections to the QES prescription by carefully including more than one saddle in the replica trick, and they too use fixed-area states to simplify the calculation enough to do so.

There are three key differences between our corrections and theirs.
One, the corrections we discuss can be $\mathcal{O}(1/G)$, not just $\mathcal{O}(1/\sqrt{G})$.
Two, our corrections can exist for an $\mathcal{O}(1)$ range of $A_2 - A_1$, a window that does not vanish as $G \to 0$.  
Finally, our corrections do not arise from fluctuations in the geometry, but rather from the bulk state affecting the boundary entropy in a different way than previously expected. In particular, the corrections in \cite{Dong:2020iod, Marolf:2020vsi} are correctly computed by the expectation of the na\"{i}ve QES prescription over all the classical geometries that can be created by the fluctuations.

We also provide a different argument justifying the use of fixed-area states in lessons about general states.
Our argument also applies to the setups in \cite{Dong:2020iod, Marolf:2020vsi}, bounding the error in some of their assumptions.

\section{Mixtures and contradictions}\label{sec:contradictions}

In this section, we further illustrate the need for a careful, refined application of the QES prescription, first generalizing the prior example by adding entanglement, then discussing the importance of the refinement for black hole entropy and the unitarity of black hole evaporation.

\subsection*{Contradiction 1: Dustball}\label{sec:dustball_example}
Our first example is the dustball geometry, which was already presented in the introduction. However, we emphasize that many of the details, as presented there, were unimportant. The contradiction can easily be generalized to higher-dimensions, to mixtures where neither state is pure, or to mixtures of a larger number of states (so long as the number is not exponential in $1/G$).

We also note that we can easily adapt this example to find a similar contradiction where the state in $b'$ is highly entangled with the state in $b$, eventually illustrating the necessity of using \emph{conditional} min- and max-entropies in \eqref{eq:refined_QES}.

\begin{figure}
\centering
\includegraphics[width = \textwidth]{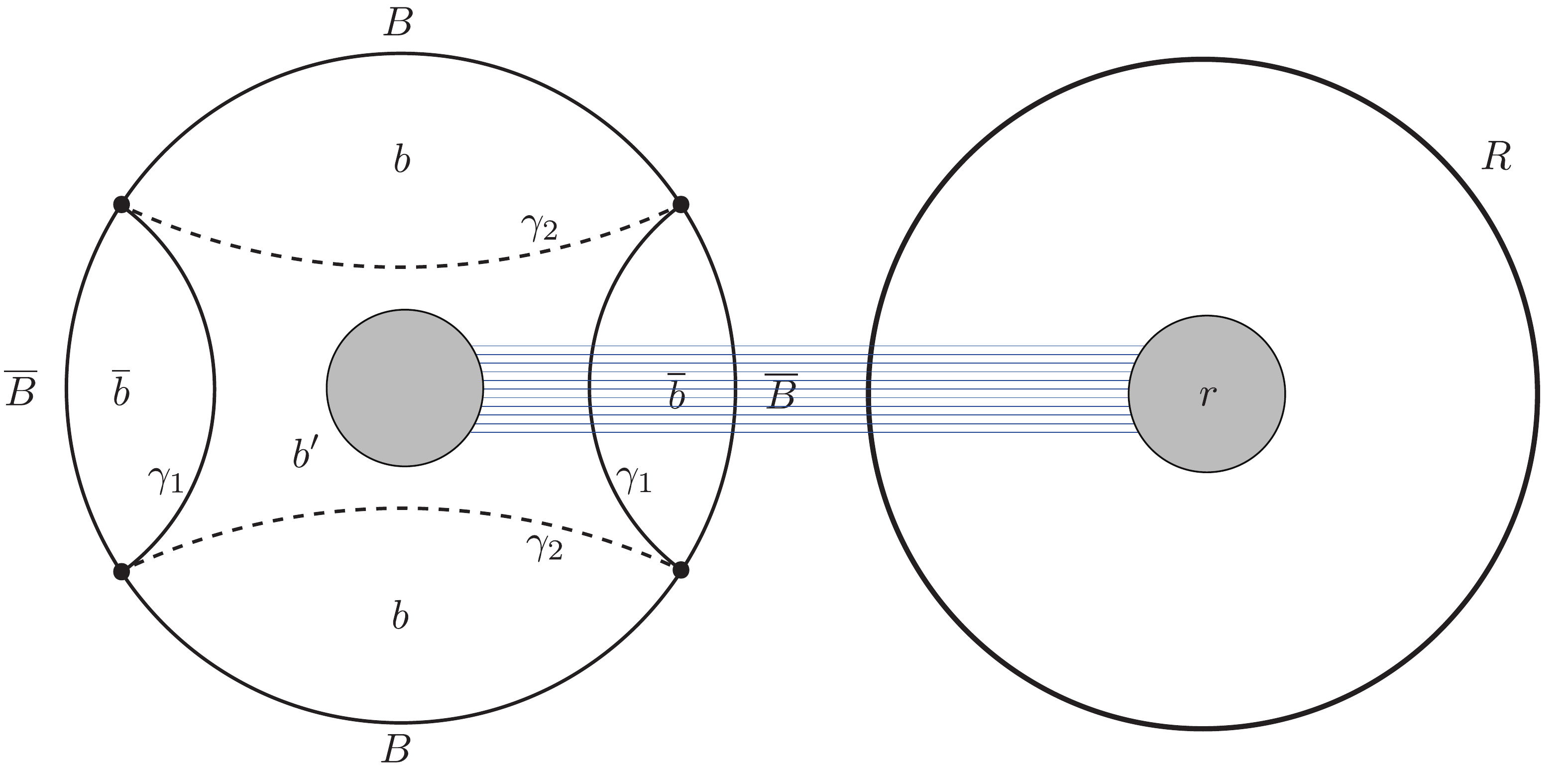}
\caption{Two entangled dustballs. Like Figure \ref{fig:dustball}, but now we consider the entropy of $\overline{B}R$, where $R$ is an entire extra copy of the boundary, dual to its own dustball. The two dustballs are in a mixture of entangled states, given by \eqref{eq:ent_dustball_state}. A na\"{i}ve application of the QES prescription gives the wrong answer for the entropy $S(\overline{B} R)$.} 
\label{fig:dustball_entangled}
\end{figure}

The first step is to consider a purification of the mixed dustball state, where the dustball is entangled with a second, identical dustball in a different bulk spacetime, as in Figure \ref{fig:dustball_entangled}. In other words, where the bulk state is
\begin{align}\label{eq:ent_dustball_state}
    \rho = p \bigg( \ket{\psi}_{b'}\bra{\psi}_{b'} \otimes \ket{\psi}_{r}\bra{\psi}_{r} \bigg) + (1-p) \ket{\Phi}\bra{\Phi}~,
\end{align}
and $\ket{\Phi}$ is a purification of $\rho_\text{thermal}$. From a boundary perspective, the mixed CFT state is purified by a second identical CFT, which we shall call the reference system $R$. Consistent with our notational conventions, we use $r$ to denote the bulk Hilbert space associated to the second CFT.

Introducing the reference system $R$ does not change the entropy $S(B)$. However, since the overall state is pure, we have $S(\overline B R) = S(B)$. The entropy $S(\overline B R)$ can also be calculated using the na\"{i}ve QES prescription. This time, the degrees of freedom in the homology region shared by both extremal surfaces (in this case $\overline b \otimes r$) are entangled with the degrees of freedom between the two surfaces (region $b'$ as before). Unsurprisingly, the na\"{i}ve QES prescription gives the same answers as before, and hence we again find a contradiction.

\subsection*{Contradiction 2: Black hole}\label{sec:bh_example}
\begin{figure}
\centering
\includegraphics[width = 0.5\textwidth]{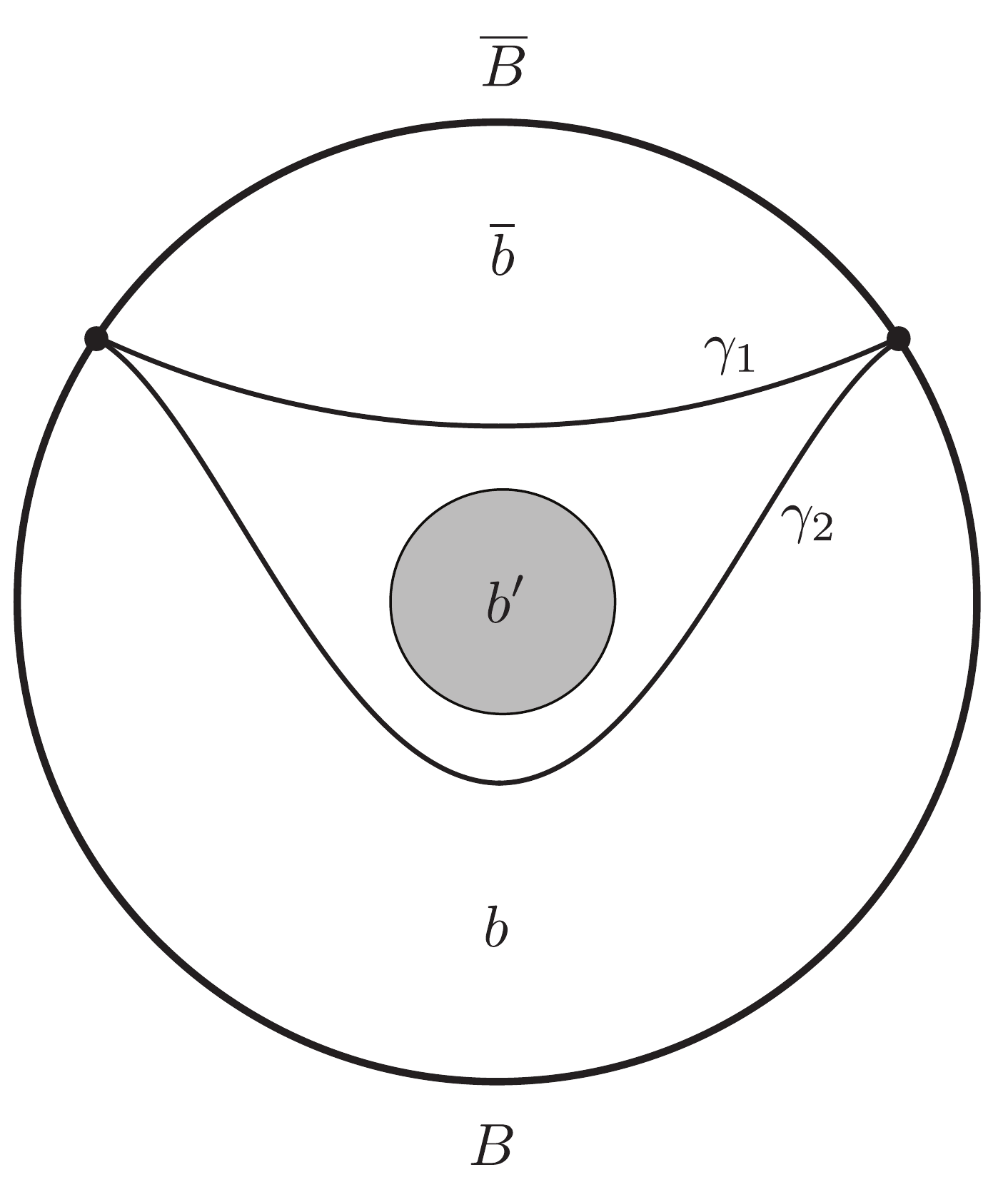}
\caption{Black hole setup in which we derive a contradiction from the QES prescription. Practically identical to the setup of Figure \ref{fig:dustball}, this setup replaces the dustball with a black hole, and its boundary regions $B$ and $\overline{B}$ are now connected.}
\label{fig:black_hole}
\end{figure}

A practically identical setup reaches the same contradiction, if we replace the dustballs with black holes \cite{Hayden:2018khn}.
See Figure \ref{fig:black_hole} for the setup with a mixed state black hole (though we also consider two entangled black holes, which would look very similar to Figure \ref{fig:dustball_entangled}). 

The advantage of this setup is that it's familiar to consider black holes with entropy growing with $1 / G$.
We can, for example, consider all states in an energy band of width $\Delta E \sim \mathcal{O}(1)$, centered on some high energy $E$. 
There are $e^{\mathcal{O}(1/G)}$ states in this subspace, and generic density matrices in this band are expected to be black holes. Additionally, unlike the dustball, we can also use a single interval (in AdS$_3$/CFT$_2$) for our boundary region, because the black hole geometry has extremal surfaces on either side of the black hole.

The big disadvantage -- indeed the reason we did not lead with this example -- is that black hole microstates seem somewhat mysterious. 
One might worry that mixtures of black hole states, like \eqref{eq:ent_dustball_state}, are secretly mixtures of classically distinct geometries, mixtures which people already expected to give averaged answers in the QES prescription. 
For example, a special case of the mixture of entangled black holes is the mixture of an energy eigenstate and the thermofield double (TFD) state.\footnote{Perhaps projected onto its dominant energy window, to fit into a finite-dimensional subspace.}
The TFD state is 
\begin{equation}
   \ket{\mathrm{TFD}} = \sum_i e^{-\beta E_i / 2} \ket{E_i}_{b'} \ket{E_i}_{r}~,  
\end{equation}
for some inverse-temperature $\beta$ and energy eigenstates $\ket{E_i}$.
Two black holes entangled like this are connected by a wormhole \cite{Maldacena:2001kr}, and hence there is a nontrivial homology constraint. This is very different from a factorized energy eigenstate, which has trivial homology constraints. The mixture of the two, 
\begin{equation}
   \text{special case:}~~~~ \rho_{b'r} = p \bigg( \ket{E}_{b'}\bra{E}_{b'} \otimes \ket{E}_{r}\bra{E}_{r} \bigg)  + (1-p) \ket{\mathrm{TFD}}\bra{\mathrm{TFD}}~,
\end{equation}
must therefore have the QES prescription applied to it with care, since it is a mixture of two distinct classical geometries.
There is a history of speculating that -- for this state -- the na\"{i}ve QES prescription gives an $S(B)$ that is indeed the average entropy $(1-p) A / 4G$ (see e.g. \cite{Almheiri:2016blp}).
The argument was that the area operator is linear, and so its expectation value in this  mixture of states must be the average of its expectation value in each. 

While that argument is fine, we emphasize that it does not explain away the contradictions we are pointing out. 
This can be made sharp using the insights from quantum error-correction in \cite{Harlow:2016vwg}.\footnote{Another way to see this is from the UV-finiteness of the generalized entropy $A/4G + S$, which implies that $G$ acts as a counterterm for the divergences in the von Neumann entropy, suggesting that as one flows to different energy scales, one changes which degrees of freedom contribute to $A/4G$ rather than $S$. We thank Netta Engelhardt for emphasizing this to us.}
From the quantum error-correction point of view, it is not necessary to count the black hole entropy as part of the ``area.'' 
A choice of code subspace that includes the black hole microstates will regard the black hole entropy as part of the matter entropy.
This would be inconsistent, giving an answer that does not equal $S(B) = (1-p)A/4G$, if the entropy of a mixture of black hole states is given by the na\"{i}ve QES answer.

\subsection*{Contradiction 3: Hawking radiation}\label{sec:radiation_example}
Our final contradiction appears in evaporating black holes.
It was shown last year that using the QES prescription allows a gravity calculation of the decrease in entropy of Hawking radiation after the Page time \cite{Penington:2019npb, Almheiri:2019psf}. This goes a long way towards resolving the famous black hole information paradox.
However, there's a lingering paradox in those calculations, if the QES prescription is applied in the na\"{i}ve way. 
We demonstrate this now.

Consider a post-Page time black hole $B$, having already emitted radiation $R$ in state $\rho_R$. 
Introduce an ancilla qubit $q$, and entangle it with $R$ in the following way.
First, put $q$ in a superposition
\begin{equation}
    \sqrt{1-p}\ket{0}_q + \sqrt{p}\ket{1}_q~.
\end{equation}
Then, perform a joint operation on $qR$, measuring the radiation if $q$ is in state $\ket{1}$, and otherwise doing nothing.
This measurement need not be complicated -- a factorized measurement on each Hawking photon is simple and will suffice.
Given measured state $\ket{\psi}_R$, the reduced state of the radiation becomes
\begin{align}\label{eq:measured_rad}
    \widetilde{\rho}_{R} = p \ket{\psi}_R \bra{\psi}_R + (1-p) \rho_R~.
\end{align}
Assuming that the evaporating black hole was following the Page curve, the entropy of the radiation, at leading order, will then be $(1-p) A_\text{hor}/4G ~(\text{$+$ subleading})$, where $A_\text{hor}$ is the area of the black hole horizon.

What does the na\"{i}ve QES prescription say that the entropy will be? As long as we don't measure the most recent Hawking quanta to escape into $R$, the locations of the quantum extremal surfaces will be unchanged. The generalized entropy of the empty surface will be $(1-p) S_\text{rad}$, where $S_\text{rad}$ is the semiclassical, thermal entropy of the radiation. The generalized entropy of the nonempty surface near the horizon will be $A_\text{hor}/4G$ as before.

As with our previous contradictions, this is just incorrect (assuming unitarity), even at leading order. The na\"{i}ve QES prescription is giving an answer that is qualitatively just as wrong as the Hawking, information-loss answer. Indeed, for small values of $(1-p)$, the na\"{i}ve QES prescription answer and the Hawking answer are the same.

A very similar contradiction can be created using purely unitary processes, without any measurements. One just creates an ancilla system $A$, in the state $\ket{0}$, that is a copy of the radiation Hilbert space $R$. Then one applies a conditional swap operator (which again factorizes into a product of local interactions) that swaps $A$ and $R$ if and only if the qubit $q$ is in the state $\ket{1}$. Assuming unitarity, the form of the reduced state on $R$ will again be given by \eqref{eq:measured_rad}. The generalized entropy of the empty surface will again be $(1-p) S_\text{rad}$, while the generalized entropy of the nonempty surface will be $A_\text{hor}/4G + p S_\text{rad}$. Again, we find a contradiction with unitarity at leading order.

\subsection*{Summary}

This section showed classes of examples in which a na\"{i}ve application of the QES prescription gets the entropy wrong at leading order. 
In Section \ref{sec:examples} we do a careful calculation that gets the entropy in these examples right, and then in Section \ref{sec:when_corrections} we describe more generally when and why there are corrections. 
First, however, we need to introduce two quantities that will characterize when the na\"{i}ve QES prescription receives these large corrections. 

\section{Smooth min- and max-entropies}\label{sec:min_max_entropy}

While na\"{i}vely the QES prescription compares only von Neumann entropies to areas, we will find that a more careful prescription compares to the area two other quantities: the smooth conditional min-entropy and smooth conditional max-entropy.
These information-theoretic quantities have historically found use in ``one-shot'' protocols, settings in which only a single copy of a quantum state is used or transferred. 

We explain these quantities now, and in all future sections refer to them heavily. 
We start with the simplest version, the classical min- and max-entropy, and gradually work up to what we really want, the (quantum) smooth conditional min- and max-entropy. 

\subsection*{Non-conditional versions}
To introduce the idea of one-shot entropies, it is helpful to temporarily forget about quantum mechanics and simply consider classical probability distributions.

Let us first recall the information-theoretic role of the Shannon entropy $S(p)$ of a classical probability distribution $p(x)$ (analogous to the von Neumann entropy in quantum mechanics).
Imagine you randomly sample from a large number of copies $n$ of the probability distribution, getting outcomes $\{x_i\}$. You, Alice, now want to communicate those outcomes to your friend Bob.

How much information do you need to send to Bob to do this? To always be successful, for any $\{x_i\}$, you need to send at least $n \log_2 d$ bits, where $d$ is the number of values $x$ can take with nonzero probability. However, if you only insist that the communication succeed with high probability (i.e. succeed for a variety of possible outcomes $\{x_i\}$ that collectively have probability $p > 1- \varepsilon$ for some small $\varepsilon$), the task becomes much easier. One can show that, at leading order for large $n$, you only need to send $n\, S(p) / \ln(2)$ bits. 
Essentially, this comes from the law of large numbers ensuring that `typical' samples from many copies of the distribution have a probability $p$ such that\footnote{The notation ``$o(n)$'' represents terms subleading to $n$, vanishing in $\lim_{n\to\infty} o(n)/n~$.}
\begin{align}
    \ln p = n\, \langle \ln p(x) \rangle_{p(x)} + o(n) = n\, S(p) +o(n)~.
\end{align}
Hence, you and Bob simply need to agree on a code, in which the $n S(p) / \ln(2)$ bits you send tell Bob which of the $e^{n S(p) + o(n)}$ ``typical strings'' you sampled.

The story in quantum mechanics is very similar: given any density matrix $\rho$, we can project $\rho^{\otimes n}$ into a `code subspace', while only changing the state a small amount. This code subspace is just built out of products of states in the Schmidt decomposition of $\rho$ that have typical entropy, as in the classical case. Such states dominate the Schmidt decomposition of $\rho^{\otimes n}$ at large $n$. 

The number of qubits needed for the code subspace grows, in the limit of large $n$, as $n S(\rho) / \ln(2)$, where $S(\rho)$ is the von Neumann entropy. If Alice has a pure state randomly sampled from $\rho^{\otimes n}$, she can therefore communicate that state to Bob with high success probability, just by sending $n S(\rho) / \ln(2)$ qubits.

However, both in classical probability and in quantum mechanics, we often (perhaps even typically) encounter situations where we don't have a large number of copies of a single density matrix or distribution. Instead, we only have a single state or distribution, which may still be very large in size. An example, of course, is holography. 
In the limit $G \to 0$, the boundary Hilbert space dimension blows up exponentially, but this does not mean we have a large number of independent copies of the same state.

In this `one-shot' setting, the von Neumann entropy does not have an important operational role.\footnote{For an exception to this general principle, see \cite{boes2019neumann}.} It is therefore somewhat surprising that the von Neumann entropy has been playing such a central role in holography, for example in determining whether entanglement wedge reconstruction is possible! 
As we shall see, the resolution is that the real quantities that are important in holography are smooth max- and min-entropies, which do have a natural operational interpretation in one-shot quantum Shannon theory. It just so happens that these `one-shot entropies' have been approximately equal to the von Neumann entropy, in most of the situations that have been considered in the literature until now.

Suppose we consider the same task as above (sending the outcome of sampling a probability distribution from Alice to Bob), but now we only sample from a single copy of the distribution. How many bits do we need to send to communicate the outcome with high probability? 
We need to be able to send a distinct message for each outcome that we want to be successfully communicated, and our success probability is maximized by choosing the outcomes with the highest probability of occurring. 
So the number of bits that need to be sent is $\log_2 N^{(\varepsilon)}$ where $N^{(\varepsilon)}$ is the smallest integer such that
\begin{align}
    \sum_{i=1}^{N^{(\varepsilon)}} p_i > 1 - \varepsilon~,
\end{align}
with the probabilities $p_i$ ordered from largest to smallest.

Again, there is an obvious quantum mechanical generalization, which gives the minimum number of qubits needed to send a quantum state, sampled from a single copy of a density matrix $\rho$, from Alice to Bob. This is given by
\begin{align} \label{eq:smoothrenyizero}
    H_0^\varepsilon(\rho) = \inf_{\lVert\widetilde \rho - \rho \rVert_1 \le \varepsilon} H_0(\widetilde \rho) = \inf_{\lVert\widetilde \rho - \rho \rVert_1 \le \varepsilon} \ln(\text{Rank}(\widetilde \rho))~.
\end{align}
Let's unpack this for a moment. We first defined the R\'{e}nyi 0-entropy (also known as the Hartley entropy) as
\begin{align}
H_0(\rho) = \ln \text{Rank}(\rho) = \lim_{\alpha \to 0} \frac{1}{1 - \alpha} \ln \tr(\rho^\alpha)~,
\end{align}
and then we `smoothed' this quantity by minimizing it over all $\widetilde \rho$ close to $\rho$ (which in this case just meant throwing away small eigenvalues). 
We measured this distance with the trace distance, or Schatten 1-norm, $||X||_1 = \tr\left( \sqrt{X^\dagger X} \right)$.

In fact, \eqref{eq:smoothrenyizero} is the original definition of the smooth max-entropy \cite{renner2005security}. It turns out however that $H_0(\rho)$ can be replaced \cite{renner2004smooth} by the R\'{e}nyi entropy
\begin{align}
    H_\alpha(\rho) = \frac{1}{1 - \alpha} \ln \tr (\rho^\alpha)~,
\end{align}
for \emph{any} $\alpha < 1$, while only changing the smooth entropy by a small amount. Specifically,
\begin{align} \label{eq:allalphasame}
    H_0^\varepsilon(\rho) \geq H_\alpha^\varepsilon (\rho) \geq H_0^{2 \varepsilon}(\rho) - \frac{1}{1 - \alpha} \ln(1/\varepsilon)~.
\end{align}
As we shall see below, $H_{1/2}(\rho)$ generalizes better to conditional entropies. It is therefore conventionally used in the modern definition of the smooth max-entropy \cite{konig2009operational},
\begin{align}\label{eq:smooth_max_ent}
    H_\text{max}^\varepsilon(\rho) = \inf_{\widetilde \rho \in \mathcal{B}^\varepsilon(\rho)} H_{1/2}(\widetilde \rho)~,
\end{align}
where we are taking an infimum over all states $\widetilde \rho$ within an $\varepsilon$-ball $\mathcal{B}^\varepsilon(\rho)$ of $\rho$.\footnote{For technical reasons, the distance measure used to define this $\varepsilon$-ball is conventionally the \emph{purified distance}, defined as the minimum trace distance between purifications of $\rho$ and $\widetilde \rho$. However, again, any reasonable distance measure will work fine (up to unimportant changes in the scaling of $\varepsilon$).}

In summary, the number of qubits needed to send Bob your quantum state with high fidelity, if you only sample the distribution one time, is the smooth max-entropy \eqref{eq:smooth_max_ent}, up to the factor of $\ln(2)$. 
The smooth max-entropy is always greater than or equal to the von Neumann entropy; sending many samples from the distribution can only improve the efficiency of the communication rate.

We can also define a complementary quantity, the \emph{smooth min-entropy}, as
\begin{equation}
\begin{split}
    H_\text{min}^\varepsilon(\rho) = \sup_{\widetilde \rho \in \mathcal{B}^\varepsilon(\rho)} H_{\infty}(\widetilde \rho)~.
\end{split}
\end{equation}
Again, $H_{\infty}(\widetilde \rho)$ could be replaced by $H_{\alpha}(\widetilde \rho)$ for any $\alpha > 1$ while changing the definition by at most $\mathcal{O}\big(\ln(1/\varepsilon)\big)$. 
It's operational interpretation is less intuitive than the smooth max-entropy, so we motivate it simply by its relationship to the conditional max-entropy, as we'll explain. 
Note that the smooth min-entropy is always less than or equal to the von Neumann entropy.

Together, these two quantities establish upper and lower bounds on the confidence interval for the value of (non-negligible) eigenvalues of $\rho$.
The smooth max-entropy encodes the size of the \emph{smallest} eigenvalues in the density matrix (which cannot be thrown away with small error), while the smooth min-entropy captures the size of the \emph{largest} eigenvalues (that cannot be thrown away).

If the spectrum is close to flat (i.e. is dominated by a small range of eigenvalues) then the smooth min- and max- entropies will be close to the von Neumann entropy (which characterizes the \emph{average} (log-)eigenvalue). In particular, thanks to the law of large numbers, this happens at leading order in $n$ when you take a large number of copies $\rho^{\otimes n}$ of a state $\rho$. This explains the importance of the von Neumann in traditional asymptotic quantum Shannon theory, which deals with exactly this limit.

It is also what has led to the success (so far) of the na\"{i}ve QES prescription; it's been used for bulk states with an (approximately) flat spectrum, where the smooth min- and max-entropy are roughly the same as the von Neumann entropy.

\subsection*{Conditional versions}

The most general quantities we will need are the smooth \emph{conditional} min- and max-entropies, which generalize the conditional von Neumann entropy. Unfortunately, the definition of these quantities is somewhat more technical, and somewhat less intuitive, than their unconditional counterparts.

The operational spirit of these quantities is the following. Let us return to the example in which Alice is trying to send a quantum state on $A$ to Bob. However, now the state is sampled from a density matrix $\rho_{AB}$, where subsystem $B$ is already held by Bob and the two subsystems may be entangled. Can this entanglement help Alice send her part of the state to Bob? It can! For a particular version of this task, called quantum state merging \cite{QI_can_be_negative}, the number of qubits that need to be sent from Alice to Bob is the smooth conditional max-entropy $H^\varepsilon_\mathrm{max}(A|B)$, which is generally less than $H^\varepsilon_\mathrm{max}(A)$. We discuss quantum state merging in detail in Section \ref{sec:EWR}.

Here are the technical definitions.
The conditional von Neumann entropy, which the conditional min- and max-entropy generalize, is normally defined as
\begin{align} \label{eq:conddiff}
    S(A|B) = S(AB) - S(B)~.
\end{align}
However, this definition does not generalize well to smooth entropies. Instead, our starting point will be a definition of the conditional entropy in terms of the relative entropy as
\begin{align} \label{eq:condrel}
    S(A|B) = -\min_{\sigma_B} D(\rho_{AB} | \mathbb{1}_A \otimes \sigma_B)~.
\end{align}
To see that this is equivalent to \eqref{eq:conddiff}, note that 
\begin{equation}
\begin{split}
    D(\rho_{AB} | \mathbb{1}_A \otimes \sigma_B) &= \tr\left(\rho_{AB} \ln \rho_{AB}\right) - \tr\left(\rho_{B} \ln \sigma_{B}\right) \\
    &= -S(AB) + S(B) + D(\rho_B | \sigma_B) \\
    &\geq -S(AB) + S(B)
\end{split}
\end{equation}
with equality if $\sigma_B = \rho_B$.

\subsubsection*{Smooth conditional min-entropy}
To generalize \eqref{eq:condrel} to a smooth conditional min-entropy, we use the fact that there is a unique quantum generalization of the classical R\'{e}nyi max-divergence $D_\infty( \rho| \sigma)$ which satisfies the data-processing inequality and additivity. This is given by
\begin{align}
    D_\infty( \rho| \sigma) = \inf \{ \lambda : \rho \leq e^\lambda \sigma \}~.
\end{align}
In words, the quantum max divergence of $\rho$ relative to $\sigma$ is the smallest number $\lambda$ such that $e^\lambda \sigma - \rho$ is positive semi-definite.

We then define the conditional min-entropy as
\begin{align}
    H_\text{min} (A| B)_\rho = -\min_{\sigma_B} D_\infty (\rho_{AB} | \mathbb{1}_A \otimes \sigma_B)~,
\end{align}
and the smooth conditional min-entropy as
\begin{align}
    H^\varepsilon_\text{min} (A| B)_\rho = \sup_{\widetilde \rho \in \mathcal{B}^\varepsilon(\rho)} H_\text{min} (A| B)_{\widetilde \rho}~.
\end{align}
We can gain some intuition by rewriting the conditional min-entropy as \cite{watrous_minmax_notes}
\begin{align}
    H^\varepsilon_\text{min}(A|B) = \inf_{\widetilde \rho \in \mathcal{B}^\varepsilon(\rho)} \left( -\ln |A| - \sup_{\Phi_B} \ln F((\mathbb{1}_A \otimes \Phi_B)\widetilde{\rho}_{AB}, \tau_{AA'})\right)~,
\end{align}
where $F$ is the fidelity $F(\rho,\tau)=\left(\tr\sqrt{\rho^{1/2}\tau\rho^{1/2}}\right)^2$, $\tau_{AA'}$ is a maximally entangled state on two copies of $A$, and $\Phi_B$ is a completely positive trace preserving map from $B$ to $A'$.
This illustrates that $H^\varepsilon_\mathrm{min}(A|B)$, in a sense, quantifies how close $\rho_{AB}$ is to a maximally entangled state, equaling its minimum $-\ln |A|$ when $A$ is maximally entangled with $B$, and its maximum $\ln|A|$ when it's completely decoupled. 

\subsubsection*{Smooth conditional max-entropy}
The smooth conditional max-entropy is most cleanly defined as a complement to the smooth conditional min-entropy. 
Recall that for any purification $\ket{\rho}_{ABC}$ of $\rho_{AB}$, we have
\begin{align}
    S(A|B) = - S(A|C)~.
\end{align}
A generalization of this equality will define the smooth conditional max-entropy.
One can show that
\begin{align}
    - H_\text{min}(A|C) = \ln|A| + \sup_{\sigma_B} \ln F\left(\rho_{AB}, \frac{\mathbb{1}_A}{|A|} \otimes \sigma_B \right)~.
\end{align}
This right hand side is a natural candidate definition for $H_\mathrm{max}(A|B)$. 
We can test this by considering the special case where subsystem $B$ is trivial (i.e. the state on $AC$ is pure). We then have
\begin{align}
      - H_\text{min}(A|C) = 2 \ln \tr \rho^{1/2} = H_{1/2}(A)~.
\end{align}
Recall that we previously used $H_{1/2}(A)$ in our formal definition of the smooth max-entropy. 
It is indeed therefore natural to define the smooth conditional max-entropy $H_\text{max}^\varepsilon(A|B)$ as
\begin{align}
    H_\text{max}^\varepsilon(A|B) = - H_\text{min}^\varepsilon(A|C) = \inf_{\widetilde \rho \in \mathcal{B}^\varepsilon(\rho)} \left( \ln |A| + \sup_{\sigma_B} \ln F(\widetilde \rho_{AB}, \frac{\mathbb{1}_A}{|A|} \otimes \sigma_B)\right).
\end{align}
This definition provides some intuition for the smooth conditional max-entropy,
as quantifying, in a sense, how close $\rho_{AB}$ is to a decoupled state $\mathbb{1}_A / |A| \otimes \sigma_B$, equaling $\ln|A|$ when $A$ is completely decoupled from $B$, and $-\ln|A|$ when it is maximally entangled with $B$. 

\section{Replica trick calculations}\label{sec:examples}

The replica trick is a standard technique for computing the von Neumann entropy $S(B)$, based on interpretting $\tr(\rho_B^n)$ as a certain observable in $n$ copies (or replicas) of the system \cite{calabrese2004entanglement}. 
We first illustrate the standard technique for computing $\tr(\rho^n)$ holographically -- using the saddle point approximation and analytically continuing the dominant saddle -- which is well-known to give the na\"{i}ve QES prescription \cite{Lewkowycz:2013nqa, Faulkner:2013ana, Dong:2017xht, Penington:2019kki,Almheiri:2019qdq}.
We then do a more careful calculation, where we analytically continue a sum over an entire family of saddles. To make this calculation analytically tractable, we make use of the fixed-area states of \cite{Akers:2018fow, Dong:2018seb}. This more careful calculation gives results that differ from the na\"{i}ve QES answer and avoid any contradictions.

\subsection{Replica trick in holography}
Given a state $\rho_{B\overline{B}}$, the goal is to compute 
\begin{equation}
    S(B) \equiv - \tr\left( \rho_B \ln \rho_B \right) = \lim_{n \to 1} \frac{1}{1-n}\ln \tr\left( \rho_B^n \right)~.
\end{equation}
The last equality is useful because $\tr(\rho_B^n)$ can be computed using a path integral.
Schematically, the Euclidean path integral preparing $\rho_{B\overline{B}}$ looks like,
\begin{equation}\label{eq:rhoBBbar_CFT}
    \includegraphics[width = 0.9\textwidth]{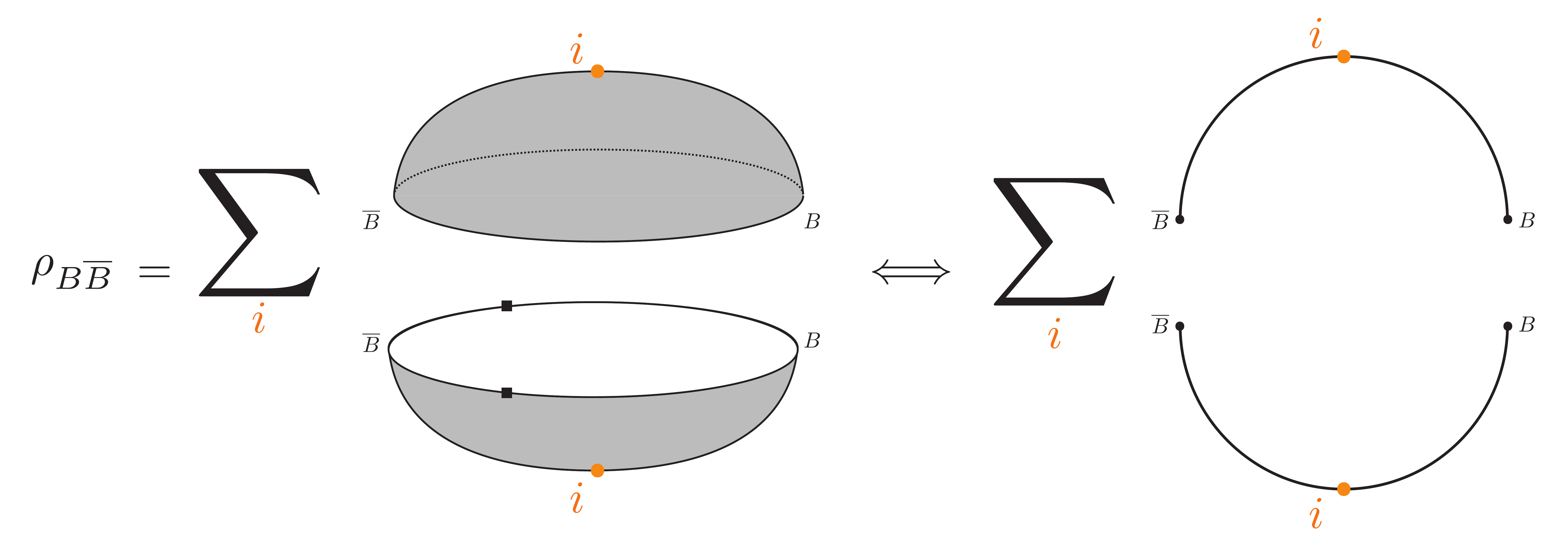}.
\end{equation}
The final picture is just a more schematic version of the first. 
The orange dots, with index $i$, label a basis of states, prepared by different boundary conditions, that are summed over. This represents the fact that a general density matrix is not just a product of a ket and a bra, but a sum of such products. 
In future diagrams, we suppress this index and the sum.

To construct the reduced density matrix $\rho_B$, we glue together $\overline{B}$ in the bras and kets:
\begin{equation}\label{eq:rhoB_CFT}
    \includegraphics[width = 0.9\textwidth]{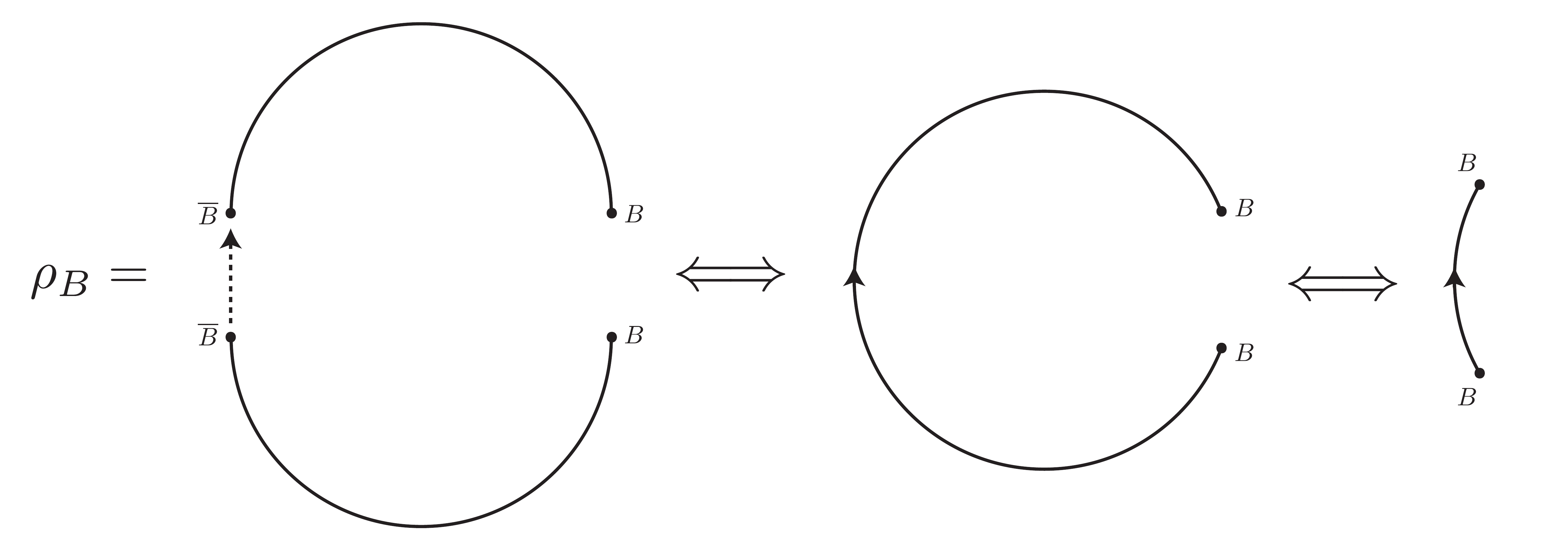}
\end{equation}
Then the path integral for e.g. $\tr\left(\rho_B^3 \right)$ involves gluing together the different copies of $B$ ``cyclically'' as 
\begin{equation}\label{eq:tr_rhoB3_CFT}
    \includegraphics[width = 0.4\textwidth]{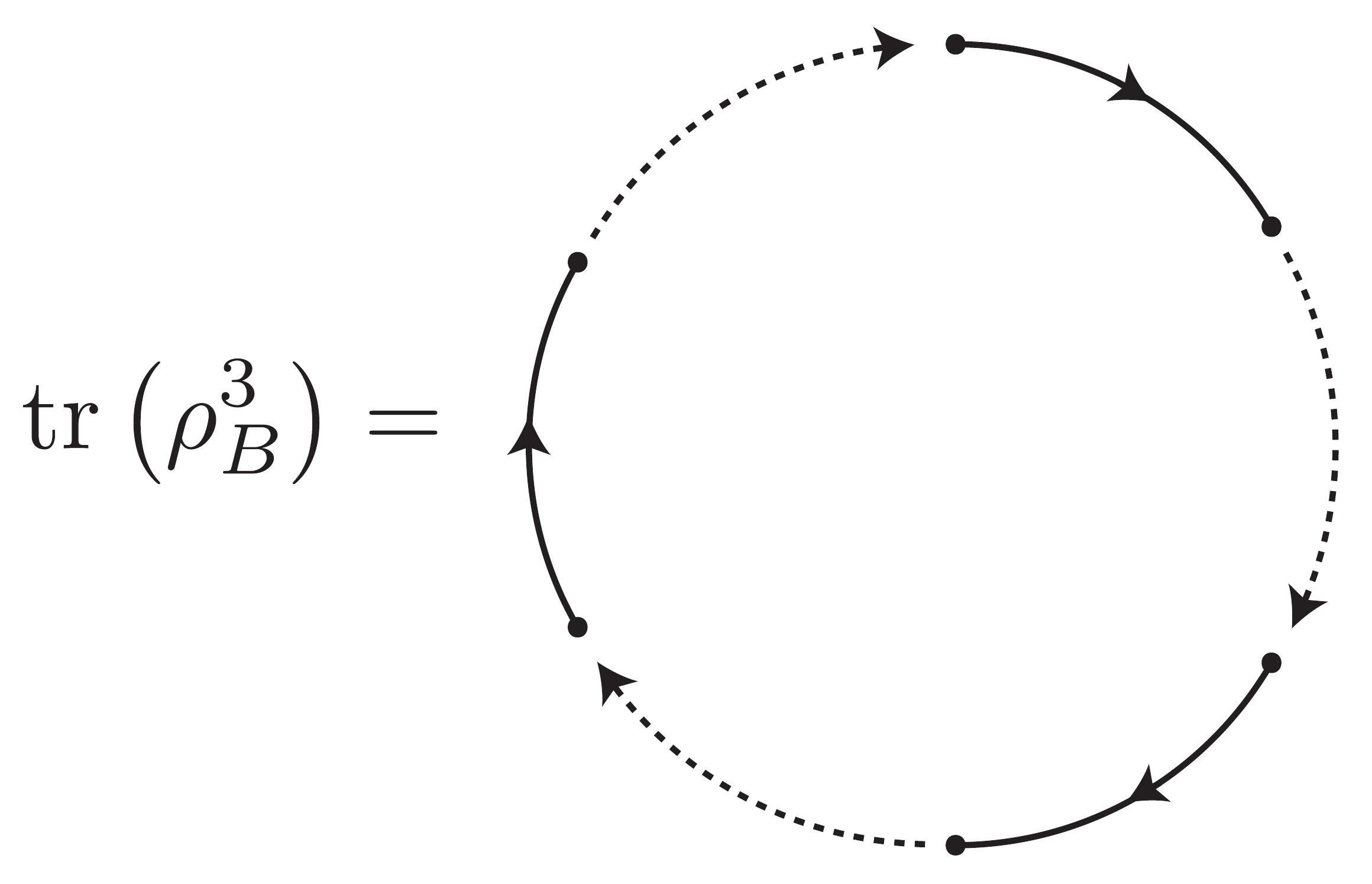}
\end{equation}
These boundary conditions can be applied with a ``twist operator'' $\tau$, which acts on $n$ copies of the $B$ Hilbert space to cyclically permute the state on each copy:
\begin{equation}
    \tr\left(\rho_B^n\right) = \tr\left(\rho_{B\overline{B}}^{\otimes n} \tau\right)~.
\end{equation}
We call this $n$-replica geometry $\mathcal{M}_n$. By evaluating the path integral on $\mathcal{M}_n$ for arbitrary $n$, and analytically continuing the answer to the limit $n \to 1$, we can compute the entanglement entropy.

We can map this boundary path integral to a bulk computation using the AdS/CFT dictionary, 
\begin{equation}
    \tr\left(\rho_{B\overline{B}}^{\otimes n}\tau\right) = \frac{Z_{B,n}}{Z_{B,1}^n}~,
\end{equation}
where $Z_{B,n}$ is the bulk partition function, defined by integating over all bulk geometries with boundary $\mathcal{M}_n$. In the semiclassical limit, this can be approximated by a sum over classical saddles. Crucially,  the saddle-point geometries are not simply $n$ copies of the original geometry glued together.
They are whatever the equations of motion provide, given that boundary data.

Partially for this reason, and partially because the number of saddles depends on $n$, this sum over saddles is generally too difficult to evaluate, let alone analytically continue.
So historically, the following trick was used \cite{Lewkowycz:2013nqa}.
{\it Assume} that a replica-symmetric configuration dominates the sum, and that all other contributions to the path integral can be ignored, such that 
\begin{equation}
    Z_{B,n} \approx e^{-I_\mathrm{grav}[g_{s,n}]}Z_{B,n}^\mathrm{mat}[g_{s,n}]~,
\end{equation}
where $g_{s,n}$ is the saddle-point metric, $I_\mathrm{grav}[g_{s,n}]$ is the gravitational action, and $Z_{B,n}^\mathrm{mat}[g_{s,n}]$ is the matter partition function on this semiclassical background. We shall call this \emph{the Lewkowycz--Maldacena (LM) assumption}.
Because the saddle is replica-symmetric, we can equivalently consider the quotient of the saddle-point geometry by the $Z_n$ replica symmetry. This is also a solution to the equations of motion, except at the fixed-points of the $Z_n$ action, where there is a conical singularity with opening angle $2 \pi/n$. 

We can now analytically continue the quotiented geometry to non-integer values of $n$. In particular, in the limit $n \to 1$, the geometry approaches the original \emph{unbackreacted} geometry, with a weak conical singularity at the $Z_n$ fixed-points. The entanglement entropy ends up being the generalized entropy of the $Z_n$ fixed-points \cite{Faulkner:2013ana}, which is forced to be a quantum extremal surface by the equations of motion \cite{Dong:2017xht,Almheiri:2019qdq,Penington:2019kki}. The dominant semiclassical saddle is the one where the QES has the smallest generalized entropy, leading to the na\"{i}ve QES prescription.

Since this traditional derivation reaches a conclusion that we have shown is contradictory, the obvious next step is to do the replica trick more carefully, without the weak link of the LM assumption.
This requires we introduce some other simplifying trick to analytically continue the sum over saddles.
This trick will involve the use of fixed-area states, which we now explain.

\subsection{Fixed-area states and their use}

The fixed-area states of \cite{Akers:2018fow, Dong:2018seb} are (approximate) eigenstates of certain area operators. 
To define such a state, consider the Euclidean path integral that prepares a particular bulk geometry, then insert into that path integral a delta function that fixes the area of some gauge-invariantly defined surfaces.\footnote{Of course, in reality the area cannot be measured exactly. Instead, we must specify to what precision we fix the area. It will be sufficient to fix it to be within a  window that is polynomially small in $G$.} We might physically prepare such a state by measuring the area of these surfaces. Saddle-points of this restricted path integral must satisfy the bulk equations of motion everywhere except at the fixed-area surfaces, where they may have a conical singularity. This is because the conical deficit angle is conjugate to the area operator and is therefore {\it undetermined} in fixed-area states, due to the uncertainty principle.\footnote{This uncertainty in the geometry makes it impossible to fix the area of two overlapping surfaces -- but there is no problem simultaneously fixing the area of two surfaces that do not cross (and therefore have commuting area operators) \cite{Bao:2018pvs, Bao:2019fpq}.}

\subsubsection*{Replica trick for fixed-area states}

Consider a state $\rho_{B\overline{B}}$ with two fixed-area surfaces, $\gamma_1$ and $\gamma_2$.
We depict its path integral as
\begin{equation}
    \includegraphics[width = 0.9\textwidth]{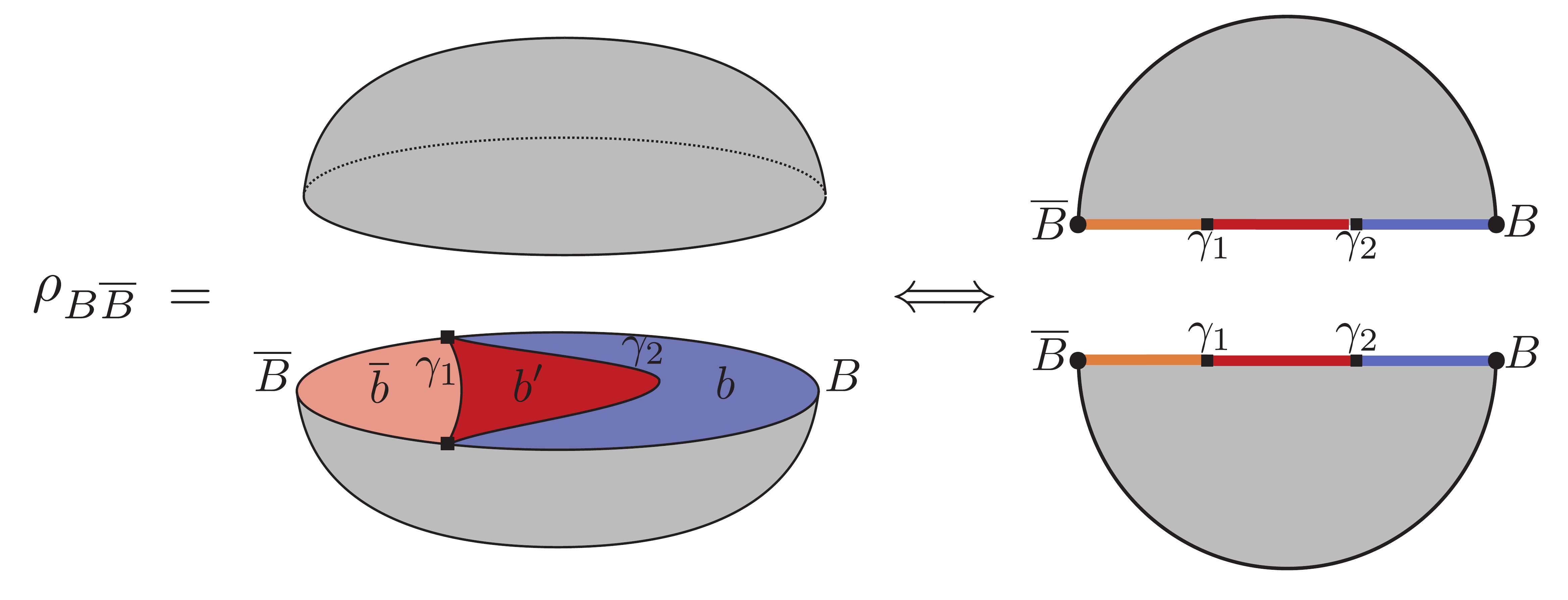}
\end{equation}

Fixing the areas in the initial state is a boundary condition and so also fixes the areas of that surface in path integrals featuring any number of replicas of that geometry.
This is what makes the sum over geometries in $Z_{B,n}$ doable.

Indeed, we can form geometries that satisfy all boundary conditions of $Z_{B,n}$ -- asymptotic boundary $\mathcal{M}_n$ plus fixed areas of all fixed-area surfaces -- simply by gluing together $n$ copies of the original $n=1$ bulk around the fixed-area surfaces.

Since we glue the boundary region $\overline B$ together in the bra and the ket path integral to make the density matrix $\rho_B$, the neighbouring bulk region $\overline b$ (shown in orange) is also always glued together
\begin{equation}\label{eq:fixed_area_rhoB}
    \includegraphics[width = \textwidth]{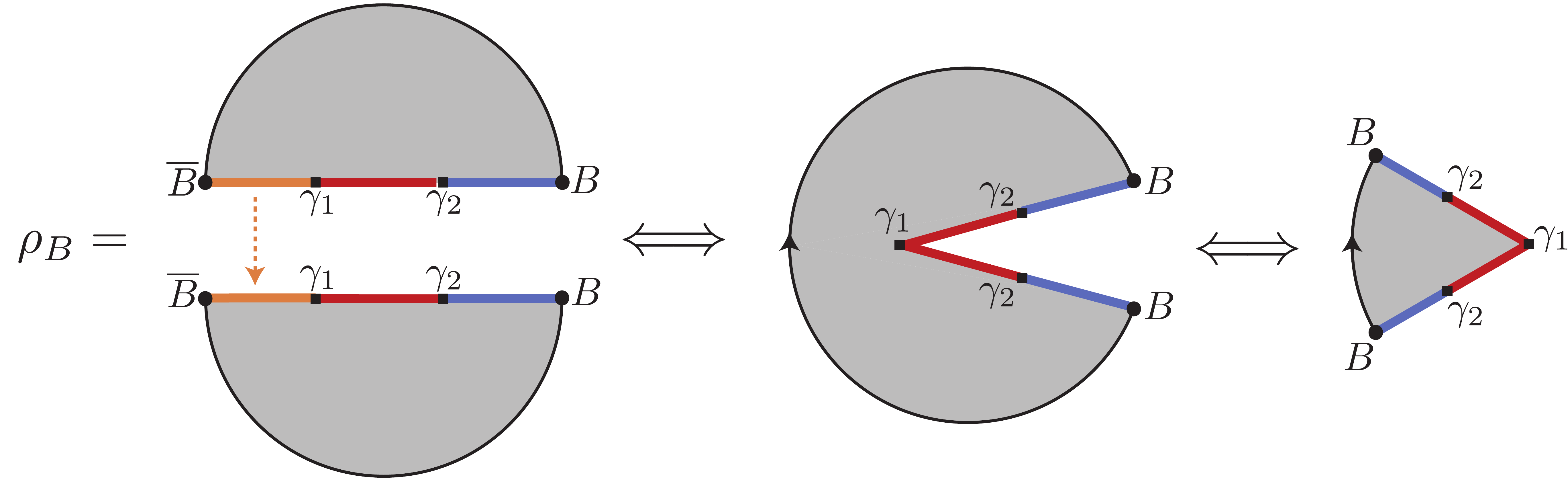}
\end{equation}
Similarly, because we glue the boundary regions $B$ together cyclically, the bulk regions $b$ get glued together cyclically. However, because we can have conical singularities at $\gamma_1$ and $\gamma_2$, the different copies of the region $b'$ can be glued together using an arbitrary permutation $\pi \in S_n$. To evaluate the full path integral, we sum over all saddles, and hence sum over all permutations $\pi$. For example,
\begin{align}
    \includegraphics[width = \textwidth]{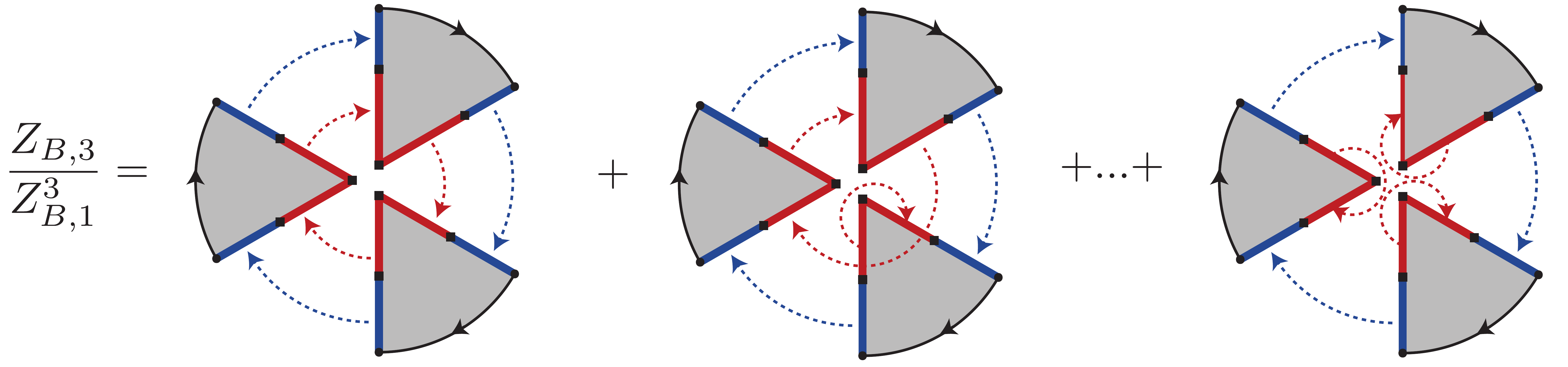}
\end{align}

Since the replica geometry consists of $n$ copies of the original unbackreacted geometry, the gravitational action away from the fixed-area surfaces cancels between the numerator $Z_{B,n}$ and the denominator $Z_{B,1}^n$. The only contribution to the gravitational action that doesn't cancel out is the contribution to the Einstein-Hilbert action from the conical singularities, which are different in the numerator and the denominator.
Each conical singularity gives a contribution equaling $(\phi - 2 \pi) A / 8 \pi G$, where $\phi$ is the opening angle of the conical singularity.

If the $b'$ regions are glued together using a permutation $\pi$, the full contribution to the action from the conical singularities in the replica geometry is therefore
\begin{equation}
    \left( n \phi_1 - 2 \pi C(\pi)  \right) \frac{A_1}{8 \pi G} + \left( n \phi_2 - 2 \pi C(\tau^{-1} \circ \pi)  \right) \frac{A_2}{8 \pi G} ~,
\end{equation}
where $C(g)$ is the number of cycles in the permutation $g$, and $\phi_1,\phi_2$ are the conical singularity angles associated to $\gamma_1,\gamma_2$ in the unreplicated geometry.
After normalization, the dependence on $\phi$ cancels. Including the matter partition function, we are then left with 
\begin{equation} \label{eq:Zneqn}
    \frac{Z_{B,n}}{Z_{B,1}^n} = \tr(\rho_B^n) = \sum_{\pi \in S_n} e^{\left(C(\pi)-n\right)A_1/4G + \left(C(\tau^{-1} \circ \pi )-n\right)A_2/4G}\tr(\rho^{\otimes n}_{bb'} \tau_{b}\pi_{b'})~.
\end{equation}
This further simplifies because we do not need to sum over all $S_n$.
Any permutation that does not maximize $C(\pi) + C(\tau^{-1} \circ \pi )$ corresponds to an action subleading by factors of the area.
The areas $A_1$ and $A_2$ are IR divergent, so those permutations are {\it infinitely} suppressed.
The remaining permutations lie on the geodesic in the Cayley graph (i.e. shortest path in permutation space, where each step is a transposition) connecting $\tau$ and the identity.  
These are the so-called ``non-crossing'' permutations $NC_n$, which all satisfy $C(\pi) + C(\tau^{-1} \circ \pi ) = n + 1$ (see e.g. \cite{non_crossings_annulus}).

Without the $\tr(\rho^{\otimes n}_{bb'} \tau_{b}\pi_{b'})$ factor, we could evaluate this sum explicitly.
The number of non-crossing permutations with $C(\tau^{-1} \circ \pi ) = k$ is the Narayana number $N(n,k)$. 
With that, we could organize the terms into a sum over $k$ and get an analytic answer in terms of hypergeometric functions.

The bulk term interferes because it depends not just on the number of cycles $C(\tau^{-1} \circ \pi )$, but also on the number of elements {\it per} cycle. 
Fortunately, there is a way, presented in \cite{Penington:2019kki}, to reorganize this sum into one over the number of elements per cycle. 

\subsubsection*{Resolvent method}
To make the calculation tractable, we will need to assume that the entropy of $b$ is small and can be ignored,\footnote{This is a stronger assumption than necessary, but is valid in the examples we'll care about. More generally, this method can work as long as either $\rho_b$ or $\rho_{\bar{b}}$ has a flat Renyi spectrum.} so that \eqref{eq:Zneqn} becomes
\begin{equation} \label{eq:Zneqnsimple}
\tr(\rho_B^n) = \sum_{\pi \in NC_n} e^{\left(C(\pi)-n\right)A_1/4G + \left(1 -C( \pi )\right)A_2/4G}\tr(\rho^{\otimes n}_{b'} \pi_{b'})~.
\end{equation}
Here, we have used the fact that $C( \pi ) + C( \tau^{-1} \circ \pi) = n+1$ for non-crossing permutations to rewrite the formula without $C( \tau^{-1} \circ \pi)$.

Define the resolvent $R_{ij}$ of $\rho_B$ as
\begin{equation}\label{eq:res_def}
    R_{ij}(\lambda) = \left( \frac{1}{\lambda \mathbb{1} - \rho_B }  \right)_{ij}
\end{equation}
This contains all the data about the eigenvalues of $\rho_B$. For example, the density $D_B(\lambda)$ of eigenvalues of $\rho_B$ is
\begin{align}
    D_B(\lambda) = -\frac{1}{\pi} \lim_{\epsilon \to 0^+} \mathrm{Im} R(\lambda + i \epsilon)~,
\end{align}
where $R$ is the trace of the resolvent. 

We compute this as follows, heavily using the fact that $\rho_{B\overline{B}}$ is a fixed-area state.
First, Taylor expand \eqref{eq:res_def} around $\rho_B = 0$ to obtain
\begin{equation}
    R(\lambda)_{ij} = \frac{1}{\lambda}\delta_{ij} + \sum_{n=1}^\infty \frac{1}{\lambda^{n+1}} \left(\rho_B^n\right)_{ij}~.
\end{equation}
We can visualize this as 
\begin{equation}
    \includegraphics[width = \textwidth]{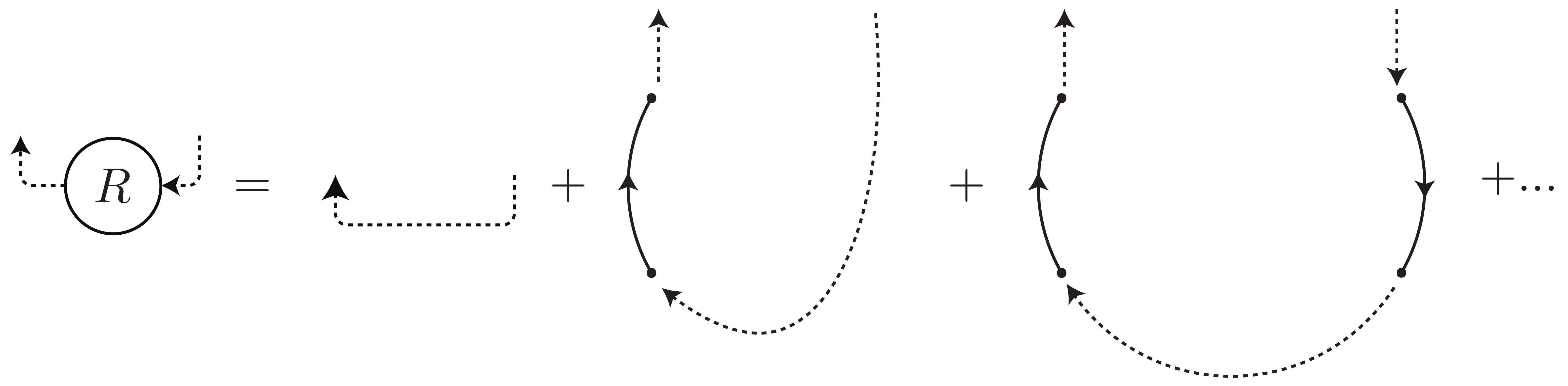}
\end{equation}
Each dashed line comes with a factor of $1 / \lambda$.
Then substitute for $\rho_B^n$ equation \eqref{eq:fixed_area_rhoB},
\begin{equation}
    \includegraphics[width = \textwidth]{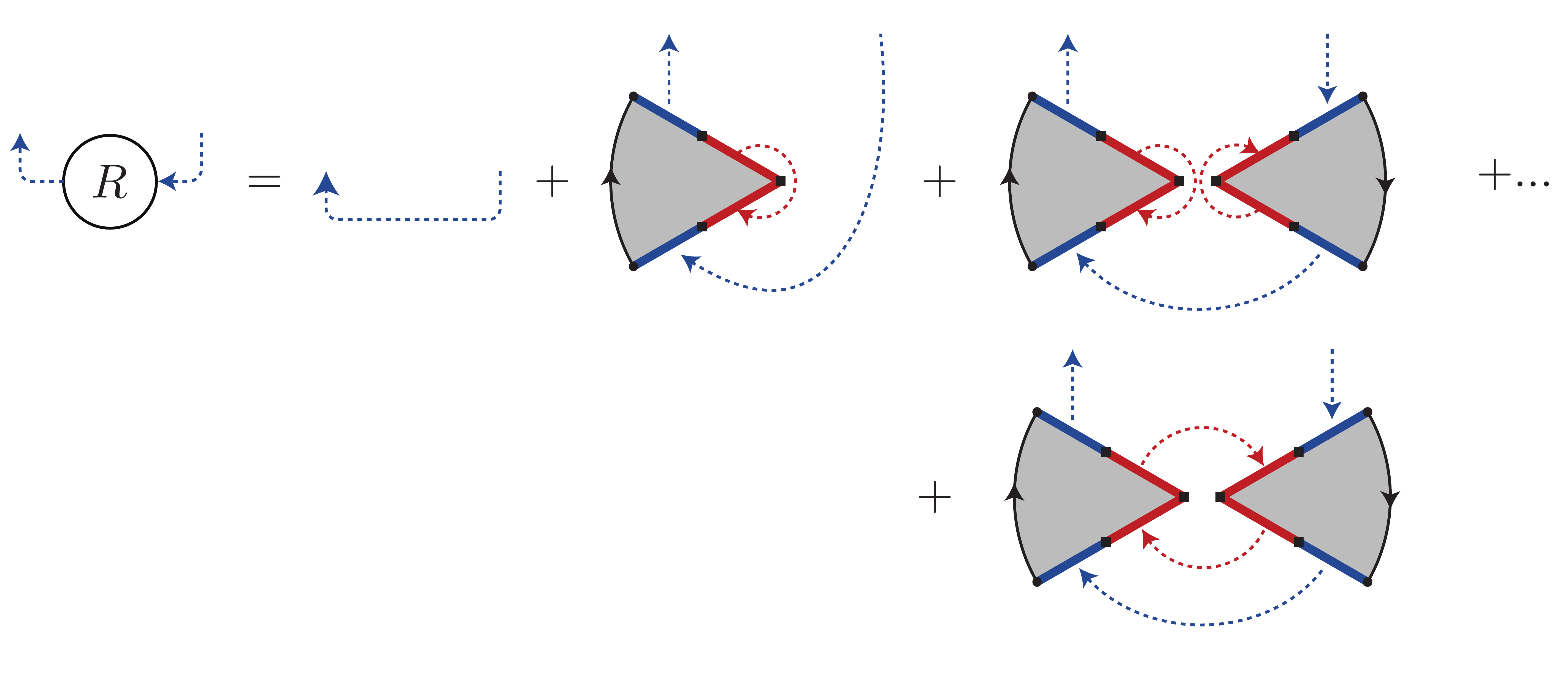}
\end{equation}
Taking the trace of this quantity -- visualized as simply connecting the dangling blue arrows into a closed loop -- gives the equation
\begin{equation}\label{eq:res_step_1}
    R = \frac{\mathrm{rank}(\rho_B)}{\lambda} + \sum_{n=1}^\infty \sum_{\pi \in NC_n} \frac{1}{\lambda^{n+1}} e^{\left(C(\pi)-n\right)A_1/4G + \left(1 - C( \pi )\right)A_2/4G}\tr(\rho^{\otimes n}_{b'} \pi_{b'})~.
\end{equation}
We can reorganize these sums in a convenient way, to get a Schwinger-Dyson equation:
\begin{equation}\label{eq:visual_SD_resolvent}
    \includegraphics[width = \textwidth]{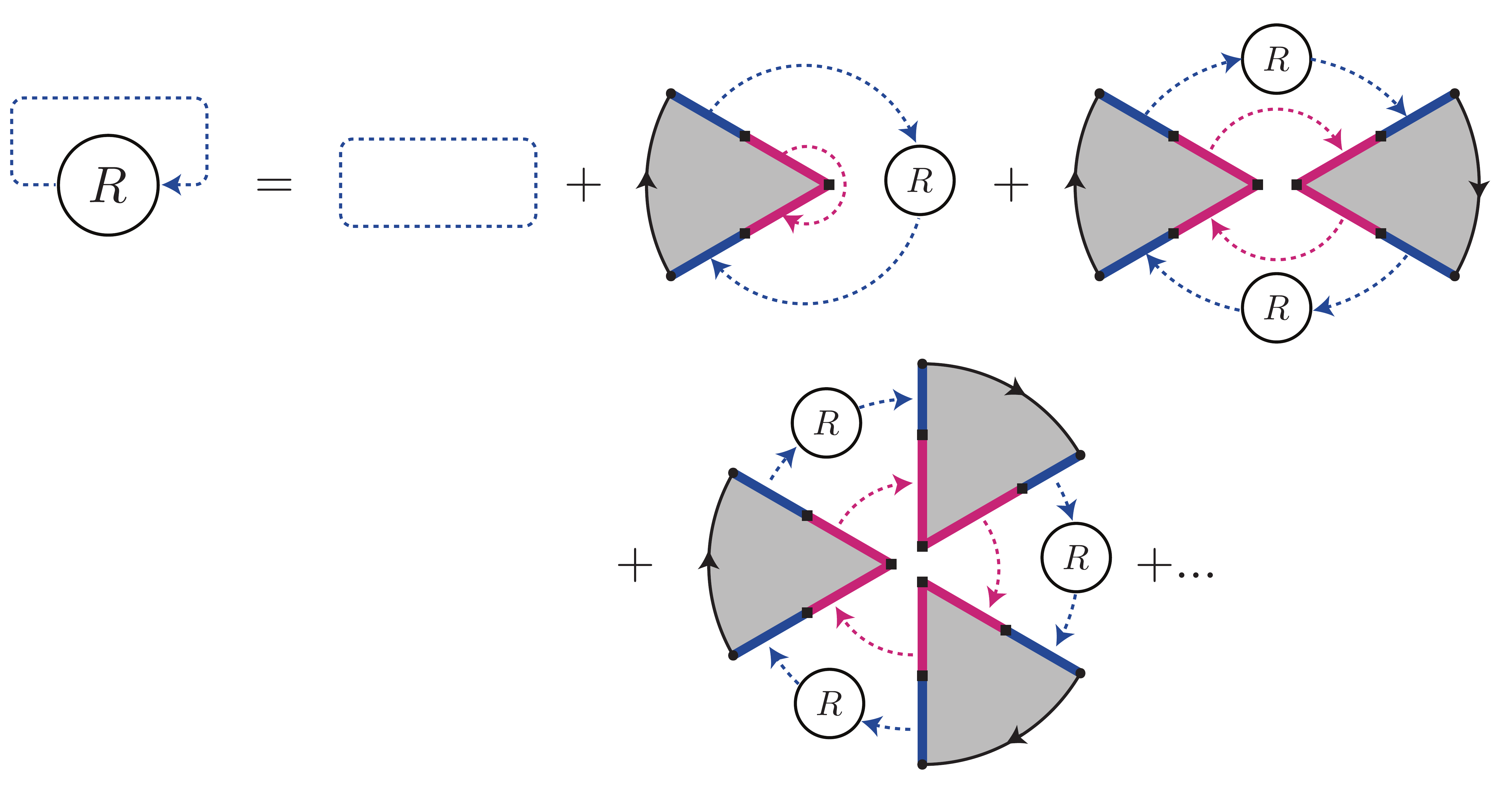}
\end{equation}
On the right hand side, the second term sums all non-crossing geometries in which the first replica of $b'$ is glued to no other replicas. The third term sums all non-crossing geometries in which the first replica of $b'$ is glued to exactly one other replica. And so on.

We now formally explain the diagrammatic expansion \eqref{eq:visual_SD_resolvent} in terms of equations. Starting with \eqref{eq:res_step_1}, decompose the sums into a sum over the number of elements $m$ in the cycle of $\pi$ that includes the first element (the ``primary cycle''), as well as the number of elements $n_i$ between the $i$th and $i+1$th element of the primary cycle:
That is, 
\begin{equation}
    \sum_{n=1}^\infty \sum_{\pi \in NC_n} \to \sum_{m=1}^\infty \sum_{n_1=0}^\infty ... \sum_{n_m=0}^\infty \sum_{\pi_{1} \in NC_{n_1}}...\sum_{\pi_{m} \in NC_{n_m}}~.
\end{equation}
The primary cycle is always cyclic, but the other permutations $\pi_i$ may not be.
We note that $n = m + \sum n_i$ and $C(\pi) = 1 + \sum C(\pi_i)~$. Also,
\begin{equation}
    \tr(\rho^{\otimes n}_{b'} \pi_{b'}) = \tr(\rho^{\otimes m}_{b'} \tau_m) \prod_{i=1}^m \tr(\rho^{\otimes n_i}_{b'} \pi_{n_i})~.
\end{equation}
Plugging these into the formula \eqref{eq:res_step_1} for the resolvent gives
\begin{equation}\label{eq:res_step_2}
\begin{split}
    R =& \frac{\mathrm{rank}(\rho_B)}{\lambda} + \\
    & \sum_{m=1}^\infty  \frac{e^{A_1/4G}\tr(\rho_{b'}^m)}{\lambda e^{m(A_1 + A_2)/4G}} \prod_{i=1}^m \left( \frac{e^{A_2/4G}}{\lambda} + \sum_{n_i = 1}^\infty \sum_{\pi_i \in NC_{n_i}} \frac{1}{\lambda^{n_i+1}}e^{\left(C(\pi_i)-n_i\right)A_1/4G + \left(1 - C(\pi_i)\right)A_2/4G}\tr(\rho^{\otimes n_i}_{b'} \pi_i)\right).
\end{split}
\end{equation}
The part in parenthesis is $R$ itself, from \eqref{eq:res_step_1}, assuming $\text{rank}(\rho_B) = e^{A_2/4G}$.\footnote{Since we can shift $R$ by $A/
\lambda$ for a real constant $A$, without changing $D(\lambda)$ away from $\lambda = 0$, we are always free to assume this.}
Therefore,
\begin{equation}\label{eq:res_step_3}
    R = \frac{e^{A_2/4G}}{\lambda} + \sum_{m=1}^\infty  \frac{e^{A_1/4G}\tr(\rho_{b'}^m) R^m}{\lambda e^{m(A_1 + A_2)/4G}}~.
\end{equation}
We are now ready to work out some specific examples.

\subsection{Examples}\label{sec:example1}

\subsection*{Example 1: Mixed states}
\subsubsection*{Setup}
Consider the setup with a dustball or a black hole, from Sections \ref{sec:intro} or \ref{sec:contradictions}, depicted in Figures \ref{fig:dustball} and \ref{fig:black_hole} respectively. 
We will simultaneously compute $S(B)$ in both cases, first modifying the setups slightly by fixing the areas of $\gamma_1$ and $\gamma_2$ to $A_1$ and $A_2$. 
All parameters we mention below apply equally well to both: e.g. $\rho_{b'}$ is the state of either the dustball or the black hole.

The same calculations also give the entropy of the \emph{black hole} in our third contradiction from Section \ref{sec:contradictions}. A particularly concrete example, where the full non-perturbative path integral can be evaluated and agrees with the answer that we find below, is the JT gravity plus end-of-the-world (EOW) brane model of \cite{Penington:2019kki}. In this case, working with fixed-area states is equivalent to working in the microcanonical ensemble (note that $A_2$ here is the horizon area of the black hole, while $A_1 = 0$), and the only bulk degrees of freedom are on the EOW brane in region $b'$, which is the assumption that we needed above to make the resolvent calculation possible.

Consider two bulk states, $\rho_{b',1} = \ket{\psi}\bra{\psi}$ pure and $\rho_{b',2}$ an arbitrary orthogonal mixed state of entropy $S$. We will assume that the state $\rho_{b',2}$ has a flat spectrum, and hence is perfectly compressible.
We will compute the entropy  $S(B)$ for their mixture,
\begin{align}\label{eq:calc1_state}
    \rho_{b'} = 
    p \ket{\psi}\bra{\psi} + (1-p) \rho_{b',2}~.
\end{align}
To keep the example as simple as possible, we assume
\begin{equation}\label{eq:scale_choice}
    (1-p)e^{-S} \ll p~.
\end{equation}
This ensures that the bulk density matrix eigenvalues $p$ and $(1-p)e^{-S}$ are separated by a large multiplicative factor.\footnote{For simplicity, we also assume $p, 1-p = \mathcal{O}(1)$, so that the corrections can be leading order. We remove this assumption in our more careful treatment in Appendix \ref{app:examples_full}.}
The von Neumann, min-, and max-entropies of this state are
\begin{equation}\label{eq:calc1_entropies}
\begin{split}
    H^\varepsilon_\mathrm{min}(b') \approx&~ - \ln(p)~, \\
    S(b') =& ~(1-p)S -p\ln(p) -(1-p)\ln(1-p)~, \\
    H^\varepsilon_\mathrm{max}(b') \approx&~ S ~.
\end{split}
\end{equation}
The na\"{i}ve QES prescription says
\begin{align}
    S_\text{na\"{i}ve}(B) = 
    \begin{cases}
       \frac{A_1}{4 G} + S({b'}) ,& S({b'}) \ll \frac{A_2-A_1}{4 G}\\
       \frac{A_2}{4 G} ,& S({b'}) \gg \frac{A_2-A_1}{4 G}~.
    \end{cases}
\end{align}
We will see that the correct answer, up to $\mathcal{O}(1)$ corrections, is
\begin{equation}
\begin{split}
    S_\mathrm{refined}(B) = 
    \begin{cases}
        \frac{A_1}{4 G} + S({b'}) ,& H^\varepsilon_\mathrm{max}(b') \ll \frac{A_2-A_1}{4 G}\\
        p\frac{A_1}{4 G} + (1-p) \frac{A_2}{4 G} ,& H^\varepsilon_\mathrm{min}(b') \ll \frac{A_2-A_1}{4 G} \ll H^\varepsilon_\mathrm{max}(b')\\
        \frac{A_2}{4 G} ,& H^\varepsilon_\mathrm{min}(b') \gg  \frac{A_2-A_1}{4  G} ~.
    \end{cases}
\end{split}
\end{equation}

\subsubsection*{Calculation}

Plug \eqref{eq:calc1_state} into \eqref{eq:res_step_3} and evaluate the two geometric sums to arrive at
\begin{align}\label{eq:res}
	\lambda R = e^{A_2/4G} +  \frac{p R}{e^{A_2/4G} - \frac{p}{e^{A_1/4G}}R} + \frac{(1-p) R}{e^{A_2/4G} - \frac{(1-p)}{e^{A_1/4G + S}}R}~.
\end{align}
The roots are the function $R(\lambda)$ that we seek.
As a cubic equation, its roots can be written analytically but are difficult to integrate to compute the entropy.
Fortunately, we can find a simple approximate solution, by using the assumption \eqref{eq:scale_choice}.

We expand \eqref{eq:res} in two different ways, which are valid at large $R$ (and hence small $\lambda$) and small $R$ (large $\lambda$) respectively. 
A full treatment, including proofs of all claims, is in Appendix \ref{app:examples_full}.
The two expansions are as follows.
\paragraph*{Expansion 1:}
For sufficiently large $R$, we have
\begin{align}\label{eq:exp1}
	\lambda R = e^{A_2/4G} - e^{A_1/4G} + \frac{(1-p) R}{e^{A_2/4G} - \frac{(1-p)}{e^{A_1/4G + S}}R} + \mathcal{O}\left( e^{A_1/4G}\frac{e^{(A_1 + A_2)/4G}}{p R} \right)~.
\end{align}

\paragraph*{Expansion 2:}
For sufficiently small $R$, we have
\begin{align}\label{eq:exp2}
	\lambda R = e^{A_2/4G} +  \frac{p R}{e^{A_2/4G} - \frac{p}{e^{A_1/4G}}R} + \frac{(1-p) R}{e^{A_2/4G}} +\mathcal{O}\left( \frac{(1-p) R}{e^{A_2/4G}}\frac{(1-p)R}{e^{(A_1 + A_2)/4G + S}} \right)~.
\end{align}

The condition \eqref{eq:scale_choice} ensures there is overlap in the conditions where the two expansions are valid, implying that some expansion is valid for all values of $R$ and $\lambda$. Each expansion gives a quadratic equation for the resolvent $R(\lambda)$, which can be easily solved  and has a single branch cut, where the eigenvalue density $D_B(\lambda)$ is nonzero. Both branch cuts are within the respective regimes of validity of the corresponding expansion, and so we find two distinct sets of eigenvalues. The eigenvalues in Expansion 1 come from the $\rho_{b',2}$, while the eigenvalues in Expansion 2 come from the $\ket{\psi}\bra{\psi}$ part of the state.

The entropy is given by
\begin{align}
    S(B) = -\int d\lambda ~ \lambda \ln(\lambda) D_B(\lambda) = \frac{1}{\pi}\int d\lambda ~ \lambda \ln(\lambda) \mathrm{Im}R(\lambda+i\varepsilon)~,
\end{align}
where we include both sets of eigenvalues in the integral. The answer depends on how $H^\varepsilon_\mathrm{min}(b')$ and $H^\varepsilon_\mathrm{max}(b')$ from \eqref{eq:calc1_entropies} compare to $\Delta A \equiv A_2 - A_1$.

\subsubsection*{Regime 1: $H^\varepsilon_\mathrm{min},H^\varepsilon_\mathrm{max} \ll \Delta A/4G $}
In this regime, the na\"{i}ve QES prescription gives the right answer.
Expansion 1 has a peak of eigenvalues at $\lambda \approx (1-p) e^{-A_1/4G - S}$, and Expansion 2 has a peak of eigenvalues at $\lambda \approx p\, e^{-A_1/4G}$. Both are within the regime of validity of their expansion.
See the top plot of Figure \ref{fig:regimes}.
Combined these peaks give entropy
\begin{equation}\label{eq:entropy_regime_1}
\begin{split}
    S(B) =& \frac{A_1}{4 G} + (1-p) S + ...~,
\end{split}
\end{equation}
where ``...'' represents terms subleading at large $S,~A_1,$ and $A_2$.
This includes the Shannon entropy term $ - p\ln\left( p \right) -  (1-p)\ln\left( 1-p\right)$.
Recall that equation \eqref{eq:entropy_regime_1} is the na\"{i}ve QES answer because $S({b'}) \lesssim H^\varepsilon_\mathrm{max}(b') \ll \Delta A / 4G$. 

\subsubsection*{Regime 2: $H^\varepsilon_\mathrm{min} \ll \Delta A / 4  G \ll H^\varepsilon_\mathrm{max}$}
Here there are large corrections to the na\"{i}ve QES prescription.
Expansion 2 describes the same peak it did in Regime 1, giving eigenvalues at $\lambda \approx p e^{-A_1 / 4G}$.
Expansion 1 now describes eigenvalues that have crossed the phase transition, which are therefore at $\lambda \approx (1-p)e^{-A_2/4G}$.
Both peaks are still well-separated, and the expansions continue to be valid at the peaks.
See the middle plot of Figure \ref{fig:regimes}.
The entropy comes out to
\begin{equation}\label{eq:entropy_regime_2}
\begin{split}
    S(B) =& p \frac{A_1}{4 G} +  (1-p) \frac{A_2}{4 G} +...~,
\end{split}
\end{equation}
and again we have dropped subleading terms, including the Shannon term.

Note that this entropy is different from the na\"{i}ve QES answer.
While the na\"{i}ve answer only cares about the relative sizes of $S({b'})$ and $\Delta A / 4 G$, this answer is independent of those relative sizes! 
Indeed, by dialing $S$, we can place $S({b'})$ on either side of $\Delta A / 4G$, as we please:
\begin{equation}
    H^\varepsilon_\mathrm{min},S({b'}) \ll \frac{\Delta A}{4 G} \ll H^\varepsilon_\mathrm{max}~,~~~~\text{     or }~~~~~ H^\varepsilon_\mathrm{min}  \ll \frac{\Delta A}{4 G} \ll S({b'}),H^\varepsilon_\mathrm{max}
\end{equation}
The entropy $S(B)$ equals \eqref{eq:entropy_regime_2} in both cases, while the na\"{i}ve QES prescription gives totally different formulas for the two cases!
In both cases, the na\"{i}ve QES prescription gives an answer that is larger (at leading order) than the correct answer.

The na\"{i}ve QES prescription failed because it treated the bulk eigenvalues in an all-or-nothing way, stubbornly refusing to acknowledge that some of the eigenvalues are much larger than the phase transition value $e^{-A_2/4G}$, even though many others are small enough to have crossed the phase transition.

\subsubsection*{Regime 3: $\Delta A/4 G \ll H^\varepsilon_\mathrm{min}, H_\mathrm{max}^\varepsilon$}
\begin{figure}[t]
\centering
\includegraphics[width = 0.7\textwidth]{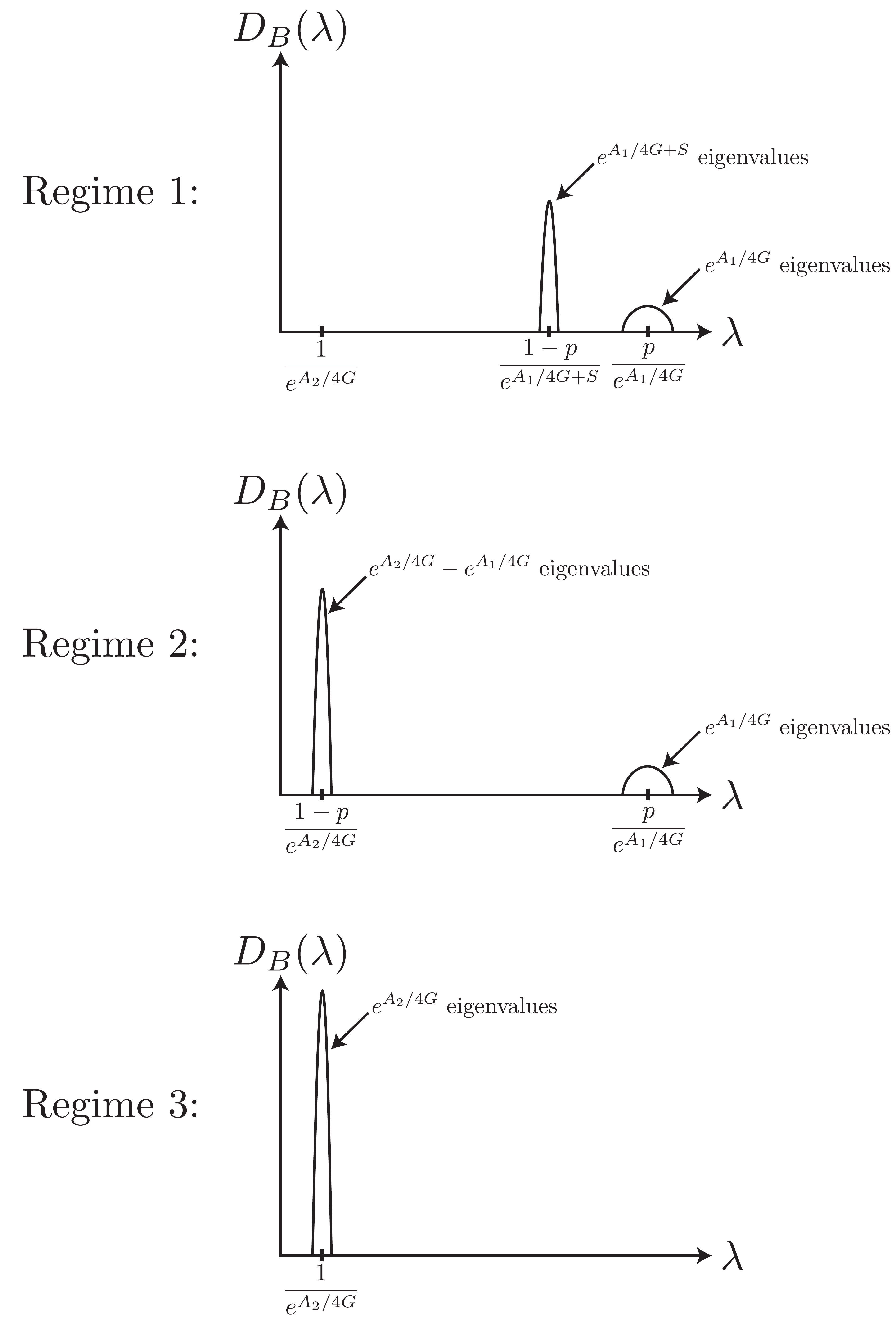}
\caption{Eigenvalue density for the three Regimes in Section \ref{sec:example1}. In Regime 1, there are two peaks of eigenvalues, each associated to one of the two states in the mixture, and each much greater than the critical value $1/e^{A_2/4G}$. Hence the na\"{i}ve QES prescription is correct. In Regime 2, one of the peaks has shifted to the critical value, while the other remained where it was, leading to large corrections in the na\"{i}ve QES prescription. In Regime 3, both peaks have moved to the critical value, and the na\"{i}ve prescription is valid again. Note the agreement with numerical results for the analogous random tensor network in Appendix \ref{app:numerics}.}
\label{fig:regimes}
\end{figure}

The na\"{i}ve QES prescription is back to receiving no corrections.
In this regime, Expansion 1 is never valid, while Expansion 2 describes a peak of eigenvalues that have crossed the phase transition, sitting at $\lambda \approx e^{-A_2/4G}$.
See the bottom plot of Figure \ref{fig:regimes}.
We obtain an entropy
\begin{equation}\label{eq:entropy_regime_3}
\begin{split}
    S(B) = \frac{A_2}{4 G} + ...~,
\end{split}    
\end{equation}
again letting ``...'' represent subleading terms. 
There is no Shannon term in this regime.
This agrees with the answer from the na\"{i}ve QES prescription because $S({b'}) \gtrsim H^\varepsilon_\mathrm{min}(b') \gg \Delta A / 4 G$.

\subsubsection*{Higher Renyis}
So far we have computed the von Neumann entropy in each regime, finding large corrections to the na\"{i}ve holographic prescription in Regime 2.
What about the higher Renyi entropies?
There is a holographic way to compute them as well \cite{Dong:2016fnf}; does that also have large corrections in some regime?

The answer is that their corrections are generally much smaller, even in Regime 2.
This is straightforward to derive with the resolvent approximations we have given.
For integer Renyi entropies with $n>1$, this is fairly self-explanatory. These can be computed directly using $n$ replicas without the need for any analytic continuation, and so can always be computed in the semiclassical limit using a saddle point approximation.

More interestingly, the corrections are also nonperturbatively small for non-integer Renyi entropies with $n > 1$ (and $n<1$), so long as $(n-1)$ is finite in the semiclassical limit. The large corrections to the von Neumann entropy come from the limit $n \to 1$ not commuting with the semiclassical limit $G \to 0$, unless we keep track of nonperturbatively small corrections.

\subsection*{Example 2: Entangled states}
\subsubsection*{Setup}
This next example demonstrates the role of  {\it conditional} min-/max-entropy. 
It closely resembles the previous one, but now the dustball or black hole $b'$ is entangled with another dustball or black hole that is always in the entanglement wedge. For evaporating black holes (or their JT + EOW brane cousins), it calculates the entropy of the Hawking radiation rather than the black hole.
The setups are detailed in Section \ref{sec:contradictions} and the dustball version is depicted in Figure \ref{fig:dustball_entangled}. 
Again those setups are simplified by fixing the areas of $\gamma_1$ and $\gamma_2$ to $A_1$ and $A_2$.

We emphasize again that this setup is, quite literally, the complement of the first one.
In that example, while we imagined a CFT $B\overline{B}$ in a mixed state, we could have instead imagined it purified by some reference system $R$.
Introducing $R$ changes nothing about that calculation.
Nonetheless, it is useful because $S(\overline{B}R) = S(B)$ regardless of the makeup of $R$.
Here we imagine $R$ to be an identical copy of $B\overline{B}$, with the same size dustball or black hole in its bulk dual $r$. 
For notational simplicity, we shall combine $R$ into $\overline{B}$, so that we are just computing $S(\overline{B})$. Similarly in the bulk we combine $r$ into $\overline b$.

We consider the following two states of the dustballs or black holes.
One is a pure, factorized state $\rho_{b'\overline{b},1} = \ket{\psi}_{b'}\bra{\psi}_{b'} \otimes \ket{\psi}_{\overline{b}}\bra{\psi}_{\overline{b}}$.
The other is a pure, maximally entangled state, $\rho_{b'\overline{b},2}$, with entanglement entropy $S$.
We will compute the entropy of their mixture,
\begin{equation}
    \rho_{b'\overline{b}} = p \bigg( \ket{\psi}_{b'}\bra{\psi}_{b'} \otimes  \ket{\psi}_{\overline{b}}\bra{\psi}_{\overline{b}} \bigg) + (1-p) \rho_{b'\overline{b},2}~.
\end{equation}
Again, we keep this example simple by assuming \eqref{eq:scale_choice}.
The conditional von Neumann, min-, and max-entropies of this state are
\begin{equation}\label{eq:calc2_entropies}
\begin{split}
    H^\varepsilon_\mathrm{min}(b'|\overline{b}) \approx& ~- S~, \\
    S(b' | \overline{b}) =& ~- (1-p)S~, \\
    H^\varepsilon_\mathrm{max}(b'|\overline{b}) \approx&~ \ln(p) ~.
\end{split}
\end{equation}
All of these are negative numbers, because of the entanglement.
The entropy $S(\overline{B})$ depends on their comparison to $(A_1 - A_2)/4G$, which is itself big and negative. 
The na\"{i}ve QES prescription says
\begin{align}
    S_\text{na\"{i}ve}(\overline{B}) = 
    \begin{cases}
       \frac{A_2}{4 G} + S(b' \overline{b}) ,& S({b'}|\overline{b}) \ll \frac{A_1-A_2}{4 G}\\
       \frac{A_1}{4 G} ,& S({b'}|\overline{b}) \gg \frac{A_1-A_2}{4 G}~.
    \end{cases}
\end{align}
The correct answer up to $\mathcal{O}(1)$, we will see, is
\begin{equation}
\begin{split}
    S_\mathrm{refined}(\overline{B}) = 
    \begin{cases}
        \frac{A_2}{4 G} ,& H^\varepsilon_\mathrm{max}(b'|\overline{b}) \ll \frac{A_1-A_2}{4 G}\\
        p\frac{A_1}{4 G} + (1-p)\frac{A_2}{4 G} ,& H^\varepsilon_\mathrm{min}(b'|\overline{b}) \ll \frac{A_1-A_2}{4 G} \ll H^\varepsilon_\mathrm{max}(b'|\overline{b})\\
        \frac{A_1}{4 G} + S(b'),& H^\varepsilon_\mathrm{min}(b'|\overline{b}) \gg \frac{A_1-A_2}{4  G} ~.
    \end{cases}
\end{split}
\end{equation}

\subsubsection*{Calculation} 

Rather than write out a resolvent like we did before, we will use a trick to compute the entropy in each of these three regimes.
Notice that these smooth conditional min- and max-entropies \eqref{eq:calc2_entropies} equal minus the max- and min-entropies \eqref{eq:calc1_entropies} respectively, from Example 1.
This was the general rule, from Section \ref{sec:min_max_entropy}: for a pure state on $ABC$, 
\begin{equation}
    H_\mathrm{min}(A|B) = - H_\mathrm{max}(A|C)~. 
\end{equation}
Since the system $b$ that we conditioned on in Example 1 was trivial, we have
\begin{align}
    H^\varepsilon_\mathrm{min}(b'|\overline{b}) =& - H_\mathrm{max}^\varepsilon(b')~,\\
    H^\varepsilon_\mathrm{max}(b'|\overline{b}) =& - H_\mathrm{min}^\varepsilon(b')~.
\end{align}
So we can compute the entropy in the three regimes as follows. 
Consider, for example, the regime in which both the conditional min- and max-entropy are less than $\Delta A / 4G \equiv (A_1 - A_2)/4G$.
This corresponds exactly to the regime in Example 1 where both min- and max-entropy were {\it greater} than $(A_2 - A_1) / 4G$.
So, using purity of $B\overline{B}$, the entropy $S(\overline{B})$ in this regime equals $S(B)$ from that regime.
Thus the entropy $S(\overline{B})$ is completely deducible from the results of Example 1. 

The key lesson is this: there is an important role played by bulk entanglement, encapsulated by the {\it conditional} min- and max-entropy.
That's the only way this setup would be consistent with the complementary answers from the previous example.

\subsubsection*{Regime 1: $H^\varepsilon_\mathrm{min}(b'|\overline{b}),H^\varepsilon_\mathrm{max}(b'|\overline{b}) \ll \Delta A/4G $}
This is Regime 3 of Example 1. 
Therefore, 
\begin{equation}\label{eq:entropy_regime_1_ex2}
\begin{split}
    S(\overline{B}) = \frac{A_2}{4 G} + ...~,
\end{split}    
\end{equation}
where ``...'' represents subleading terms.
The na\"{i}ve QES prescription gives the same answer because $S(b'|b) \lesssim H^\varepsilon_\mathrm{max}(b'|\overline{b}) \ll \Delta A/4G$.

\subsubsection*{Regime 2: $H^\varepsilon_\mathrm{min}(b'|\overline{b}) \ll \Delta A / 4  G \ll H^\varepsilon_\mathrm{max}(b'|\overline{b})$}
This is Regime 2 of Example 1, so again here there are large corrections to the na\"{i}ve QES prescription.
The entropy comes out to
\begin{equation}\label{eq:entropy_regime_2_ex2}
\begin{split}
    S(\overline{B}) =& p \frac{A_1}{4 G} +  (1-p) \frac{A_2}{4 G} +...~,
\end{split}
\end{equation}
and again we have dropped subleading terms, including the $\mathcal{O}(1)$ Shannon term.
This entropy is different than the na\"{i}ve QES answer.

\subsubsection*{Regime 3: $\Delta A/4 G \ll H^\varepsilon_\mathrm{min}(b'|\overline{b}), H_\mathrm{max}^\varepsilon(b'|\overline{b})$}
This is Regime 1 of Example 1, and hence the na\"{i}ve QES prescription is back to receiving no corrections.
The entropy is
\begin{equation}\label{eq:entropy_regime_3_ex2}
\begin{split}
    S(\overline{B}) =& \frac{A_1}{4 G} + (1-p) S + ...~,
\end{split}
\end{equation}
where again ``...'' includes the Shannon term.
This matches the na\"{i}ve QES answer because $S(b'|b) \gtrsim H^\varepsilon_\mathrm{min}(b'|\overline{b}) \gg \Delta A/4 G$.

\subsection*{Example 3: Arbitrary Entanglement Spectra}
What about more general bulk states $\rho_{b'}$, which aren't simply the mixture of two states with (approximately) flat entanglement spectra (again forgetting about entanglement, for now)? 
For an arbitrary bulk state $\rho_{b'}$ with eigenvalue density $D_{b'}(\lambda_{b'})$, the resolvent recursion relation \eqref{eq:res_step_3} becomes
\begin{align} \label{eq:res_gen_spec}
    \lambda R = e^{A_2/4G} + \int d \lambda_{b'} \,\frac{\lambda_{b'} D_{b'}(\lambda_{b'}) R}{e^{A_2/4G} - \lambda_{b'}\, e^{-A_1/4 G} R}~.
\end{align}
We will not be able to solve this equation as precisely as we were able to calculate the resolvents in the preceding examples, but we will still have sufficient control to calculate the von Neumann entropy up to $\mathcal{O}(1)$ corrections.

\subsubsection*{Calculation}
Our strategy will closely mirror the strategy used to calculate corrections to the von Neumann entropy near the Page transition in \cite{Penington:2019kki}, and we refer the reader to that paper (and in particular Appendix F) for more detailed justifications. We will perturbatively approximate the resolvent for $\lambda \gg e^{-A_2/4G}$, and argue that there are no eigenvalues with $\lambda \ll e^{-A_2/4G}$ whenever the smooth max-entropy is sufficiently large. Combined these two results will enable us to calculate the von Neumann entropy up to $\mathcal{O}(1)$ corrections.

For $\lambda \gg e^{-A_2/4G}$, we treat the second term in \eqref{eq:res_gen_spec} as a small perturbation. At leading order, the resolvent is given by $R_0(\lambda) = e^{A_2/4G}/\lambda$. The leading contribution to the density of states comes from the first perturbative correction
\begin{align}
    R_1(\lambda) = e^{A_1/4G} \int \, d \lambda_{b'} \, \frac{\lambda_{b'} D_{b'}(\lambda_{b'})}{\lambda \,(e^{A_1/4G}\, \lambda - \lambda_{b'})}~.
\end{align}
To justify this perturbative approximation, we assume $\lambda$ has a small imaginary part $i \epsilon$, with $\lambda \gg \epsilon \gg e^{-A_2/4G}$. Hence
\begin{align}
    |R_1(\lambda)| \leq \int d \lambda_{b'} \, \frac{\lambda_{b'} D_{b'}(\lambda_{b'})}{\epsilon \lambda }~ = \frac{1}{\epsilon \lambda} \ll R_0(\lambda),
\end{align}
as desired. 

Suppose we are in Regime 1, where the smooth max-entropy $H_\text{max}^\varepsilon(b') \ll (A_2 - A_1)/ 4G$. Then we can approximate our bulk state by a nearby state with $D_{b'}(\lambda_{b'}) = 0$, except when $\lambda_{b'} \gg e^{-A_2/4G}$.\footnote{The effect of this $O(\varepsilon$) approximation to the state on the von Neumann entropy is controlled by Fannes inequality \cite{fannes1973continuity}. See \eqref{eq:fannes}.} For $\lambda \lesssim e^{-A_2/4G}$, we can ignore the first term in the denominator of \eqref{eq:res_gen_spec} to get the self-consistent approximation
\begin{align} \label{eq:R_smooth_max_small}
    R \approx \frac{1}{\lambda}\left[e^{A_2/4G} - \int \, d \lambda_{b'} \,D_{b'}(\lambda_{b'})\right]~.
\end{align}
Note that going to higher orders in perturbation theory will not introduce a nonzero eigenvalue density, because there are no poles in \eqref{eq:res_gen_spec} for these values of $\lambda$. We conclude that in Regime 1 we have $ D(\lambda) = e^{A_1/4G} D_{b'}(e^{A_1/4G} \lambda),$
and hence $S(B) = \frac{A_1}{4 G} + S(b')$. 

What about when $H_\text{max}^\varepsilon(b') \gg (A_2 - A_1)/ 4G$? We want to argue that there are no eigenvalues with $\lambda \ll \varepsilon e^{-A_2/4G}$, and hence that $R(\lambda)$ is negative and real. To do so, we rewrite \eqref{eq:res_gen_spec} to give $\lambda$ as a function of $R$
\begin{align} \label{eq:as_func_lambda}
    \lambda = \frac{e^{A_2/4G}}{R} + \int d \lambda_{b'} \,\frac{\lambda_{b'} D_{b'}(\lambda_{b'})}{e^{A_2/4G} - \lambda_{b'}\, e^{-A_1/4 G} R}~.
\end{align}
For small negative $R$, $\lambda$ is large and negative, since the first term dominates. When $R$ is very large and negative however, the second term dominates (thanks to our assumptions about the smooth max-entropy), and so $\lambda$ is positive. There will be  some intermediate $R$ where $\lambda$ is maximal, which gives the bottom of the entanglement spectrum.

To lower bound this maximum, we choose some $R \gg e^{A_2/2G}$. Then the second term in \eqref{eq:as_func_lambda} dominates and we find
\begin{align}
    \lambda \gtrsim e^{-A_2/4G} \int_{\lambda_{b'} \ll e^{(A_1 + A_2)/4G} /R} d \lambda_{b'} \,\lambda_{b'} D_{b'}(\lambda_{b'}) \gtrsim \mathcal{O}(\varepsilon \,e^{-A_2/4G})~.
\end{align}
The last approximation again follows from our assumption about the size of the smooth max-entropy. We therefore conclude that there are no eigenvalues with $\lambda \ll e^{-A_2/4G}$, as expected.

We can now calculate the entropy $S(B)$. Since we know the eigenvalue density for both $\lambda \gg e^{-A_2/4G}$ and $\lambda \ll \varepsilon e^{-A_2/4G}$, we know the remaining eigenvalues must all have $\varepsilon\, e^{-A_2/4G} \lesssim \lambda \lesssim e^{-A_2/4G}$. Up to $\mathcal{O}(\ln \varepsilon)$ corrections, this means that
\begin{align}
    S(B) = \frac{A_1}{4 G} + \int d \lambda_{b'} \,\lambda_{b'} D_{b'}(\lambda_{b'}) \min (\ln \lambda'_{b'}, \frac{A_2 - A_1}{4 G})~.
\end{align}
When $H_\text{min}^\varepsilon(b') \gg (A_2 - A_1)/ 4G$, we can ignore the first term in the minimization and we recover the na\"{i}ve QES prescription result $S(B) = A_2/ 4G$. However, when $H_\text{min}^\varepsilon(b') \ll (A_2 - A_1)/ 4G$, we find leading order corrections. Our results agree with the refined QES prescription in all three regimes.

\subsection*{Summary} 
Let us summarize what we learned in this section.
Doing a careful calculation -- without the LM assumption -- reveals a refinement of the na\"{i}ve QES formula, which can differ from the na\"{i}ve one at leading order.
This refined QES prescription compares the smooth conditional min- and max-entropies to the difference in areas. 
We have seen this in three examples, all of which were fixed-area states, and all of which had particular simple bulk states. Unfortunately, states where there is a large amount of entropy in all the bulk regions and an arbitrary entanglement structure, and states where the areas are not fixed, are beyond the current technology we have for computing the replica trick without using the LM assumption.
However, in the next section, we will derive the QES refinement more generally, beyond these particular bulk states and beyond fixed-area states, by using a more indirect approach.

\section{Corrections in general holographic states}\label{sec:when_corrections}

We start by arguing that the na\"{i}ve QES prescription \emph{is} valid, whenever the smooth conditional min- and max-entropy are safely on the same side of $(A_2 - A_1)/4G$. 
This generalizes half the pattern from our examples, now showing that for \emph{any} state on $bb'\bar{b}$ the na\"{i}ve QES prescription can be trusted when the min- and max-entropy are on the same side of the area difference, though we emphasize that we still limit ourselves to two competing fixed-area QES, as in Figure \ref{fig:dustball}.

Then in Section \ref{sec:necessity_large_corrections} we prove that there \emph{are} generally large corrections to the na\"{i}ve QES prescription in the regime where we did not prove the corrections are small. 
I.e. there are large corrections when the min- and max-entropy are on different sides of the area difference.

Finally, in Section \ref{sec:fixedarea_to_general} we remove the fixed-area requirement, demonstrating that more general geometries follow the same pattern, up to a relatively small difference, $\mathcal{O}(\ln G)$. 

Altogether, our argument shows that there are large corrections in general holographic states if and only if the bulk min- and max-entropy straddle the difference in areas between the two competing QES.

\subsection{The regime of validity of na\"{i}ve QES in fixed-area states} \label{sec:fixed_area_conditional}
We first argue that the na\"{i}ve QES prescription is valid, up to $o(1)$ corrections, for general fixed-area states with two extremal surfaces, so long as either
\begin{align} \label{eq:QESgamma2}
    \frac{A_2}{4 G} \ll \frac{A_1}{4 G} + H_\text{min}^\varepsilon(b'|b)~,
\end{align}
in which case the minimal QES is the surface $\gamma_2$, or
\begin{align}
    \frac{A_2}{4 G} \gg \frac{A_1}{4 G} + H_\text{max}^\varepsilon(b'|b)~,
\end{align}
in which case the minimal QES is the surface $\gamma_1$. By ``much greater than'', $\gg$, we mean a difference that is much larger than $\mathcal{O}(\ln G)$. 
Therefore, large corrections can only exist if
\begin{align}\label{eq:correction_conditions}
   \frac{A_1}{4 G} + H_\text{max}^\varepsilon (b'|b) \gtrsim \frac{A_2}{4 G} \gtrsim \frac{A_1}{4 G} + H_\text{min}^\varepsilon (b'|b)~.
\end{align}
We will later argue in Section \ref{sec:necessity_large_corrections} that significant corrections (at least $\mathcal{O}(1)$ in size) in fact \emph{always} exist when 
\begin{align}
   \frac{A_1}{4 G} + H_\text{max}^\varepsilon (b'|b) \gg \frac{A_2}{4 G} \gg \frac{A_1}{4 G} + H_\text{min}^\varepsilon (b'|b)~.
\end{align}

Our strategy will be to make use of the correspondence between the nonperturbative corrections to the replica trick entropy in a) fixed-area states in gravity and b) single-tensor random tensor networks (RTNs) \cite{Hayden:2016cfa}, a nonperturbative equivalence first noted in \cite{Penington:2019kki}.

Let us start by reviewing that correspondence. We have already evaluated $\tr(\rho^n)$ for fixed-area states in Section \ref{sec:examples}. We found that for a general normalized bulk state $\rho_{b b' \bar{b}}$, the dual normalized boundary state $\rho_{B \overline{B}}$ satisfied
\begin{align}
    \tr(\rho_B^n) = \sum_\pi \tr(\tau_b\, \pi_{b'} \rho_{b b' \bar b}^{\otimes n}) \exp\left(\left[C(\tau^{-1} \circ \pi) - n\right] \frac{A_2}{4 G} + \left[C(\pi) - n\right] \frac{A_1}{4 G}\right)~,
\end{align}
where the fixed permutation $\tau$ is cyclic, the sum is over permutations $\pi$ that maximize $C(\tau^{-1} \circ \pi) + C(\pi)$, and the operators $\tau_b,\, \pi_{b'}$ permute the $n$ copies of their respective subsystem.

\begin{figure}
\centering
\includegraphics[width = 0.5\textwidth]{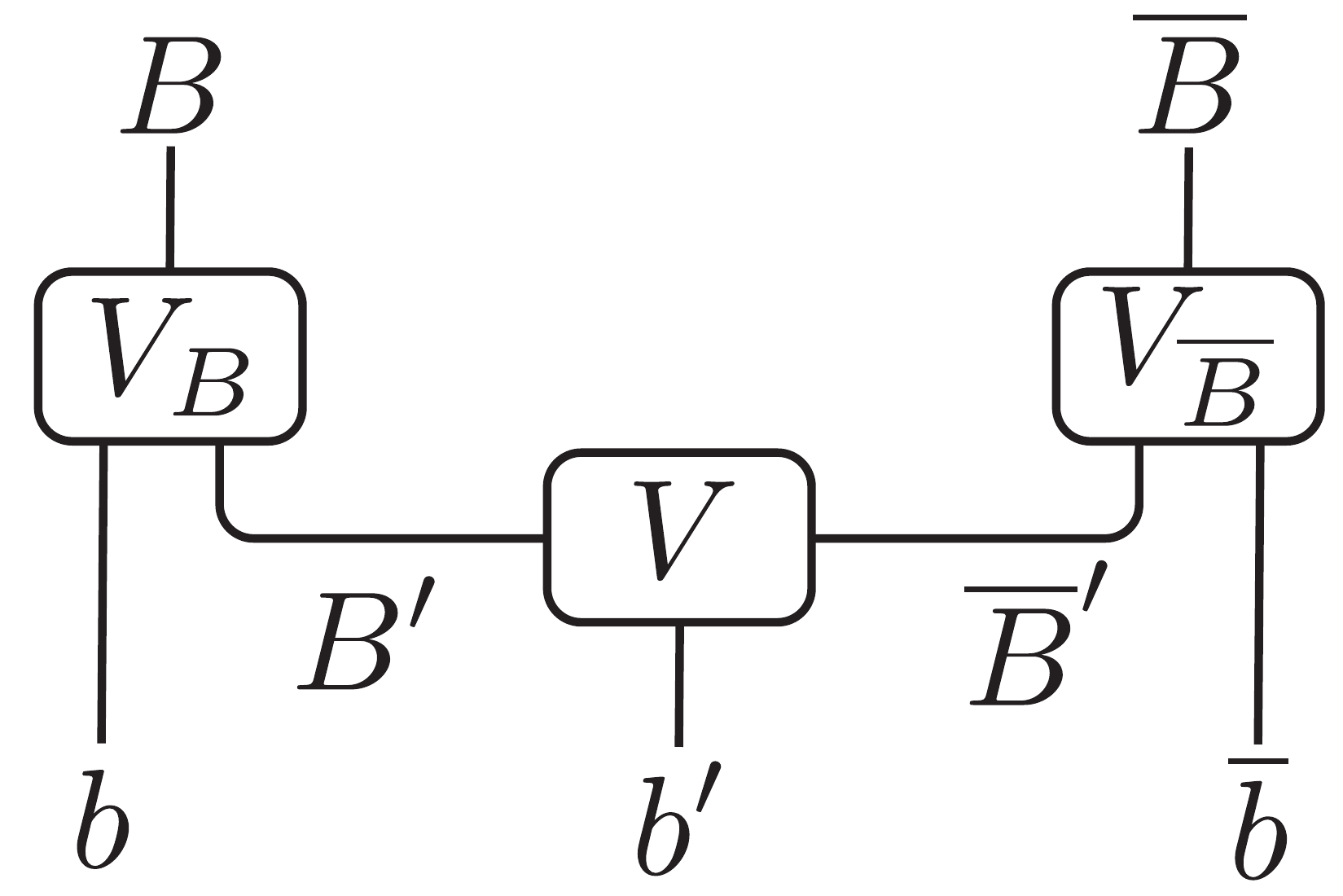}
\caption{Simple random tensor network analogous to Figure \ref{fig:dustball} with fixed areas. The ``bulk legs'' $b, b', \bar{b}$ are related by isometries $V, V_B, V_{\overline{B}}$ to ``boundary legs'' $B, \overline{B}$. The (log of the) dimensions of the legs $B'$ and $\overline{B}'$ play the role of the areas in Figure \ref{fig:dustball}.}
\label{fig:RTN}
\end{figure}

We want to show that one finds the same formula in RTNs, where 
\begin{align}
\rho_{B \overline{B}} = V_B V_{\overline{B}} V \rho_{b b' \bar b}  V^\dagger V_{\bar B}^\dagger V_B^\dagger
\end{align}
and $V: b' \to B' \otimes \overline{B}'$, $V_B: b \otimes B' \to B$ and $V_{\overline{B}} : \bar b \otimes \overline{B}' \to \overline{B}$ are random isometries. This is shown graphically in Figure \ref{fig:RTN}. Here the subsystems $B'$ and $\overline{B}'$ have dimensions $d_{B'} = \exp(A_2/4 G)$ and   $d_{\overline{B}'} = \exp(A_1/4 G)$ respectively.

We can write \cite{Hayden:2016cfa}
\begin{align}
    \tr(\rho_B^n) = \tr(\tau_b \tau_{B'} (U V_0\rho_{b b'}V_0^\dagger U^\dagger)^{\otimes n})~.
\end{align}
Here we have written $V = U V_0$ for a fixed isometry $V_0$ and a Haar random unitary $U$. 
Now, we can use the formula \cite{harrow2013church} 
\begin{align}
    \int dU U_{i_1 j_1}\dots U_{i_n j_n} U^\dagger_{j'_1 i'_1} \dots U^\dagger_{j'_n i'_n} = d^{-n}\sum_{\pi} \delta_{i_1 i'_{\pi(1)}} \delta_{j_1 j'_{\pi(1)}} \dots \delta_{i_n i'_{\pi(n)}} \delta_{j_n j'_{\pi(n)}} + \mathcal{O}(d^{-n-1})~,
\end{align}
where $\pi(i) \in \{1,...,n\}$ represents the arbitrary permutation $\pi$ acting on the $i$-th element of an $n$-element set.
Combine with $d = d_{B'} d_{\overline{B}'}$ to obtain
\begin{align}
    \tr(\rho_B^n) = d^{-n} \sum_{\pi} \tr(\tau_b \pi_{b'} \rho_{b b'}) \tr(\tau_{B'} \pi_{B'}) \tr(\pi_{\overline{B}'}) = \sum_{\pi} \tr(\tau_b \pi_{b'} \rho_{b b'}) d_{B'}^{C(\tau^{-1} \circ \pi) - n} d_{\overline{B}'}^{C(\pi) - n}~.
\end{align}
This is exactly the result that we found in the gravity calculation.

Since we can (in principle if not in practice) calculate the entropy $S(B)$ simply by analytically continuing $\tr(\rho_B^n)$, the RTN must have the same entanglement entropy as the gravity calculation.
Armed with this knowledge, we can calculate the entanglement entropy in the RTN, \emph{using any techique we want}, and thereby find the gravitational answer as well.

The key result that we will use is the one-shot decoupling theorem, Theorem III.1 of \cite{Berta_2011} (see Appendix \ref{app:one-shot_decoupling} for our summary of the proof), which says that for $V_0 \rho_{b b'} V_0^\dagger$, then so long as
\begin{align} \label{eq:mininequality}
   \ln d_{B'} < \ln d_{\overline{B}'} + H_\text{min}(b' | b)_\rho - 2 \ln \frac{1}{\varepsilon}~,
\end{align}
we have
\begin{align} \label{eq:haaraveragething}
    \int dU \left\lVert \tr_{\overline{B}'}(U V_0 \rho_{b b'} V_0^\dagger U^\dagger) - \rho_b \otimes \frac{\mathbb{1}_{B'}}{d_{B'}} \right\rVert_1 \leq \varepsilon~.
\end{align}
What does this theorem mean? 
It states that \eqref{eq:mininequality} is a sufficient condition to ensure 
\begin{align} \label{eq:approximatedecoupling}
    \rho_B \approx \widetilde{\rho}_B = V_B\rho_b \otimes \mathbb{1}_{B'}/d_{B'} V_B^\dagger~.
\end{align}
Moreover, the condition \eqref{eq:mininequality} can be weakened, replacing the min-entropy $H_\text{min}(b' | b)_\rho$ by its smooth version $H_\text{min}^\varepsilon(b' | b)_\rho$, with only a small degradation in the quality of the approximation, as follows from the definition of smoothing.
 
The state on the right hand side of \eqref{eq:approximatedecoupling} has two essential features. The first is that it depends only on the reduced state $\rho_b$, and is completely independent of $b'$. The second is that its entropy
\begin{align}
    S(\widetilde{\rho}_B) = \ln d_{B'} + S(\rho_b)
\end{align}
corresponds in gravity to the generalized entropy of the surface $\gamma_2$.

We want to use this to bound the entropy of the state $\rho_B$ itself. To do so, we need the Fannes' inequality \cite{fannes1973continuity}
\begin{align} \label{eq:fannes}
    |S(\rho) - S(\sigma)| \leq \frac{1}{2}\lVert \rho - \sigma \rVert_1 \ln d + S_2\left(\frac{1}{2} \lVert \rho - \sigma \rVert_1\right).
\end{align}
Here $S_2(p) = - p \ln p - (1-p) \ln (1-p)$ is the Shannon entropy of the probability distribution $(p, 1-p)$.

Applying this inequality to the states in \eqref{eq:haaraveragething}, we find
 that
 \begin{align} \label{eq:SrhoBblah}
     S(\rho_B) = S(\widetilde{\rho}_B) + \mathcal{O}(\varepsilon \ln d_B') + \mathcal{O}(\varepsilon \ln \varepsilon) = \frac{A_2}{4 G} + S(\rho_b) + \mathcal{O}\left(\frac{\varepsilon}{4G}\right) + \mathcal{O}(\varepsilon \ln \varepsilon)~.
 \end{align}
 If we take $\varepsilon$ (in both \eqref{eq:haaraveragething} and the smooth min-entropy) to be polynomially small in $G$ (say $\mathcal{O}(G^2)$), then \eqref{eq:mininequality} is satisfied whenever 
 \begin{align}
      \frac{A_1}{4 G} + H_\text{min}^\varepsilon(b'|b) - \frac{A_2}{4 G}\gg 0
\end{align}
and this difference grows faster than $\ln(1/G)$ in the semiclassical limit $G \to 0$. Moreover, \eqref{eq:SrhoBblah} says $S(\rho_B)$ is given by the generalized entropy of the quantum extremal surface $\gamma_2$, up to a perturbatively small ($\mathcal{O}(G)$) correction.

We also want to show the QES prescription is valid, this time with minimal QES $\gamma_1$, so long as
\begin{align} \label{eq:maxconditionholding} 
    \frac{A_1}{4 G} + H_\text{max}^\varepsilon(b'|b) \ll \frac{A_2}{4 G}~.
\end{align}
It turns out that this follows by the same arguments used above, applied to the complementary boundary region.

We first consider some arbitrary purification $\ket{\psi}_{b b' \bar b R}$ of the bulk state $\rho_{b b' \bar b}$. We now want to calculate the entanglement entropy $S(\overline{B} R)$ of the corresponding pure boundary state. This is, of course, equal to the entropy $S(\rho_B)$ that we are really interested in.

As can be seen immediately from the tensor network picture, this is exactly the same situation that we considered before, except that $B$ has been replaced by $\overline{B} \otimes R$, $b$ has been replaced by $\bar b \otimes R$, and the areas $A_1$ and $A_2$ have been exchanged. It follows that
\begin{align}
    S(B) = S(\overline{B} R) \approx \frac{A_1}{4 G} + S(\bar b R)_\psi = \frac{A_1}{4 G} + S(b b')_\rho~,
\end{align}
so long as
\begin{align}
      \frac{A_2}{4 G} + H_\text{min}^\varepsilon(b'|\bar b R)_\psi - \frac{A_1}{4 G}\gg 0~.
\end{align}
Since $\ket{\psi}_{b b' \bar b R}$ is pure, $H_\text{min}^\varepsilon(b'|\bar b R)_\psi = - H_\text{max}^\varepsilon(b'|b)_\rho$ and so this is exactly \eqref{eq:maxconditionholding}.

It is worth briefly commenting on whether an equivalent formula to \eqref{eq:haaraveragething} could be directly shown in gravity. The proof of \eqref{eq:haaraveragething} is reviewed in Appendix \ref{app:one-shot_decoupling} and involves calculating the Hilbert-Schmidt distance between the entangled and product states, and then using the Hilbert-Schmidt norm to bound the trace-norm. Since the Hilbert-Schmidt norm can be computed directly using a path integral (without analytic continuation), it should in principle be possible to evaluate in gravity. Just like for R\'{e}nyi entropy calculations, for fixed-area states the gravity answer should agree with the random tensor network answer. However, in order to derive a gravitational decoupling theorem, one would still need to use some quantum information tricks (basically a clever application of the Cauchy-Schwarz inequality) in order to eventually bound the trace-norm. The derivation would therefore still not be a completely direct gravity calculation.

\subsection{The regime where na\"{i}ve QES fails}\label{sec:necessity_large_corrections}
So far we have only argued that there \emph{aren't} significant correction to the QES prescription, so long as we \emph{don't} have
\begin{align}
    \frac{A_1}{4G} + H^\varepsilon_\text{max}(b'|b) \gtrsim \frac{A_2}{4 G} \gtrsim \frac{A_1}{4G} + H^\varepsilon_\text{min}(b'|b)~.
\end{align}
In this section, we argue that there \emph{do} exist significant corrections (at least $\mathcal{O}(1)$ in size) when
\begin{align} \label{eq:violationcorrections}
    \frac{A_1}{4G} + H^\varepsilon_\text{max}(b'|b) \gg \frac{A_2}{4 G} \gg \frac{A_1}{4G} + H^\varepsilon_\text{min}(b'|b)~.
\end{align}
Here, $\varepsilon$ can be relatively small, but should be parametrically $\mathcal{O}(1)$ in the semiclassical limit. Note that we do not have a general argument that these corrections need to be \emph{leading order}, although we strongly expect that this is the case so long as \eqref{eq:violationcorrections} holds at leading order, as we found in the simple examples in Section \ref{sec:examples}. Finding a proof that this is true in full generality is an important task for future work.

Our main tool will be the converse one-shot decoupling theorem of \cite{dupuis2014one}. Applied to the random tensor network in Figure \ref{fig:RTN}, this says that, if \eqref{eq:violationcorrections} holds, then
\begin{align}
    \lVert \rho_{\overline B' \overline b R} - \rho_{\overline B'} \otimes \rho_{\overline b R}\rVert_1 = \mathcal{O}(1)
\end{align}
and
\begin{align}
    \lVert \rho_{B' b } - \rho_{B'} \otimes \rho_{b}\rVert_1 = \mathcal{O}(1)~.
\end{align}
Using Pinsker's inequality, these lower bound the relative entropies 
\begin{align}
    S(\rho_{B' b} || \rho_{B'} \otimes \rho_{b}) = I(B':b)
\end{align}
and
\begin{align}
    S(\rho_{\overline B' \overline b R} || \rho_{\overline B'} \otimes \rho_{\overline b R}) = I(\overline B':\overline b R)~.
\end{align}
There is therefore at least an $\mathcal{O}(1)$ amount of mutual information both between $B'$ and $b$ and betwen $\overline B'$ and $\overline b \otimes R$. Since there is no upper bound on the relative entropy from the trace distance, the mutual information can, of course, be parametrically larger than $\mathcal{O}(1)$, which we expect to happen when the inequalities holds at leading order.

The na\"{i}ve QES prescription says that
\begin{align}
    S(B) =S(B'b) = S(\overline B R) =S(\overline B' \overline b R) = \min \bigg(\ln d_{B'} + S(b),\ln d_{\overline B'} + S(b b') \bigg)~.
\end{align}
However
\begin{align}
    \ln d_{B'} + S(b) > S(B') + S(b) = S(B'b) + I(B':b)~,
\end{align}
and
\begin{align}
    \ln d_{\overline B'} + S(b b') > S(B') + S(\overline b R) \geq S(B'b) + I(\overline B':\overline b R)~.
\end{align}
Since there is $\mathcal{O}(1)$ mutual information in each case, we find that the na\"{i}ve QES prescription receives at least $\mathcal{O}(1)$ corrections. Since the same replica trick calculation gives the entropy of both the random tensor network and the corresponding fixed-area state, the same is true for fixed-area states.

\subsection{From fixed-area states to general holographic states}\label{sec:fixedarea_to_general}
Of course, most of the time we are not interested in fixed-area states. Instead the states of interest (vacuum-AdS, perturbative excitations above the vacuum, thermofield double states etc.) generally have small ($\mathcal{O}(\sqrt{G})$) fluctuations in the area of the extremal surface(s).

In this section, we argue that, up to small $\mathcal{O}(\ln G)$ corrections, the entropies of such states can be calculated by expanding the state as a superposition of fixed-area states, and then taking an expectation of the entropies of the states in the superposition.\footnote{For simplicity, in this section we only consider pure states. If the state of interest is mixed, one can simply first purify it using a reference system and then replace all references to the complementary region $\overline{B}$ in the argument below by $\overline{B} \otimes R$.}

From our point of view, the primary importance of this result is that, at leading order, the entanglement entropy of a generic holographic state is the same as the entropy of a fixed-area state with the same classical area.
Therefore general holographic states inherit the leading order corrections we found for fixed-area states. It also shows that the corrections to the na\"{i}ve QES prescription are small, for general holographic states, so long as
\begin{align}
    \frac{A_2 -A_1}{4G} - H_\text{max}^\varepsilon(b'|b) \gg \sqrt{\frac{\ln G}{G}}~,
\end{align}
and similarly for $H_\mathrm{min}^\varepsilon(b'|b)$.

That said, our argument also has other technical applications, for example bounding the error in the assumptions used in \cite{Marolf:2020vsi} to calculate the $\mathcal{O}(1/\sqrt{G})$ corrections to the entanglement entropy near a QES phase transition.

Our starting point is that the general holographic state $\ket{\psi}$ can be written as a superposition over fixed-area states $\ket{A_1, A_2}$ as
\begin{align}
    \ket{\psi} = \sum_{A_1, A_2} \sqrt{p(A_1,A_2)} \ket{A_1, A_2}~.
\end{align}
The fluctuations in the area are Gaussian (in the semiclassical limit) with width $\mathcal{O}(\sqrt{G})$ (see \cite{Marolf:2020vsi} for detailed calculations), so we can approximate the state up to any polynomially small error (w.r.t $G$) by a state with support only on an $\mathcal{O}(\sqrt{G \ln G})$ range of values.\footnote{The error in neglecting the tail of a Gaussian outside a window $\Delta x$ goes like $e^{-{\mathcal{O}\left(\Delta x/ \sigma \right)^2}}$. So if $\Delta x$ equals $k$ standard deviations, the error goes like $e^{-\mathcal{O}(k^2)}$. Hence the $\sqrt{G}$ is for the standard deviation, and the $\sqrt{\ln G}$ ensures we capture a greater number of standard deviations as $G \to 0$, such that the error tends to zero polynomially in $G$.}

As with any continuously valued measurement operator, it is not well defined to measure the area exactly. Instead, the area operator should be viewed as a projection-valued measure (PVM), and the states $\ket{A_1, A_2}$ should be viewed as the outcome of measuring the area to some precision $\delta$. We shall take $\delta$ to be polynomially small with respect to $G$.

It follows from the preceding two paragraphs that the number of distinct fixed-area states in the superposition scales as $\mathcal{O}(G \ln G /\delta^2)$. (Note that we are taking a superposition over states with both $A_1$ and $A_2$ fixed, which squares the number of terms in the superposition.) Crucially this means that the number of states is polynomial in $1/G$.

We are now almost ready to consider the reduced density matrix $\rho_B = \tr_{\overline{B}} \ket{\psi} \bra{\psi}$. However, as an intermediate step we first consider taking a superposition over only states with different values of $A_1$, for some fixed $A_2$. In other words, we have
\begin{align}
    \rho_B (A_2) = \sum_{A_1, A_1'} \sqrt{p(A_1|A_2) p(A_1'| A_2)} \tr_{\overline{B}} \bigg( \ket{A_1, A_2} \bra{A_1', A_2} \bigg)~.
\end{align}
The first thing to observe is that the bulk operator $\hat A_1$ is always reconstructable on the boundary region $\overline{B}$. Hence the only terms that survive the partial trace have $A_1 = A_1'$.

We therefore find that $\rho_B (A_2)$ can be written as the incoherent mixture
\begin{align}
    \rho_B (A_2) = \sum_{A_1} p(A_1|A_2) \tr_{\bar B} \bigg( \ket{A_1, A_2} \bra{A_1, A_2} \bigg)~.
\end{align}
However, as discussed in Section \ref{sec:contradictions}, we can bound the entropy of such a mixture from above and below by
\begin{align} \label{eq:boundsonA1fluct}
    \sum_{A_1} p(A_1|A_2) S(B)_{\ket{A_1, A_2}} \leq S(\rho_B (A_2)) \leq \sum_{A_1} p(A_1|A_2) S(B)_{\ket{A_1, A_2}} - p(A_1|A_2) \ln p(A_1|A_2)~.
\end{align}
The difference between the upper and lower bounds is an $O(\ln G)$ entropy of mixing term (because there were $\mathcal{O}(\sqrt{G \ln G}/\delta)$ distinct states in the superposition) and hence can be ignored at leading order (and for calculating the $O(1/\sqrt{G})$ corrections discussed in \cite{Dong:2020iod,Marolf:2020vsi}).

Now we need to take a superposition over different values of $A_2$. Because all the states involved are pure, $S(\rho_B) = S(\rho_{\overline{B}})$, and, for any $A_2$, $S(\rho_B (A_2)) = S(\rho_{\overline{B}} (A_2))$. We can therefore compute the entropy of the reduced state on $\overline{B}$ rather than $B$.

Since $A_2$ can always be reconstructed on $B$, this is again an incoherent mixture
\begin{align}
    \rho_{\overline{B}} = \sum_{A_2} p(A_2) \rho_{\overline{B}}(A_2).
\end{align}
Hence we have
\begin{align} \label{eq:boundsonA2fluct}
    \sum_{A_2} p(A_2) S(\rho_B(A_2)) \leq S(\rho_B) \leq \sum_{A_2} p(A_2) S(\rho_B(A_2)) - p(A_2) \ln p(A_2).
\end{align}
Again the difference between the lower and upper bounds is $O(\ln G)$ and so can be ignored in lower order calculations.

Altogether, we therefore find
\begin{align}
    S(\rho_B) = \sum_{A_1, A_2} p(A_1,A_2) S(B)_{\ket{A_1, A_2}} + O(\ln G)~,
\end{align}
which is exactly what we set out to show. In particular, the QES prescription is valid at leading order for general holographic states, whenever it is valid for the corresponding fixed-area states. Moreover, the QES prescription receives leading order corrections, whenever there are leading order corrections to the entropy of corresponding fixed-area states. When the difference in areas is smaller than the fluctuations in this difference, we also find the ($\mathcal{O}(\sqrt{1/G})$) corrections from \cite{Penington:2019kki,Dong:2020iod, Marolf:2020vsi}.

There's one remaining remark to make. The fluctuations in the areas $A_1$ and $A_2$ are formally divergent when we take the radial cut-off to infinity. This leads to a natural question of whether we were justified in treating the potential entropy of mixing terms as smaller than $\mathcal{O}(1/G)$ but non-divergent corrections.

The short answer is that this subtlety does not matter for our purposes.
The IR fluctuations create a \emph{constant} (divergent) difference between the entropy in fixed-area states and entropy in general states, independent of which QES is dominant, independent of the bulk state. 
Hence the entropy $S(B)$ in a general bulk state can be computed as the expectation of the entropies of the fixed-area states in its superposition, as we already argued, \emph{plus} a constant shift. 
This shift is currently underappreciated and deserves more study, but it does not affect our ability to infer general corrections from fixed-area states.

The longer answer is as follows.
First note that the IR pieces of $A_2$ and $A_1$ are the same, so the fluctuations of $A_1$, in states where the area of $A_2$ is fixed, do not diverge \cite{Marolf:2020vsi}.
This implies the difference between the lower and upper bounds in \eqref{eq:boundsonA1fluct} is genuinely finite.
Moreover, this implies that $A_2 - A_1$ is independent of this IR subtlety, implying the condition for corrections \eqref{eq:correction_conditions} remains well-defined. 

The important effect of this IR subtlety is in the entropy of mixing term in \eqref{eq:boundsonA2fluct}, and is indeed divergent. 
However, it represents a large constant shift -- not a large window -- because the lower bound can be strengthened to include this divergence as well. 
This works as follows.
Let there be some fixed radial cutoff $\varepsilon$, such that $A_2$ diverges in the $\varepsilon \to 0$ limit. 
Group fixed-area states into blocks corresponding to some $\mathcal{O}(1)$ range of areas. There are a polynomial in $1/G$ number, $\mathcal{O}(1/\delta)$, of fixed-area states in each block. This number grows as $G \to 0$.
As the IR cutoff is taken away, the number of such blocks grows as $1/\varepsilon$ to some power.

We can separate the Shannon term associated to the mixing of these blocks from the Shannon term associated to the mixing of the $\mathcal{O}(1/\delta)$ states within each block. 
The first Shannon term does not depend on $G$, only on $\varepsilon$.

This IR Shannon term, crucially, can be included in the above lower bound of \eqref{eq:boundsonA2fluct}.
The resulting inequality is true because the different blocks are distinguishable on both $B$ \emph{and} $\overline{B}$.
Indeed, $A_1$, a quantity known to $\overline{B}$, takes on vastly different values in the different blocks (because its IR value matches that of $A_2$). 

Therefore, the entropy of mixing associated to the IR fluctuations of area can be understood as a constant shift to the entropy, present even if there is just a single fixed-area surface.
This concludes the argument. 

\section{Entanglement wedge reconstruction}\label{sec:EWR}

This refinement of the QES prescription brings with it a refinement of the condition for entanglement wedge reconstruction (EWR).

We show in Section \ref{sec:state_dep_EWR} the refined conditions are the following.
A region $B$ of the boundary will be able to reconstruct the state of a region $b'$ of the bulk, as in Figure \ref{fig:dustball}, given a bulk state $\rho$, if and only if\footnote{We discuss setups with more than two candidate QES in Section \ref{sec:refined_prescription}.}
\begin{equation}\label{eq:EWR_condition}
    \frac{A_2}{4G}  \gg \frac{A_1}{4G} + H^\varepsilon_\mathrm{max}(b'|b)_\rho~.
\end{equation}
This condition is similar to that from Hayden and Penington\footnote{See also Dong, Harlow, and Wall \cite{Dong:2016eik}, which first derived EWR in settings with small code subspaces, where the minimal QES is determined up to perturbative corrections by the area term.} \cite{Hayden:2018khn} (see also \cite{Akers:2019wxj}), but builds on it in a key way. The similarity is that both depend at some level on the comparison between $\Delta A / 4 G$ and  $H_\mathrm{max}(b')_\rho$ (though \cite{Hayden:2018khn} did not say it this way).

The key difference is that \eqref{eq:EWR_condition} tells you whether $B$ can reconstruct \emph{the particular state} $\rho$. 
The condition from \cite{Hayden:2018khn} tells you whether there exists a single reconstruction procedure that works for \emph{any state} in a code subspace that contains $\rho$.

This difference shows up in two places: the smoothing of $H_\mathrm{max}$, and the conditioning on $b$. The smoothing allows us to only care about the \emph{approximate} dimension of $\rho_{b'}$, formalizing the intuitive notion that we can ignore small pieces of the wavefunction and still approximately reconstruct the state.
The conditioning on $b$ quantifies how entanglement in $\rho$ helps $B$ reconstruct $b'$, formalizing the intuition that bulk entanglement between $b$ and $b'$ can aid reconstruction. 

In Section \ref{sec:EWR_state_merging}, we explain that this new, state-specific formulation of EWR \eqref{eq:EWR_condition} is equivalent to a well-known quantum information task, one-shot quantum state merging. 
Furthermore, we explain that the AdS/CFT dictionary performs this task \emph{maximally efficiently}. 
EWR is just very efficient one-shot quantum state merging.\footnote{Let us make a helpful distinction. The term ``entanglement wedge reconstruction'' usually means two things at the same time: the task of encoding $b'$ into $B$ (and then decoding), and also the particular protocol implicit in the AdS/CFT dictionary, the protocol that performs the task. The task, we will explain, is a special case of quantum state merging. The protocol, we will argue, is a very efficient way to perform quantum state merging.}

\subsection{State-specific EWR}\label{sec:state_dep_EWR}
Let us first carefully define what we mean by EWR for an arbitrary, single bulk state $\rho$.

Traditionally, EWR has been defined not for a single (mixed) state $\rho$, but for a \emph{code subspace} of states $\mathcal{H}_\text{code}$. There are then two definitions of what it means for EWR to be possible, depending on whether we work in the Schr\"{o}dinger or Heisenberg picture. In the Schr\"{o}dinger picture, we need to find a quantum channel $\mathcal{R}: B \to b \otimes b'$ that recovers the reduced bulk state on $b \otimes b'$ from the reduced boundary state on $B$, for any state in the code subspace. In the Heisenberg picture, for any bulk operator acting on $b \otimes b'$, we need to find an operator $O_B$, acting only on the boundary region $B$, whose action is the same as the action of the bulk operator, when applied to any state in the code subspace.\footnote{We also require that, when the bulk operator is Hermitian, the boundary reconstruction is also Hermitian, and, when the bulk operator is unitary, the boundary operator is also unitary.}

We can replace this definition with a definition that considers only a single state, by utilizing a canonical purification $\ket{\psi}_{b b' \bar b R}$ of the maximal mixed state within the code subspace. 
In this language, EWR is possible if and only if, for any bulk operator on $b \otimes b'$, there exists an operator reconstruction on $B$ that has the correct action on $\ket{\psi}_{b b' \bar b R}$. 
Similarly, in the Schr\"{o}dinger picture, EWR is possible -- in this single-state language -- if and only if it is possible to recover a canonical purification of the bulk state on $\bar b \otimes R$ from the boundary state on region $B$.

If EWR were exact, this single-state definition would be exactly equivalent to the traditional, code subspace definition. However, because EWR is in practice only approximate, there is a slight difference. In the traditional definition, the error is commonly defined as the `worst-case' error, i.e. the largest output error for any input state. The error when acting on a maximally entangled state is more like an `average-case' error: the reconstruction can do a lot worse on particular input states, as long as it does well for most input states. (See e.g. the discussion in \cite{Chen:2019gbt}.)

An advantage of this new, single-state definition is that it very naturally generalizes to mixed bulk states $\rho$ with an entanglement spectrum that isn't flat. Again, we simply say that EWR is possible if a boundary operator exists with the correct action on a (canonical) purification of the bulk state $\rho$. When the state $\rho$ is unentangled, this just means that we are taking a `weighted-average' error, where $\rho_{b'}$ tells us how different states should be weighted. However, when the state $\rho_{b b'}$ is entangled, we can take advantage of that entanglement to make reconstruction easier. This has no classical analogue.

When is EWR of region $b'$ -- using this more general definition -- possible? We start by considering the tensor network shown in Figure \ref{fig:RTN}. In this setup, a necessary and sufficient condition for EWR is \emph{approximate decoupling} \cite{hayden2008decoupling, dupuis2010decoupling}. Namely, that
\begin{align} \label{eq:approximate_decoupling}
    \left\lVert \tr_{B'} \left(V \rho_{b' \overline b R} V^\dagger\right) - \rho_{\overline b R} \otimes \frac{\mathbb{1}_{\overline B'}}{d_{\overline B'}} \right\rVert_1 \leq \varepsilon~.
\end{align}
Roughly speaking, the intuition for this is that all purifications are equivalent up to unitaries, and $B$ purifies $\overline B' \otimes \overline b \otimes R$. It follows that, if (and only if) the reduced state on $\overline B' \otimes \overline b \otimes R$ is (approximately) the product of a state on $\overline B'$ and a state on $\overline b \otimes R$, then we can extract a purification of $\overline b \otimes R$ from $B$. As discussed above, this is just the Schr\"{o}dinger picture definition of EWR.

As discussed in Section \ref{sec:fixed_area_conditional}, \eqref{eq:approximate_decoupling} holds if and only if 
\begin{align} \label{eq:EWRcond}
    - H_\text{min}^\varepsilon(b'| \bar b \otimes R) = H_\text{max}^\varepsilon(b'|b) \ll \ln \frac{d_{B'}}{d_{\overline B'}} = \frac{A_2 - A_1}{4 G}.
\end{align}
In other words, the condition for EWR of region $b'$ is exactly the condition for the QES prescription to be valid, with minimal QES $\gamma_1$ (and hence region $b'$ is `in the entanglement wedge').

We would like to show that the same condition holds for EWR in gravity. Given our discussion in Section \ref{sec:fixed_area_conditional} about the close connections between random tensor networks and fixed-area states, it should be unsurprising that this indeed the case.

The simplest argument for this is to use the Petz map reconstruction \cite{barnum2002reversing,ohya2004quantum,Cotler:2017erl}. This is a explicit general-purpose construction for reconstructing operators that is known to be close to optimal. Specifically, using the Petz map (with reference state $\rho_{b b'} \otimes \sigma_{\overline b}$ for any full-rank state $\sigma_{\overline b}$ ) will give a reconstruction error that is at most twice the optimal error \cite{barnum2002reversing, Chen:2019gbt}. Hence, for the random tensor network the Petz map reconstruction will work with small error, if and only if \eqref{eq:EWRcond} holds.

However, Petz map matrix elements can be computed using a replica trick \cite{Penington:2019kki}. And, as for the von Neumann entropy, the replica trick calculation is identical for both fixed-area states and random tensor networks \cite{Penington:2019kki, Jia:2020etj}. We can therefore use the known results for random tensor networks to do the analytic continuation and conclude that the Petz map reconstruction succeeds (and hence EWR is possible at all) if and only if \eqref{eq:EWRcond} holds.

What about EWR in states where the extremal surface areas are not fixed? Since the area operator $A_2$ can always be measured on $B$, we are free to consider states of fixed $A_2$. If entanglement wedge reconstruction is possible for all values of the area $A_2$, it must also be possible for states that involve superpositions over $A_2$, because we can reconstruct an operator $\phi_{b'}$ as
\begin{align}
    \phi_B = \sum_{A_2} \Pi_{A_2} \phi_B^{(A_2)} \Pi_{A_2}~,
\end{align}
where the sum is over possible values of the area $A_2$, $\phi_B^{(A_2)}$ is a reconstruction of $\phi_{b'}$ for states with area $A_2$, and $\Pi_{A_2}$ is a projector onto the area being $A_2$. 

In general, we can't do the same thing for the area $A_1$, since it is not always measurable from $B$. However, if the region $b'$ is reconstructable on $B$ for all states in the superposition, then $A_1$ can be reconstructed in $B$ for all the states, and we can use exactly the same argument to contruct operators that work for superpositions of eigenstates of $A_1$. We therefore conclude that entanglement wedge reconstruction is possible so long as
\begin{align}
    \frac{A_2 - A_1}{4 G} - H_\text{max}^\varepsilon(b'|b) \gg \mathcal{O}\bigg(\sqrt{\frac{\ln G}{G}}\bigg)~.
\end{align}

The above argument was somewhat sloppy. Our previous argument for EWR of the region $b'$, in fixed-area states, involved operators that acted within a single fixed-area code subspace (as in the tensor networks). The operator $A_1$ instead compares code subspaces with different areas. How do we know that it has the same reconstruction conditions?

Again, we can turn to the Petz map. To reconstruct the operator $A_1$ using the Petz map, we need to consider a reference state that involves a mixture of states with different areas $A_1$. In the replica trick calculation of the Petz map matrix elements, the area $A_1$ in each replica has to be the same, whenever the different replicas are glued together at the surface $\gamma_1$. If this is the case, the parts of the mixed reference state with the `wrong' area $A_1$ will not contribute to the operator action, and the reconstruction will succeed. If some of the replicas are instead glued together at the surface $\gamma_2$, then the areas $A_1$ do not need to be the same, and the reconstruction will fail.

The same statement is also true for the Petz reconstruction of ordinary bulk operators in region $b'$ \cite{Penington:2019kki}: the reconstruction succeeds if and only if the contribution from saddles where replicas are glued together at $\gamma_2$ is small (and so can be safely ignored while doing the analytic continuation). We already argued that those reconstructions succeed when \eqref{eq:EWRcond} holds. Hence, when \eqref{eq:EWRcond} holds, gluing at $\gamma_1$ must dominate the analytic continuation, and hence the operator $A_1$ must also be reconstructible.

\subsection{EWR as one-shot quantum state merging}\label{sec:EWR_state_merging}
This single-state reformulation of EWR is a special case of a ubiquitous information-theoretic task, known as one-shot quantum state-merging \cite{QI_can_be_negative, horodecki2007quantum, berta2009single}.

In quantum state-merging, Alice and Bob share a quantum state. This state is chosen from some arbitrary ensemble of pure states with density matrix $\rho_{AB}$. Alternatively, we can consider a single purification $\ket{\psi}_{ABR}$ of $\rho_{AB}$. The objective of the task is to transfer Alice's part of the state to Bob while sending as few qubits from Alice to Bob as possible. In other words, to produce an output state $\ket{\psi'}_{A'BR} \approx \ket{\psi}_{ABR}$ where the $A'$ and $B$ subsystems are both held by Bob. Equivalently, the average error between the initial state (shared between Alice and Bob) and the final state (held only by Bob), should be small, where the average is over the pure states in the ensemble with density matrix $\rho_{AB}$.

It should be clear that this task is closely related to state-specific EWR. There too, the bulk state either is chosen from some ensemble $\rho_{b b' \bar b}$, or is purified by a reference system as $\ket{\psi}_{b b' \bar b R}$. The part of the state that is held by Bob corresponds to the part of the state that is encoded in the boundary region $B$. The bulk region $b$ is always encoded in the boundary region $B$; this corresponds to the part of the state that is \emph{initially} held by Bob.

In the one-shot setting (where Alice and Bob are trying to merge a single copy of the state), it is known that the minimum number of qubits required for state merging is $H_\text{max}(A|B) / \ln(2)$. 
Remarkably, this is exactly how many qubits gravity seems to require!
The number of qubits from region $b'$ that can be decoded in region $B$ is $\Delta A/ 4\mathrm{ln}(2) G$, as stated in \eqref{eq:EWRcond}. 
Hence it seems that EWR can be explained not just as a special case of quantum state merging, but as an optimal implementation of it!

However, there is an important caveat that we have ignored until now. In quantum state merging as traditionally defined, it is crucial that unlimited \emph{classical} information can be sent from Alice to Bob \cite{QI_can_be_negative}. Without this classical communication, significantly more quantum communication would be required. 

Holography does \emph{not} transfer large amounts of classical information from $b'$ to $B$. Indeed, the amount of transferred classical information is bounded by the Holevo information, which is also equal to $\Delta A/ 4 G$ \cite{Bao:2017guc}.
That is, the total number of transferred qubits \emph{plus} bits is bounded by $\Delta A / 4 G$.
There is no additional classical communication that can make state merging achievable.

So if EWR is accomplishing state merging, why did our results from Section \ref{sec:state_dep_EWR} suggest that we only need
\begin{align}
\frac{\Delta A} {4 G} > H^\varepsilon_\text{max}(A|B)\label{eq:deltaAsmoothmax}
\end{align}
for EWR to be possible? It turns out that the full power of classical communication is unnecessary for quantum state merging. Instead, a weaker communication primitive, known as zero-bit communication, is sufficient \cite{hayden2017alphabits}. The number of zero-bits communicated from region $b'$ encoded in region $B$ is not constrained by $\Delta A$, and it is this additional information that allows the state merging protocol to succeed when \eqref{eq:deltaAsmoothmax} holds.

To understand this, we start with the resource inequality governing a highly efficient, rather general quantum protocol, the ``one-shot mother protocol,'' also known as (one-shot) quantum state transfer or fully quantum Slepian-Wolf \cite{abeyesinghe2009mother, Datta_2011}. 
The inequality states that 
\begin{equation} \label{eq:mother}
\begin{split}
    \langle \psi_{ABR}\rangle + \frac{\left[H_0^{\varepsilon}(A)_\psi + H^{\varepsilon}_{\text{max}}(A | B)_{\psi}\right]}{2\ln(2)}
    \, \mathrm{qubits} \geq \frac{\left[H_0^{\varepsilon}(A)_\psi - H^{\varepsilon}_{\text{max}}(A | B)_\psi\right]}{2\ln(2)} \,\mathrm{ebits} + \langle \psi_{A'BR} \rangle~.
\end{split}
\end{equation}
At first glance, this inequality is somewhat terrifying. Let's take some time to unpack it. The whole statement relates the relative usefulness of different quantum communication resources. On the left, we start with the state $\ket{\psi}$, which is shared between Alice, Bob, and the reference $R$. Alice also has the ability to send $
    \left[H_0^{\varepsilon}(A)_\psi + H^{\varepsilon}_{\text{max}}(A | B)_{\psi}\right]/2\ln(2)$ qubits to Bob.
    
The claim is that this is more useful to Alice and Bob than the resources on the right hand side, because the resources on the left can be used to \emph{create} the resources on the right (up to some small error). What are the resources on the right? We still have the state $\ket{\psi}$, but it has now been successfully `merged,' so that everything except the reference is now in system $A'B$, held entirely by Bob. Alice and Bob have also gained $\left[H_0^{\varepsilon}(A)_\psi + H^{\varepsilon}_{\text{max}}(A | B)_{\psi}\right]/2\ln(2)$ Bell pairs or `ebits'. 
    
    For clarity of presentation, we dropped additional terms in \eqref{eq:mother} of size $\mathcal{O}(\ln \varepsilon)$, terms correcting the number of qubits required and ebits produced. These corrections are subleading for appropriate choices of $\varepsilon$ in the limit where the entropies are large. We note that the inequality is optimal in the following sense: in any protocol for one-shot quantum state transfer, the number of qubits communicated, minus the ebits of entanglement gained, will be at least
    $$
    H^{\varepsilon'}_{\text{max}}(A | B)_{\psi} + \mathcal{O}(\ln(\varepsilon')),
    $$
    for a particular $\varepsilon'$ that is controlled by the protocol error.
    
    How does this relate to quantum state merging? In the language of resource inequalities, \emph{quantum teleportation} states that
    \begin{align} \label{eq:cbittele}
        1 \,\text{ebit} + 2 \,\text{cbits} \geq 1\, \text{qubit}~,
    \end{align}
    where a cbit is a classical bit.
    Substituting this inequality into \eqref{eq:mother}, and recalling that classical communication is free in traditional quantum state merging, we find that the number of qubits that need to be sent is $H_\text{max}^\varepsilon(A|B)$.
    Hence unlimited classical communication does allow Alice to give her state to Bob, just by using the mother protocol and transferring $H^\varepsilon_\mathrm{max}(A|B)$ qubits.
    
    As an aside: note that quantum conditional entropies can be negative. What does it mean if only a negative number of qubits need to be sent from Alice to Bob? The answer is that the communication cost in state merging is defined \emph{catalytically}. If the protocol produces Bell pairs, these can be stored, ready to use, together with the free classical communication, to produce quantum communication in the future. We can end up with more ability to communicate than we started with!
    
    Returning to the main point, we emphasize that classical bits are not actually required to do teleportation. Zero-bits are sufficient. We have
    \begin{align} \label{eq:0bittele}
        1 \,\text{ebit} + 2\, \text{zero-bits} \eqa 1\, \text{qubit}~,
    \end{align}
    where the $(a)$ means that \eqref{eq:0bittele} only holds at leading order in the limit where we have a large number of each type of bit. Note that, unlike \eqref{eq:cbittele}, \eqref{eq:0bittele} is an equality, not an inequality. Zero-bits are the minimal resource required for teleportation.
    
    Therefore, with enough zero-bits communicated from Alice to Bob, Alice can give Bob her state with just $H^\varepsilon_\mathrm{max}(A|B) / \ln(2)$ qubits, using the mother protocol.
    To see this, substitute \eqref{eq:0bittele} into \eqref{eq:mother}, finding that
    \begin{align} \label{eq:zerobitmother}
        \langle \psi_{ABR}\rangle  + \frac{H^{\varepsilon}_{\text{max}}(A | B)_{\psi}}{\ln(2)}\, \mathrm{qubits} + \frac{\left[H_0^{\varepsilon}(A)_\psi -H^{\varepsilon}_{\text{max}}(A | B)_{\psi}\right]}{\ln(2)}\, \text{zero-bits} \geq \langle \psi_{A'BR} \rangle~.
    \end{align}
    State merging is just as easy with free zero-bit communication as with free classical communication.
    
    How many zero-bits \emph{are} communicated from $b'$ to $B$?
    More broadly, what is the total amount of information about $b'$ encoded in region $B$?
    These questions were answered in \cite{Hayden:2018khn}.\footnote{The total information was correctly computed in \cite{Hayden:2018khn}, even though they used the na\"{i}ve prescription, because the relevant state for computing the transferred information is the maximally-mixed state, which is perfectly compressible.}
    For $\Delta A > 0$, region $B$ encodes the `$\alpha$-bits' of region $b'$ for
    \begin{align}
        \alpha = \frac{\Delta A}{4 G S_0}.
    \end{align}
    Here $S_0 = \ln(d_{b'})$ is the thermodynamic entropy in region $b'$. So, for example, when the code space states in region $b'$ are the possible microstates of a black hole with horizon area $A_\text{hor}$, we have $\alpha = \Delta A/ A_\text{hor}$.
    
    We can convert $\alpha$-bits into a mixture of qubits and zero-bits using another resource equality from \cite{hayden2017alphabits}, namely
    \begin{align}
        1 \,\,\alpha\text{-bit} = \alpha \,\, \text{qubits} + (1 - \alpha) \,\,\text{zero-bits}~.
    \end{align}
    We therefore find that region $B$ can receive
    \begin{align} \label{eq:totalinfo}
        \frac{1}{\ln(2)}S_0 \,\,\alpha\text{-bits} = \frac{\Delta A}{4 \ln(2) G} \,\, \text{qubits} + \frac{1}{\ln(2)}\left[ S_0 - \frac{\Delta A}{4 G} \right] \,\,\text{zero-bits}
    \end{align}
    from region $b'$. 
    This is worth emphasizing: the AdS/CFT dictionary transfers more than $\Delta A / 4 \ln(2) G$ qubits of information from $b'$ to $B$. 
    It also transfers many zero-bits, precisely $\left[ S_0 - \frac{\Delta A}{4 G} \right]/\ln(2)$.

    That was for $\Delta A > 0$; what about $\Delta A < 0$? In this case, region $B$ encodes no physical information about region $b'$ (if $b'$ is not heavily entangled with $b$).
    Nonetheless, the right hand side of \eqref{eq:totalinfo} still formally defines the amount of information from $b'$ accessible in $B$.  
    This is important, for example, if we start adding bulk entanglement, as in the following scenario.
    Imagine that more than $|\Delta A|/4 \ln(2) G$ Bell pairs are shared between regions $b'$ and $b$. Then the zero-bits of the remaining degrees of freedom in $b'$ \emph{will} be encoded in $B$. This follows from the associated phase transition in the minimal QES.
    This phase transition is reflected in \eqref{eq:totalinfo} in the following way. Converting qubits into ebits and zero-bits using \eqref{eq:0bittele}, the right hand side of \eqref{eq:totalinfo} says that $|\Delta A|/4 \ln(2) G$ ebits allow $S_0 - |\Delta A|/ 4 \ln(2) G$ zero-bits to be transferred from $b'$ to $B$, which is exactly what we just found. (Any additional ebits will continue to combine with those zero-bits to form qubits of communication, reflecting the fact that adding more and more entanglement between $b'$ and $b$ allows $B$ to recover larger and larger subspaces of $b'$.)

    As an aside, we emphasize that \eqref{eq:totalinfo} allowing additional zero-bits (on top of $\Delta A/4 \ln(2) G$ qubits) from region $b'$ to be encoded in region $B$ is not some strange phenomenon that only happens in quantum gravity. Instead, it happens very generically whenever you have a noisy quantum channel. 
    Consider the well-known properties of the quantum capacity of a channel, i.e. the number of qubits that can be communicated through that channel.
    The quantum capacity of a noisy channel is given by the so-called maximal regularized coherent information. However, the \emph{entanglement-assisted} quantum capacity is given by half the maximal mutual information, and is generically strictly larger. The difference comes from the channel having an additional zero-bit capacity. Free entanglement allows the zero-bits to be `upgraded' to qubits, giving additional qubit capacity. 
 
     Let's see how the same phenomenon manifests itself in gravity. Suppose we have $S_0 > \Delta A/4G$. In this case, without using entanglement, we can learn, at most, $\Delta A/4\ln (2) \,G$ qubits in $b'$ from $B$. Not all the information is encoded there. However, let's imagine we entangle $(S_0 - \Delta A/4 G)/2 \ln 2$ Bell pairs between region $b$ and $b'$. If we do this, all the information about the remaining $(S_0 + \Delta A/4 G)/2\ln 2$ qubits in region $b'$ will be successfully encoded in region $B$ (the entanglement wedge will have expanded to include $b'$). By using entanglement between $b$ and $b'$, we have increased the amount of information about region $b'$ that is accessible in region $B$.
    This increase in information capacity from entanglement assistance comes from the extra zero-bits in \eqref{eq:totalinfo}.
    
    Having understood the information transferred from bulk to boundary, we are now ready to interpret the conditions for EWR that we found in Section \ref{sec:state_dep_EWR}. We first note that for any state $\ket{\psi}$, we have
    \begin{align}
        H_0^\varepsilon(b')_\psi \leq H_0(b')_\psi =  \ln  \text{Rank}(\psi_{b'}) \leq \ln d_{b'} = S_0.
    \end{align}
    It therefore follows from \eqref{eq:zerobitmother} and \eqref{eq:totalinfo} that there are sufficient qubits and zero-bits for state merging, and hence the encoding (and reconstruction) of region $b'$ from region $B$ using \emph{any} protocol, if and only if 
    \begin{align}\label{eq:EWR_condition_2}
        H_\text{max}^\varepsilon(b'|b)_\psi < \frac{\Delta A}{4 G}~.
    \end{align}
    This is exactly what we found in Section \ref{sec:state_dep_EWR}. 
    
    To summarize, we noted that the task of encoding $b'$ in $B$ is the same as the task in quantum state merging. 
    This simply followed from definitions.
    We were then led to ask how \emph{efficiently} AdS/CFT \emph{performs} this task, requiring us to carefully account for exactly how much information is transferred from $b'$ to $B$ by the AdS/CFT dictionary.
    The total information, we noted, is $\Delta A / 4 \ln(2) G$ qubits \emph{plus} additional zero-bits \eqref{eq:totalinfo}.
    This is just enough transferred information for the most efficient state-merging protocol (the mother protocol) to work.
    I.e. one could not transfer the bulk information in $b'$ to $B$ using any fewer resources.
    It's remarkable that AdS/CFT encodes $b'$ in $B$ exactly when just enough information is transferred from $b'$ to $B$ for \emph{any} protocol to do it. 
    EWR is a maximally efficient state merging protocol. 

    In contrast, the na\"{i}ve QES prescription suggests that AdS/CFT exceeds the maximal efficiency bound, performing state merging as though every state were perfectly compressible.\footnote{This was the realization that led to this work. From \cite{Hayden:2018khn} we knew the amount of information being transferred from bulk to boundary. Seemingly in contradiction was the fact that the na\"{i}ve QES prescription implies EWR for any state with small enough von Neumann entropy \cite{Almheiri:2014lwa,Dong:2016eik,Harlow:2016vwg}. (Taking into account reconstruction errors means that this is only true using the state-specific definition from Section \ref{sec:state_dep_EWR}.) The resolution is that the na\"{i}ve prescription needs to be refined.}

    We emphasize that the arguments in this section should not be interpreted as an independent proof of the results from Section \ref{sec:when_corrections}. A channel having sufficient capacity to carry out some task does not automatically mean that any (possibly inefficient) protocol using that channel will actually perform the task. Conversely, one could worry that region $B$ might encode some other form of information about region $b'$, distinct from both qubits and zero-bits, which could help make state merging possible even when the zero-bits and qubits alone would be insufficient. 
    
    Instead, our point was to make precise the relationship between entanglement wedge reconstruction (and other questions in AdS/CFT) and standard protocols in quantum information, such as state merging, which may not have been clear to members of either community.
    
    In particular, we want to emphasize that the relevant quantum information protocols are always \emph{one-shot} protocols. After all, in AdS/CFT, one only typically considers a single copy of a holographic state, rather than a large number of identical copies. The only reason that the von Neumann entropy has proven relevant is that until now people have generally only considered states where the von Neumann entropy is equal to the one-shot entropies, at least at leading order. Once you consider states where this is not the case, it should not be surprising that it is one-shot entropies which play the crucial role.
    
\section{Beyond two extremal surfaces}\label{sec:refined_prescription}
So far we have presented refined conditions for the QES prescription when there are exactly two competing surfaces, \eqref{eq:refined_QES}.
In this section, we discuss the natural generalization of this rule which considers \emph{all} bulk surfaces homologous to $B$.

The upshot is that the condition for large corrections is no longer two simple inequalities; it becomes a family of inequalities. Together these inequalities determine what information is actually transmitted to $B$.

All the claims about reconstruction in this section can be shown in random tensor networks using a careful application of the one-shot decoupling theorem. We expect based on our arguments from Section \ref{sec:when_corrections} that they should also be true in AdS/CFT.

\subsection{Applying the refined prescription}
The refined way to find the entanglement wedge (EW) is as follows.\footnote{These are the refined conditions for moments of time symmetry. We expect there exists a covariant generalization, in the way HRT \cite{Hubeny:2007xt} generalized RT \cite{Ryu:2006bv}. This may well require use of the maximin formalism \cite{Wall:2012uf, Akers:2019lzs}. }

\subsubsection*{Step 1: find the \emph{max}-entanglement wedge (max-EW)}
The max-EW is intuitively the bulk region that $B$ can definitely reconstruct with small error. In this sense, it most closely resembles the traditional operational definition of the entanglement wedge.

We define the max-EW as the largest region $b$ that satisfies all of the following inequalities:
\begin{align}\label{eq:max-EW_condition}
    \forall b' \subset b,~ H^\varepsilon_\mathrm{max}(b-b' | b') < \frac{A(b') - A(b)}{4G}~, 
\end{align}
where $b-b'$ is the complement of $b'$ in $b$.

This definition implicitly assumes that there exists some `largest' region satisfying \eqref{eq:max-EW_condition} that contains all other regions satisfying \eqref{eq:max-EW_condition}. We shall prove in the next subsection that this is indeed the case. The essential intuition is that, if we can reconstruct region $b_1$, and we can reconstruct region $b_2$, then we should also be able to reconstruct their union. Having access to additional degrees of freedom can only make reconstruction easier.

In principle, \eqref{eq:max-EW_condition} requires checking infinitely many subregions $b'$. However, in practice, except in situations where the bulk entropy gradients can become very large (such as evaporating black holes) it should be sufficient to only check regions where $\partial b'$ is perturbatively close to a classical extremal surface. This is because the classical area gradient must be $\mathcal{O}(G)$ at minima of $(A(b') - A(b))/4G - H^\varepsilon_\mathrm{max}(b-b' | b')$. This justifies the simple conditions given  in \eqref{eq:refined_QES} when only two extremal surfaces exist.

\subsubsection*{Step 2: find the \emph{min}-entanglement wedge (min-EW)}
The min-EW is the complement of region $B$ definitely knows no information about. In other words, it is the region that region $B$ may know at least some information about. For pure states, it is the complement of the max-EW of $\overline{B}$. For mixed states, it can be smaller.

We define the min-EW as the smallest region $b$ that satisfies all of the following inequalities, for $\bar{b}$ the complement of $b$:
\begin{align}\label{eq:min-EW_condition}
    \forall \bar{b}' \subset \bar{b},~ H^\varepsilon_\mathrm{min}(\bar{b}' | b) > \frac{A(b) - A(b\bar{b}')}{4G}~, 
\end{align}
where $b\bar{b}'$ is the union of $b$ and $\bar{b}'$. Again, the existence of a smallest such region is nontrivial, and is equivalent to the existence of a max-EW for the purification of $B$, namely $\overline B R$.

\subsubsection*{Step 3: define EW as min-EW $=$ max-EW}
In general, the max-EW is contained in the min-EW, as we will prove in the next subsection.
In the special case in which they are the same, we can define the EW to be equal to both of them, and the entropy $S(B)$ equals the generalized entropy of this EW.

However, if the min-EW contains a region that the max-EW doesn't, then $B$ may have partial information about that region. In general in such cases, the entanglement entropy $S(B)$ will not be equal to the generalized entropy of any single surface.

\subsection{Properties of the min-EW and max-EW}\label{sec:properties_min-EW_max-EW}

In this subsection, we prove several important properties of the min-EW and max-EW. To do so, we will need certain inequalities that are satisfied by smooth min- and max-entropies.
The first is that both the min- and max-entropies satisfy strong subadditivity
\begin{align}
    H_\text{min/max}^\varepsilon (A | B) \geq H_\text{min/max}^\varepsilon (A | BC)~.
\end{align}
Secondly, the smooth min- and max-entropies satisfy a number of approximate chain rule inequalities \cite{vitanov2013chain}. Most importantly for our purposes, we have
\begin{align} \label{eq:chain}
    H_\text{max}^\varepsilon (AB | C) \leq H_\text{max}^\varepsilon (A |B C) + H_\text{max}^\varepsilon (B | C) + \mathcal{O}(\ln \varepsilon)~.
\end{align}

\subsubsection*{Property 1: existence of the min/max-EW}
We will show that, given any two regions satifying \eqref{eq:max-EW_condition}, their union will also satisfy \eqref{eq:max-EW_condition}. This immediately implies the existence of the max-EW, and implies the existence of the min-EW by the equivalence with the min-EW of $\overline B R$.
\begin{figure}
\centering
\includegraphics[width = 0.5\textwidth]{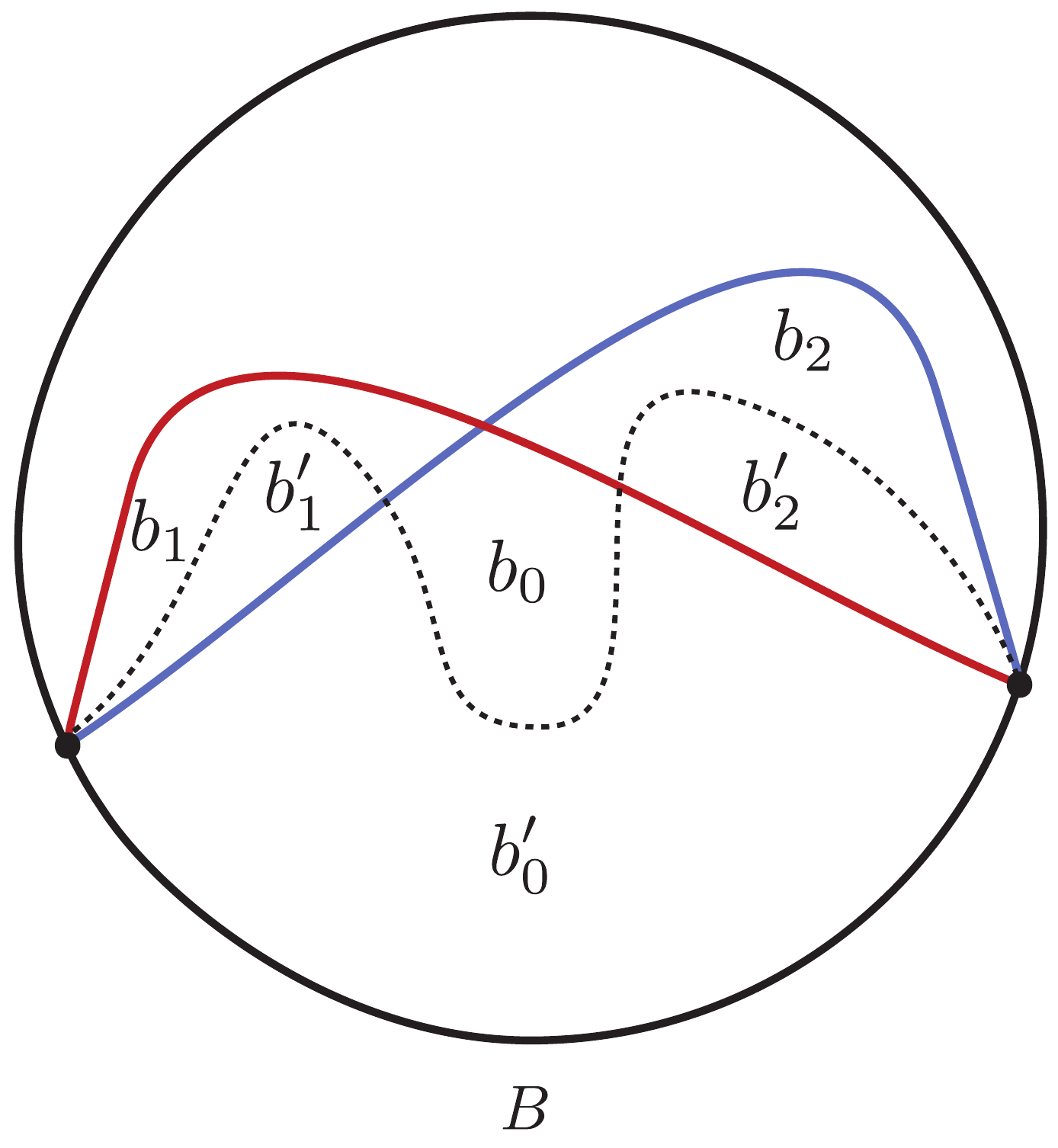}
\caption{Setup in the proof of Property 1. Both the region bounded by the blue surface and that bounded by the red surface are assumed to satisfy \eqref{eq:max-EW_condition}. We prove that therefore the union of those regions satisfies \eqref{eq:max-EW_condition}, by showing it to hold for an arbitrary choice $b' = b_0' b_1' b_2'$, depicted here bounded by the dashed black line. }
\label{fig:min_max_EW_existence}
\end{figure}
To prove this, we need to consider three overlapping regions: the two original regions, and an arbitrary subregion $b'$ of their union. These three overlapping regions can be decomposed into six disjoint regions, which we label $b_0$, $b_0'$, $b_1$, $b_1'$, $b_2$, $b_2'$, as shown in Figure \ref{fig:min_max_EW_existence}. The original two regions are given by $b_0 b_0' b_1 b_1'$ and $b_0 b_0' b_2 b_2'$. Their union is then $b = b_0 b_0' b_1 b_1' b_2 b_2'$. We need to show that
\begin{align}
    H_\text{max}^\varepsilon (b -b'|b') < \frac{A(b') - A(b)}{4G}
\end{align}
for the arbitrary region $b' = b_0' b_1' b_2'$ in $b$.

Because the original two regions satisfied \eqref{eq:max-EW_condition}, we know that
\begin{align}
    H_\text{max}^\varepsilon (b_0 b_i|b'_0 b'_i) < \frac{A(b'_0 b'_i) - A(b_0 b'_0 b_i b_i')}{4G}~,
\end{align}
for $i = 1 \text{ or }2$, as well as
\begin{align}
    H_\text{max}^\varepsilon (b_i|b_0 b'_0 b'_i) < \frac{A(b_0 b'_0 b'_i) - A(b_0 b'_0 b_i b_i')}{4G} ~.
\end{align}
Adding together these four inequalities (two for each of the two regions) and comparing the area terms, we find
\begin{align}
    H_\text{max}^\varepsilon (b_0 b_1|b_0' b_1') + H_\text{max}^\varepsilon (b_1|b_0 b'_0 b'_1) + H_\text{max}^\varepsilon (b_0 b_2 |b_0' b'_2) + H_\text{max}^\varepsilon (b_2 |b_0 b'_0 b_2') \leq 2 \frac{A(b') - A(b)}{4 G}~.
\end{align}
We can then simplify the left hand side, using
\begin{align}
    H_\text{max}^\varepsilon (b_0 b_1|b_0' b_1') + H_\text{max}^\varepsilon (b_2 |b_0 b_0' b_2') &\geq H_\text{max}^\varepsilon (b_0 b_1 |b_0' b_1' b_2') + H_\text{max}^\varepsilon (b_2 | b_0 b_0' b_1 b_1' b_2') \\&\geq H_\text{max}^\varepsilon (b_0 b_1 b_2 |b_0' b_1' b_2') + \mathcal{O}(\ln \varepsilon)~.
\end{align}
The first inequality uses SSA and the second uses the chain rule \eqref{eq:chain}. Together with a similar set of inequalities with $1$ and $2$ exchanged, this gives
\begin{align}
    2 H_\text{max}^\varepsilon (b_0 b_1 b_2 |b_0' b_1' b_2') + \mathcal{O}(\ln \varepsilon) \leq 2 \frac{A(b') - A(b)}{4 G}~.
\end{align}
The max- and min-EW are therefore well-defined, up to $\mathcal{O}(\ln \varepsilon)$ corrections (which is the same entropy difference that was required for EWR and the QES prescription to hold safely, anyway).

\subsubsection*{Property 2: min-/max-EW nesting}

Almost the exact same argument shows that the max-EW and min-EW satisfy nesting.
That is, a boundary region $B_1 \subseteq B_2$ must have a max-EW (min-EW) that is entirely contained in the max-EW (min-EW) of $B_2$. 

To prove this for the max-EW, once again let the regions $b_0$, $b_0'$, $b_1$, $b_1'$, $b_2$, $b_2'$ be disjoint, with the max-EW of $B_1$ given by $b_0 b_0' b_1 b_1'$ and the max-EW of $B_2$ given by $b_0 b_0' b_2 b_2'$.
Their union is $b = b_0 b_0' b_1 b_1' b_2 b_2'$.
We need to show that
\begin{align}
    H_\text{max}^\varepsilon (b -b'|b') < \frac{A(b') - A(b)}{4G}
\end{align}
for an arbitrary region $b' = b_0' \cup b_1' \cup b_2'$ in $b$. This will imply that the max-EW of $B_2$ should have included $b_1 b_1'$ since the beginning.
The proof, given this setup, is identical to the previous one.

The proof for the min-EW follows from nesting of the max-EW of the complement plus a puryifying reference system.

\subsubsection*{Property 3: max-EW $\subseteq$ min-EW}

The max-EW is always contained in the min-EW.
Intuitively this must be true if, as we claim, the max-EW characterizes the region that $B$ has (approximately) \emph{all} information about, while the min-EW characterizes the region that $B$ has \emph{any} information about.

To prove this, we assume for contradiction that there is some region $b'$ that is contained in the max-EW, but not in the min-EW. Let $b$ be the intersection of the max- and min-EWs, let $\overline b'$ be the region contained in the min-EW, but not the max-EW and let $\overline b$ be the complement of the union of the two wedges.

Then it must both be true that
\begin{equation}
    H^\varepsilon_\mathrm{max}(b'|b) < \frac{A(b) - A(b' b)}{4G}~,
\end{equation}
and that
\begin{equation}
    H^\varepsilon_\mathrm{min}(b'|b \overline b') > \frac{A(b \overline b' ) - A(b \overline b' b')}{4G}~.
\end{equation}
However,
\begin{align}
    \frac{A(b) - A(b'  b)}{4G} \leq \frac{A(b \overline b') - A(b \overline b' b')}{4G}~,
\end{align}
while
\begin{align}
    H^\varepsilon_\mathrm{max}(b'|b) \geq H^\varepsilon_\mathrm{max}(b'|b \overline b') \geq H^\varepsilon_\mathrm{min}(b'|b \overline b').
\end{align}
We therefore have our desired contradiction.

\subsubsection*{Property 4: max-EW $=$ min-EW only at minimal generalized entropy surfaces}
In the special case that the min-EW and max-EW equal the same region $b$, they must be bounded by a surface that minimizes $A(b)/4G + S(b)$. 

Consider a general deformation $b' = b_0 b_2$ of $b = b_0 b_1$. 
We want to show 
\begin{align}\label{eq:min_entropy_from_min_equals_max}
    \frac{A(b_0 b_2)}{4G} + S(b_0 b_2) > \frac{A(b_0 b_1)}{4G} + S(b_0 b_1)~.
\end{align}
From \eqref{eq:max-EW_condition} we know 
\begin{equation}\label{eq:this_max_EW_condition}
     \frac{A(b_0) - A(b_0 b_1)}{4G} \ge H^\varepsilon_\mathrm{max}(b_1|b_0)~.
\end{equation}
This implies
\begin{align}
    \frac{A(b_0 b_2) - A(b_0 b_1 b_2)}{4G} &> H^\varepsilon_\mathrm{max}(b_1 | b_0 b_2) \\ 
    &\ge S(b_0 b_1 b_2) - S(b_0 b_2)~,
\end{align}
where in the first line we used SSA of both area and max-entropy, and in the second line we used $H^\varepsilon_\mathrm{max}(A|B) \ge S(A|B)$.
Meanwhile, \eqref{eq:min-EW_condition} tells us
\begin{align}\label{eq:this_min_EW_condition}
    \frac{A(b_0 b_1) - A(b_0 b_1 b_2)}{4G} &< H^\varepsilon_\mathrm{min}(b_2 | b_0 b_1) \\
    &\le S(b_0 b_1 b_2) - S(b_0 b_1)~,
\end{align}
where in the second line we used $H^\varepsilon_\mathrm{min}(A|B) \le S(A|B)$.
Combining these two inequalities gives \eqref{eq:min_entropy_from_min_equals_max}, where the inequality must be strict for non-trivial $b_1 b_2$ because \eqref{eq:this_max_EW_condition} and \eqref{eq:this_min_EW_condition} are strict for non-trivial $b_1$ and $b_2$ respectively. This is what we set out to show.

The converse is not true.
A minimal generalized entropy surface will not in general satisfy all of \eqref{eq:max-EW_condition} and \eqref{eq:min-EW_condition}.
However, if all states were perfectly compressible, then this converse would be true, and therefore the na\"{i}ve QES prescription would hold. 
Indeed, (conditional) perfect compressibility implies $H^\varepsilon_\mathrm{min} = S = H^\varepsilon_\mathrm{max}$, and equations \eqref{eq:max-EW_condition} and \eqref{eq:min-EW_condition} would both be satisfied only by the minimal generalized entropy surface.

This is one way to understand the refined conditions for the QES prescription.

\subsection{Full reconstruction outside the max-EW}
Everything in the max-EW can be fully reconstructed from $B$. Similarly no information reaches $B$ from degrees of freedom outside the min-EW. However, the converses of these statements are not necessarily true. There can be regions outside the max-EW which can be fully reconstructed; and regions inside the min-EW that cannot. Nonetheless, when the min-EW and max-EW are not equal, there is always \emph{some} nonempty intermediate region that is partially, but not fully, reconstructible.

\begin{figure}
\centering
\includegraphics[width = \textwidth]{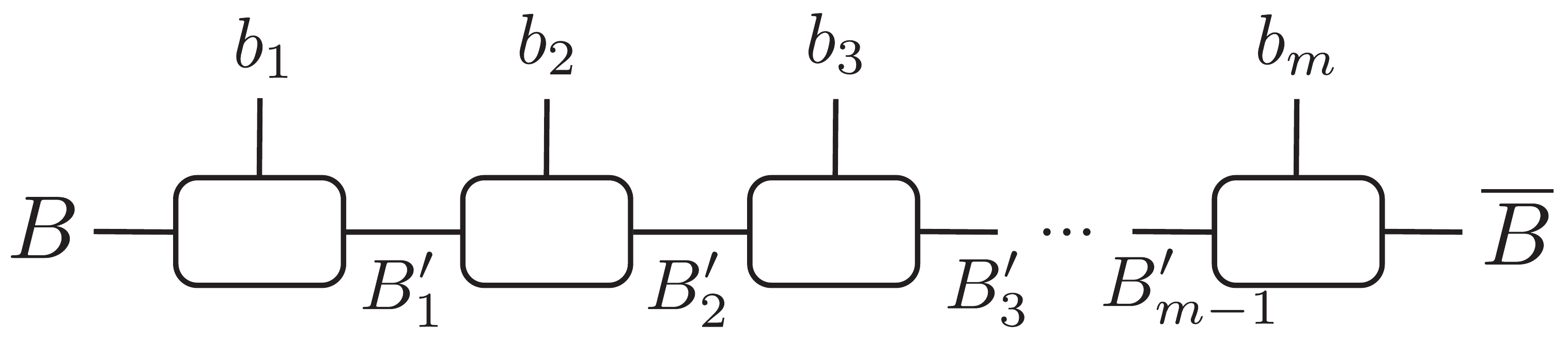}
\caption{Random tensor network used to demonstrate reconstruction outside the max-EW. Each box is a tripartite random tensor with ``bulk'' leg $b_i$, connected to the two tensors on either side of it with maximally entangled ``in-plane'' legs $B_{i-1}'$ and $B_i'$. The tensors on the ends each have a single ``boundary'' leg, $B$ or $\overline{B}$. Which bulk legs are reconstructable on $B$ depends on two things: the dimensions of the in-plane legs and the state on the bulk legs. A lower bound on the size of this reconstructable region is the max-EW, but it could be larger, depending on how much helpful information there is from the legs in the min-EW.}
\label{fig:RTN_line}
\end{figure}

A tensor network example will make this clearer. 
Consider $m$ tripartite random tensors arranged in a line, each with bulk leg $b_{i}$ for $i \in \{1,...,m\}$, each connected to the tensors to its left and right by maximally entangled ``in-plane'' legs $B'_{i}$, of dimension $e^{A_i/4G}$, except for the first and last tensor, which have one dangling in-plane leg each (the ``boundary'' legs). 
Let $B$ be the name of the left boundary leg and $\overline{B}$ be the right one, with dimensions much larger than any $e^{A_i/4G}$.
See Figure \ref{fig:RTN_line}.

Consider a bulk state that is a mixture of a) a pure state with a large amount of entanglement (with entanglement entropy $S$) between $b_2$ and $b_3$ and b) a pure state on $b_2$ and a highly mixed state (with entropy $S$) on $b_3$. If the extremal surface areas satisfy, 
\begin{align}
    \frac{A_3}{4G} - S \ll \frac{A_2}{4 G} \ll \frac{A_3}{4G} \ll \frac{A_1}{4G} \ll \frac{A_2}{4G} + S, 
\end{align}
then we find that the max-EW is $b = b_1$, while the min-EW is $\hat b = b_1 b_2 b_3$.

However, the two states in the mixture are perfectly compressible, with EWs that consist of $b_1 b_2 b_3$ and $b_1 b_2$ respectively. Hence $b_2$ is reconstructible in both. And it is easy to check that the two states must be close to orthogonal on $B$ (e.g. their entropies differ at $\mathcal{O}(1/G)$ and both have approximately flat spectra). So $b_2$ must be reconstructible for a mixture of the two states.

How is this possible? The answer is that the max-EW is the largest region that can be reconstructed without knowing anything about the state outside that region. However, because (and only because) the min-EW is larger than the max-EW, some partial information from outside the max-EW makes it through the tensor network legs. In particular, some information from $b_3$ makes it through the in-plane leg $B_2'$ (namely the part of the state which is entangled with $b_2$). And this additional information makes it possible for all the information in $b_2$ to reach $B$.

More formally, even though 
\begin{align} \label{eq:simple_cond_max}
H_\text{max}^\varepsilon(b_2|b_1) \gg (A_1 - A_2)/4G~,
\end{align}
we have
\begin{align} \label{eq:tensor_cond_max}
    H_\text{max}^\varepsilon(b_2 B_2'|b_1) \ll \frac{A_1}{4G},
\end{align}
where the state on $B_2'$ is the state produced by the entire network to its right (and then tracing over $\overline B$). If the min-EW did not contain anything outside $b_1 b_2$, this state would be maximally mixed and \eqref{eq:tensor_cond_max} would reduce to \eqref{eq:simple_cond_max}. When this is not the case, the in-plane legs can expand the fully reconstructable region.

While sensible in a tensor network, it is not clear how we should define a quantity analogous to $H^\varepsilon_\mathrm{max}(b_2 B'_2|b_1)$ in AdS/CFT, except by explicitly converting the calculation into one involving tensor networks.
Hence if the min-EW and max-EW are not equal, it may be hard to identify with certainty the full region of the bulk where everything can be reconstructed in $B$.
It will be at least as big as the max-EW, defined by \eqref{eq:max-EW_condition}, but could be larger (because of additional information from outside the max-EW).
Likewise (by looking at the complementary region in a purification, as usual), if the min-EW and max-EW are not equal, the full region of the bulk that $B$ has any information about may be smaller (but not larger) than the min-EW, defined by \eqref{eq:min-EW_condition}.

\section{Discussion}\label{sec:discussion}

\subsubsection*{Refining the QES prescription}
In this paper, we have introduced a refinement of the usual QES prescription. This refinement is both necessary for the boundary entanglement entropies to be self-consistent, and follows from careful application of the replica trick. Without our refinements, the QES prescription would only be valid for the limited subclass of states that are perfectly compressible.

Specifically, we have strengthened the conditions required for the entropy $S(B)$ to be given by the generalized entropy of the minimal QES. In the language of Section \ref{sec:refined_prescription}, this is only true when the max- and min-entanglement wedges coincide (perhaps up to perturbative corrections). When the two wedges do not coincide, the entropy $S(B)$ is much more complicated. This is closely related to the breakdown of complementary reconstruction, with a large region that cannot be fully reconstructed from either region $B$ or from its complement.

\subsubsection*{Fundamental lesson: EWR as one-shot state merging}

In many ways, this second point about entanglement wedge reconstruction (EWR) is the more fundamental one.
 For pedagogical reasons, our presentation was, in a certain sense, inverted. We led by demonstrating the large corrections to the QES prescription, in Sections \ref{sec:contradictions} through \ref{sec:when_corrections}. Only then in Section \ref{sec:EWR} did we explain that EWR should be understood through the lens of one-shot quantum state merging, necessitating the refined conditions for reconstruction.

The QES prescription is just a rule for computing one particular boundary quantity (the von Neumann entropy of a reduced state). This is just one measure of the boundary entanglement structure (albeit a very simple and useful one). EWR is stronger, telling us a deep fact about how information in the bulk is distributed on the boundary that is \emph{independent} of the particular measure (Petz map operators, relative entropies, modular flows etc.) that one might use to probe it.

As we argued in Section \ref{sec:EWR}, the information theoretic task of encoding the bulk into the boundary is manifestly a form of one-shot state merging, albeit one that uses zero-bits rather than the traditional classical bits. 
Just from this, one can see that the na\"{i}ve QES prescription implied EWR conditions that were too powerful.
There simply is not enough information transferred from the bulk to the boundary via the AdS/CFT dictionary. No quantum information protocol could encode the bulk in the boundary in the way implied by the na\"{i}ve QES prescription; 
it is incompatible with quantum Shannon theory.

This reinterpretation of EWR in terms of one-shot quantum state merging seems likely to have important future consequences. For one thing, it opens the door to connecting QES and quantum error-correction \cite{AkersPeningtonInProgress}, providing an understanding of the QES prescription that doesn’t come from the Euclidean path integral. This might shed light on how to modify Hawking’s calculation of non-unitary black hole evaporation. Indeed — the new arguments from the QES prescription \cite{Penington:2019npb, Almheiri:2019psf, Penington:2019kki, Almheiri:2019qdq} give a unitary answer, but  — unlike Hawking — make vital use the Euclidean path integral. A Hilbert space understanding of the QES prescription may connect the calculations.

\subsubsection*{Generalized min-/max-entropy}
That we know of, these refinements are the first example of a generalization of the generalized entropy that replaces the von Neumann entropy by a new entropy measure (in this case the smooth min-/max-entropy). 
The generalized entropy of a codimension-2 surface, defined as the area plus matter von Neumann entropy,\footnote{Or more generally, the gravitational entropy \cite{wald1993black, Dong:2013qoa} plus matter von Neumann entropy.} is believed to be a well-defined continuum quantity, having passed many non-trivial checks.
It is UV finite, is scheme independent, and seems to correctly generalize the classical area in many classical general relativity theorems \cite{Wall:2011hj, Engelhardt:2014gca, Bousso:2015mna, Bousso:2019dxk}.
It therefore made perfect sense to promote extremal area surfaces to extremal generalized entropy surfaces, in the na\"{i}ve QES prescription.

In contrast, the refined QES prescription asks us to do something new:
to add the smooth min-entropy or max-entropy of the bulk fields to the area.
The arguments from this paper suggest that this must be equally well-defined. In particular, there should be an appropriate renormalization procedure that makes these differences UV-finite.

The leading UV-divergence in the smooth min- and max-entropy of a subregion in quantum field theory is proportional to the area (just like for the von Neumann entropy). This is essentially because the UV-divergent parts of the subregion states are thermal Rindler modes, and hence are perfectly compressible.

However, the difference between the von Neumann entropy and the smooth min-/max-entropy will still be $\mathcal{O}(\sqrt{S})$ and hence UV-divergent \cite{HaydenUnpublished,Bao:2018pvs}. This means that the smooth min-/max-entropies cannot be renormalized by the same quantity as the von Neumann entropy.

This is OK. As discussed in Section \ref{sec:fixedarea_to_general}, the relevant area difference is not the expectation of the difference in area, but a \emph{lower confidence bound} on the difference in areas. This differs from the expectation of the difference by $\mathcal{O}(\sqrt{G})$, which is the correct scaling to renormalize the difference between von Neumann entropies and min/max-entropies. 

It is therefore natural to hope that the generalized smooth min- and max-entropies, defined as \begin{align}
    H^\varepsilon_\text{min/max}(b) + \frac{A^\varepsilon_\text{min/max}(b)}{4G}
\end{align}
with $A^\varepsilon_\text{min/max}$ respectively lower and upper-confidence bounds on the area, should be a UV-finite quantity.

If this is indeed the case, we should also expect that the conditional generalized smooth max-entropy
\begin{align}
    H^\varepsilon_\text{max}(b'|b) + \frac{\left[A(b b') -A(b)\right]^\varepsilon_\text{max}}{4G}
\end{align}
should also be UV-finite (note $\left[(A(b b') -A(b)\right]^\varepsilon_\text{max}$ is again an upper confidence bound on $A(b b') -A(b)$). There are two sets of modes that give divergent contributions to $H^\varepsilon_\text{max}(b'|b)$: modes near the boundary of $b b'$, and modes near the boundary between $b$ and $b'$. The contribution to the divergence of $H^\varepsilon_\text{max}(b'|b)$ from UV-modes near $\partial (b b')$ will be the same as for the smooth max-entropy $H^\varepsilon_\text{max}(b' b)$ (because these modes are unentangled with $b$). Meanwhile the divergence from UV-modes near $\partial b$ will be the same as for the smooth min-entropy $H^\varepsilon_\text{min}(b'\overline b)$, except with the opposite sign, because $H^\varepsilon_\text{max}(b'|b) = - H^\varepsilon_\text{min}(b'|\overline b)$. Hence we should expect the total divergence to be renormalized by $\left[A(b b') - A(b)\right]^\varepsilon$.

As usual, the UV-finiteness of the smooth conditional generalized min-entropy also follows by considering complementary subsystems.

Note that the conditional generalized smooth min- and max- entropies should also be IR finite (just like the conditional generalized entropy $(A(b b') -A(b))/4G+ S(b'|b)$). This follows from $\left[A(b b') -A(b)\right]^\varepsilon$ and $H^\varepsilon_\text{max}(b'|b)$ being separately IR-finite. The refined conditions for the QES prescription \eqref{eq:refined_QES} can therefore be written in terms of the sign of the (finite) conditional generalized smooth min- and max-entropies. 
So instead of \eqref{eq:refined_QES}, we should really write
\begin{equation}\label{eq:very_refined_QES}
    S(B)_\mathrm{refined} = 
    \begin{cases}
        \langle A_1 \rangle/4G + S(bb') ,& H^\varepsilon_\mathrm{max}(b'|b) + \frac{[A_1 - A_2]^\varepsilon_\mathrm{max}}{4G} \le 0 \\
        \text{(depends on details)} ,& H^\varepsilon_\mathrm{min}(b'|b) + \frac{[A_1 - A_2]^\varepsilon_\mathrm{min}}{4G} \le 0 \le H^\varepsilon_\mathrm{max}(b'|b) + \frac{[A_1 - A_2]_\mathrm{max}^\varepsilon}{4G} \\
        \langle A_2 \rangle /4G + S(b) ,& H^\varepsilon_\mathrm{min}(b'|b) + \frac{[A_1 - A_2]^\varepsilon_\mathrm{min}}{4G} \ge 0 ~.
    \end{cases}
\end{equation}
This formulation naturally unifies the corrections discussed in this paper with the corrections from \cite{Dong:2020iod, Marolf:2020vsi}, which considered situations in which $H^\varepsilon_\mathrm{max} = H^\varepsilon_\mathrm{min} = 0$, while $[A_1 - A_2]^\varepsilon_\mathrm{min} < 0 < [A_1 - A_2]^\varepsilon_\mathrm{max}$.

\subsubsection*{Bit threads}
The bit threads paradigm \cite{Freedman:2016zud}, to the extent that it continues to be useful with large bulk entropies, should have a matching refinement. A good first step towards finding it is to incorporate bulk entropy, possibly by allowing threads to end on a ``reference system'' understood to purify the bulk matter. 

A more sophisticated modification that is sometimes mentioned is to allow threads to `pass through' entanglement, effectively using the bulk Bell pairs as `Planckian wormholes.' For this to be consistent with our refinement of the QES prescription, the number of bit threads that can pass through these Planckian wormholes should be controlled by the conditional min- and max-entropy, not the von Neumann entropy.

It would also be interesting to incorporate zero-bits into this framework, allowing bit threads to more precisely depict the total flow of information in AdS/CFT.

\subsubsection*{Other future work}

We have not given a direct path integral argument for these QES refinements for general bulk states. Our argument was more indirect: We proved it for RTN using linear algebra. Because the RTN entropy can be computed using the replica trick, the replica trick must enforce these refinements. The RTN replica trick is identical to the fixed area state replica trick, and so the same results must be true in fixed-area states. More typical (non fixed-area) states have the same entropy as the average of fixed area states that comprise them, plus subleading corrections. Although we think this argument is compelling, it is very indirect. There should be some way to relate bulk min- and max-entropy to the holographic calculation, allowing a direct replica trick proof of our result. In particular, we should be able to directly see why they are the quantities that determine whether the LM assumption is valid.

These results should also be generalized to von Neumann algebras.
We have discussed subregions ($b$, $b'$, etc.) instead of subalgebras, only for simplicity.
The smooth conditional min- and max-entropy admit algebraic definitions, which is a better language for bulk reconstruction.

We also didn't give any description as convenient as the na\"{i}ve QES prescription when $H^\varepsilon_\mathrm{min} < \Delta A/4G < H^\varepsilon_\mathrm{max}$. We did provide useful {\it bounds}, for example that the entropy of a state in that regime is less than the average of the entropies of any mixture comprising that state, plus $\mathcal{O}(\ln d)$, where $d$ is the number of states in the mixture.
But getting something stronger, such as an explicit formula, may be too much to hope for. Any formula would need to encode the details of the entanglement structure of the mixed state $\rho_{b b'}$. This is known to be very hard to characterize.

\section*{Acknowledgements}
It's a pleasure to thank Netta Engelhardt, Daniel Harlow, Patrick Hayden, Tom Faulkner, Pratik Rath, and Arvin Shahbazi-Moghaddam for discussions.
CA is supported by the US Department of Energy grants DE-SC0018944 and DE-SC0019127,  and also the Simons foundation as a member of the It from Qubit collaboration. 
GP is supported in part by AFOSR award FA9550-16-1- 0082 and DOE award {DE-SC0019380}.

\appendix

\section{Detailed evaluation of the mixture resolvent}\label{app:examples_full}
Here we elaborate on the calculations in Section \ref{sec:example1}, detailing how to go from the cubic resolvent
\begin{align}\label{eq:res_app}
	\lambda R = e^{A_2/4G} +  \frac{p R}{e^{A_2/4G} - \frac{p\lambda_1}{e^{A_1/4G}}R} + \frac{(1-p) R}{e^{A_2/4G} - \frac{(1-p)\lambda_2}{e^{A_1/4G}}R}~,
\end{align}
to the eigenvalues in each regime. 
Note the differences between this resolvent and \eqref{eq:res}.
Here, the bulk state is
\begin{equation}
    \rho_{b'} = 
    \begin{pmatrix}
        p \lambda_1 \mathbb{1}_{\frac{1}{\lambda_1}} & 0 \\ 
        0 & (1-p) \lambda_2 \mathbb{1}_{\frac{1}{\lambda_2}}
    \end{pmatrix}~.
\end{equation}
This is a slight generalization of the state from Section \ref{sec:example1}, in that we don't require one state to be pure. However, we still require both to have flat spectra.
We recover the state from Section \ref{sec:example1} by setting $\lambda_{1} =  1$ and  $\lambda_2 = e^{-S}$.

Recall that we assumed, for simplicity,
\begin{equation}\label{eq:scales_app}
    (1-p)\lambda_2 \ll p\lambda_1~,~~~~~~A_1 \ll A_2~.
\end{equation}
This condition ensures that our small and large $R$ expansions have overlapping regimes of validity. Unlike in Section \ref{sec:example1}, we will not assume that $p, 1-p = \mathcal{O}(1)$. This will require us to introduce a third expansion that is valid for sufficiently small $R$ and very small $p$.

Here are the three expansions we use, plus details about their associated spectra, along with information that will be useful in evaluating their regime of validity. 
These details are computed with the help of Appendix \ref{app:resfacts}.

\subsection*{The Expansions}

\paragraph*{Expansion 1:}
Consider the large $R$ expansion
\begin{align}\label{eq:app_exp1}
	\lambda R = e^{A_2/4G} - \frac{e^{A_1/4G}}{\lambda_1} + \frac{(1-p) R}{e^{A_2/4G} - \frac{(1-p)\lambda_2}{e^{A_1/4G}}R} + \mathcal{O}\left( \frac{e^{A_1/4G}}{\lambda_1}\frac{e^{(A_1 + A_2)/4G}}{p \lambda_1 R} \right)~.
\end{align}
Using the results of Appendix \ref{app:resfacts}, this leads to 
\begin{equation}
    \text{Number of eigenvalues} =
    \begin{cases}
                e^{A_2/4G} - \frac{e^{A_1/4G}}{\lambda_1} ,& \frac{1}{\lambda_2}  \gg e^{(A_2 - A_1)/4G} \gg  \frac{1}{\lambda_1}\\
        \frac{e^{A_1/4G}}{\lambda_2} ,& \frac{1}{\lambda_2} \ll e^{(A_2 - A_1)/4G} \gg \frac{1}{\lambda_1}
    \end{cases}
\end{equation}
of average size
\begin{equation}
    \lambda_\mathrm{avg} =
    \begin{cases}
        (1-p)e^{-A_2/4G} ,& \frac{1}{\lambda_2}  \gg e^{(A_2 - A_1)/4G}\\ 
        (1-p)\lambda_2e^{-A_1/4G}\left(1 - \frac{e^{-(A_2-A_1)/4G}}{\lambda_1}\right),& \frac{1}{\lambda_2} \ll e^{(A_2 - A_1)/4G}~.
    \end{cases}
\end{equation}
To analyze when this expansion is valid, it is useful to know the value of the resolvent.
At $\lambda_\mathrm{avg}$, the resolvent is 
\begin{equation}
    R(\lambda_\mathrm{avg}) =
    \begin{cases}
        -i \frac{e^{(A_1 + A_2)/4G}}{(1-p)\lambda_2}\sqrt{\lambda_2 e^{(A_2 - A_1)/4G} - \frac{\lambda_2}{\lambda_1}} + \frac{e^{2 A_2 / 4G} - e^{(A_1 + A_2)/4G}/\lambda_1}{2(1-p)} + ...,& \frac{1}{\lambda_2} \gg e^{(A_2 - A_1)/4G}\\ 
        \frac{e^{(A_1 + A_2)/4G}}{(1-p)\lambda_2}\left(1 - i\left(\lambda_2 e^{(A_2 - A_1)/4G} - \frac{\lambda_2}{\lambda_1}\right)^{-1/2} + ... \right),& \frac{1}{\lambda_2} \ll e^{(A_2 - A_1)/4G}~.
    \end{cases}
\end{equation}


\paragraph*{Expansion 2:}
Consider the small $R$ expansion
\begin{align}\label{eq:app_exp2}
	\lambda R = e^{A_2/4G} +  \frac{p R}{e^{A_2/4G} - \frac{p\lambda_1}{e^{A_1/4G}}R} + \frac{(1-p) R}{e^{A_2/4G}} +\mathcal{O}\left( \frac{(1-p) R}{e^{A_2/4G}}\frac{(1-p)\lambda_2 R}{e^{(A_1 + A_2)/4G}} \right)~.
\end{align}
This results in 
\begin{equation}
\begin{split}
    &\text{Number of eigenvalues} =~ 
    \begin{cases}
    e^{A_2/4G} ,& \frac{1}{\lambda_1} \gg e^{(A_2 - A_1) / 4G} \\
    \frac{e^{A_1/4G}}{\lambda_1} ,& \frac{1}{\lambda_1} \ll e^{(A_2 - A_1) / 4G} 
    \end{cases}
\end{split}
\end{equation}
of average size
\begin{equation}
\begin{split}
    \lambda_\mathrm{avg} = 
    \begin{cases}
    e^{-A_2/4G} ,& \frac{1}{\lambda_1} \gg e^{(A_2 - A_1) / 4G} \\
   p \lambda_1 e^{-A_1/4G} + (1-p)e^{-A_2/4G} ,& \frac{1}{\lambda_1} \ll e^{(A_2 - A_1) / 4G}~.
    \end{cases}
\end{split}
\end{equation}
At $\lambda_\mathrm{avg}$, the resolvent is
\begin{align}
    R(\lambda_\mathrm{avg}) = 
    \begin{cases}
    -i\frac{e^{(A_1 + A_2)/4G}}{p\lambda_1}\sqrt{\lambda_1 e^{(A_2 - A_1)/4G}} + \frac{e^{2 A_2/4G}}{2p} + ...,& \frac{1}{\lambda_1} \gg e^{(A_2 - A_1) / 4G} \\
    \frac{e^{(A_1 + A_2)/4G}}{p\lambda_1}\left( 1 - i \left(\lambda_1 e^{(A_2 - A_1)/4G}\right)^{-1/2}+...\right),& \frac{1}{\lambda_1} \ll e^{(A_2 - A_1) / 4G}~.
    \end{cases}
\end{align}

For very small $p$, the second term on the right hand side of \eqref{eq:app_exp2} can become smaller than the terms that were dropped. It is therefore helpful to use a slightly adapted version of Expansion 2, namely
\begin{align}\label{eq:exp2adapt}
	\lambda R = e^{A_2/4G} +  \frac{p R}{e^{A_2/4G} - \frac{p\lambda_1}{e^{A_1/4G}}R} + \frac{e^{ -A_2/4G}(1-p) R}{1 - \frac{(1-p) \lambda_2 }{p\lambda_1}} +\mathcal{O}\left( \frac{(1-p)^2\lambda_2 R \left(R - \frac{e^{(A_1 + A_2)/4G}}{p\lambda_1}\right)}{e^{(A_1 + 2A_2)/4G}} \right)~.
\end{align}
The only effect of this change is that now
\begin{align}
    \lambda_\mathrm{avg} = p \lambda_1 e^{-A_1/4G} + \frac{(1-p)e^{-A_2/4G}}{1 - \frac{(1-p)\lambda_2}{p \lambda_1}}.
\end{align}
Finally, we note that for $1/\lambda_1 \ll e^{(A_2 - A_1)/4G}$, and for values of $\lambda$ where $D(\lambda) \neq 0$, we have
\begin{align} \label{eq:resolventshift}
    R - \frac{e^{(A_1+A_2)/4G}}{p \lambda_1} \leq \mathcal{O}\left(\frac{e^{( A_1 + A_2)/8G} }{ \lambda_1^{1/2}}\right)~.
\end{align}
This will again be important when considering small values of $p$.

\paragraph*{Expansion 3:} 
Finally, we can use an alternative small $R$ expansion, where we expand \emph{both} the $\lambda_1$ and $\lambda_2$ terms up to $\mathcal{O}(R^2)$,
\begin{align}\label{eq:exp3}
	\lambda R = e^{A_2/4G} +  \frac{R}{e^{A_2/4G}} + \frac{p^2 \lambda_1 + (1-p)^2 \lambda_2}{e^{(A_1 + 2 A_2)/4G}}R^2 + \mathcal{O}\left( \frac{(p^3 \lambda_1^2 + (1-p)^3 \lambda_2^2) R^3}{e^{(2 A_1 + 3 A_2)/4G}} \right)~.
\end{align}
This results in
\begin{align}
    D(\lambda) = \frac{e^{(A_1 + 2 A_2)/4G}}{2 \pi(p^2 \lambda_1 + (1-p)^2 \lambda_2)}\sqrt{ \frac{p^2 \lambda_1 + (1-p)^2 \lambda_2}{e^{(A_1+A_2)/4G}} - (\lambda - e^{A_2/4G})^2},
\end{align}
which gives $e^{A_2/4G}$ eigenvalues with average eigenvalue $\lambda_\text{avg} = e^{-A_2/4G}$. Finally,
\begin{align}
R(\lambda_\text{avg}) = i e^{A_2/4G}\sqrt{\frac{e^{(A_1+A_2)/4G}}{p^2 \lambda_1 + (1-p)^2 \lambda_2}}.
\end{align}
This expansion is important because when $p$ is very small, the $\mathcal{O}(R^2)$ correction from the $\lambda_2$ term may be larger than the corresponding correction from the $\lambda_1$ term, even though $p \lambda_1 \gg (1-p)\lambda_2$.

\subsection*{The Regimes}
Here are the three regimes, each defined by the relative size of $\Delta A / 4 G \equiv (A_2 - A_1)/4G$ and
\begin{equation}
\begin{split}
    H^\varepsilon_\mathrm{min}(b') =& \ln\left(\frac{1}{p\lambda_1}\right) ~,\\
    H^\varepsilon_\mathrm{max}(b') \approx& \ln\left(\frac{1}{\lambda_2}\right)~.
\end{split}
\end{equation}
There are corrections to the na\"{i}ve QES prescription only in Regime 2, when $H^\varepsilon_\mathrm{min}(b')$ and $H^\varepsilon_\mathrm{max}(b')$ are on different sides of $\Delta A / 4 G$.

\subsubsection*{Regime 1: $H^\varepsilon_\mathrm{min},H^\varepsilon_\mathrm{max} \ll \Delta A/4 G $}
In this regime, Expansion 1 is always valid at is eigenvalue peak, which is at $\lambda_\text{avg} = p \lambda_2 e^{-A_1/4G}$. Expansion 2 is valid at its eigenvalue peak, with $\lambda_\text{avg} = (1-p) \lambda_1 e^{-A_1/4G}$, unless $p$ is very small, in which case we need to use the adapted version of Expansion 2. This only has a small effect on the eigenvalue peak.
Thus for all parameter values, assuming \eqref{eq:scales_app}, the entropy is given by
\begin{equation}\label{eq:entropy_regime_1_app}
\begin{split}
    S(B) =& \frac{A_1}{4  G} + p\ln\left( \frac{1}{p\ln(\lambda_1)} \right) +  (1-p)\ln\left( \frac{1}{(1-p)\ln(\lambda_2)} \right) + ...~,
\end{split}
\end{equation}
where we have suppressed terms that vanish in the limits we've taken.
The na\"{i}ve quantum extremal surface prescription gives the correct answer.

\begin{proof}
    Consider Expansion 1.
    The resolvent evaluated at $\mathcal{O}(\lambda_\mathrm{avg})$ is approximately $$\mathcal{O}(R(\lambda_\text{avg})) = \mathcal{O}\left(\frac{e^{(A_1 + A_2)/4G}}{ (1-p) \lambda_2}\right)~.$$
    Therefore the dropped terms at $\lambda = \mathcal{O}(\lambda_\mathrm{avg})$ are $\mathcal{O}((1-p)\lambda_2 e^{A_1/4G} / p \lambda_1^2)$. 
    The smallest kept term is $\mathcal{O}(e^{A_1/4G}/\lambda_1)$. 
    Therefore the ratio dropped/kept is $\mathcal{O}((1-p)\lambda_2 / p \lambda_1)$, which is small given our choice \eqref{eq:scales_app}. 
    
    Now consider Expansion 2. 
    The resolvent at $\lambda = \mathcal{O}(\lambda_\mathrm{avg})$ is $$\mathcal{O}(R(\lambda_\text{avg})) = \mathcal{O}\left(\frac{e^{(A_1 + A_2)/4G}}{p \lambda_1}\right)~.$$
    The largest dropped term, at $\mathcal{O}(\lambda_\mathrm{avg})$, is given by plugging this into the dropped term in \eqref{eq:app_exp1}.
    The smallest kept term is either the second term, with size $\mathcal{O}(e^{A_1/4G}/\lambda_1)$, or the third term with size $\mathcal{O}((1-p) e^{A_1/4G}/p\lambda_1)~.$ 
    In the latter case, the ratio dropped/kept equals $\mathcal{O}((1-p)\lambda_2/p\lambda_1)$.
    This is small given \eqref{eq:scales_app}.
    In the former case, the ratio equals $\mathcal{O}((1-p)^2 \lambda_2 / p^2 \lambda_1 )$. This is small, unless $p$ itself is very small.
    
    For small $p$, we need to be a bit more careful, recognizing that the second term in Expansion 2 only becomes important near the eigenvalue peak where its denominator is small, and also to make use of the adapted version \eqref{eq:exp2adapt} of Expansion 2. Using this adapted version, we find that the ratio of the dropped term to the second term is
    \begin{align}
        \mathcal{O}\left(\frac{(1-p)^2 \lambda_2 \lambda_1 (R - \frac{e^{(A_1+A_2)/4G}}{p \lambda_1})^2}{e^{(2 A_1 + 2A_2)/4G}}\right) = \mathcal{O}\left(\frac{(1-p)^2 \lambda_2 e^{(A_1-A_2)/4G}}{p^2 \lambda_1^2 }\right)~.
    \end{align}
    Going from the left hand side to the right hand side, we used the fact that \eqref{eq:resolventshift} holds near the eigenvalue peak. This ratio is small so long as $(1-p) \lambda_2 \ll p \lambda_1$ and $p \lambda_1 \gg e^{(A_1 - A_2)/4G}$.
    
    It is now a simple matter to compute the entropy using the eigenvalues from Expansion 1 and (the adapted) Expansion 2 to get \eqref{eq:entropy_regime_1_app}.
\end{proof}

\subsubsection*{Regime 2: $H^\varepsilon_\mathrm{min} \ll \Delta A / 4  G \ll H^\varepsilon_\mathrm{max}$}
This is the regime in which there are large corrections to the na\"{i}ve QES prescription.
Again, Expansions 1 and 2 are valid at their eigenvalue peak, unless $p$ is small where we need to use the adapted version of Expansion 2. However, Expansion 1 now gives an eigenvalue peak at $\lambda_\text{avg} = (1-p)e^{-A_2/4G}$

Thus, assuming \eqref{eq:scales_app}:
\begin{equation}\label{eq:entropy_regime_2_app}
\begin{split}
    S(B) =& p \frac{A_1}{4  G} + p\ln\left( \frac{1}{p\ln(\lambda_1)} \right) +  (1-p) \frac{A_2}{4  G} + (1-p)\ln\left(\frac{1}{1-p}\right)+...~,
\end{split}
\end{equation}
and again we have dropped terms that vanish in the limits we've taken.

\begin{proof}
    Expansion 2 works identically to Regime 1, so we only consider Expansion 1. 
    The largest dropped term for $\lambda = \mathcal{O}(\lambda_\text{avg})$ is $\mathcal{O}((1-p)\lambda_2 e^{A_1/4G}/ p \lambda_1^2 \sqrt{\lambda_2 e^{(A_2 - A_1)/4G}})$.
    The smallest kept term is $e^{A_1/4G}/\lambda_1$. The ratio dropped/kept is $$\mathcal{O}\left(\frac{(1-p)\sqrt{\lambda_2}}{p\lambda_1 e^{(A_2 - A_1)/4G}}\right)~.$$
    This is small so long as $(1-p)\lambda_2 \ll p \lambda_1$ and $p \lambda_1 \gg e^{(A_1 - A_2)/4G}$.

\end{proof}

\subsubsection*{Regime 3: $\Delta A/4 G \ll H^\varepsilon_\mathrm{min}, H^\varepsilon_\mathrm{max}$}
In this regime, the na\"{i}ve QES prescription does not receive large corrections.
This regime is interesting because it requires $p \lambda_1 e^{(A_2 - A_1)/4G} \ll 1$, which can be achieved whether or not $\lambda_1 e^{(A_2 - A_1)/4G}$ is greater or less than 1. 
If $\lambda_1 e^{(A_2 - A_1)/4G} \gg 1$, then Expansions 1 and 2 are valid, so long as $p^2 \lambda_1^2 e^{(A_2 - A_1)/4G} \gg (1-p)^2 \lambda_2$. However the entropy calculation gives a different answer:
\begin{equation}\label{eq:entropy_regime_3_app}
\begin{split}
    S(B) \approx&\frac{A_2}{4 G} + ...~,
\end{split}    
\end{equation}
where ``...'' represents terms that vanish in the limits we've taken.
We get the same answer in a different way if $\lambda_1 e^{(A_2 - A_1)/4G} \ll 1$.
In this parameter range, Expansion 1 is never valid.
Expansion 2 is valid at its eigenvalue peak, and gives \eqref{eq:entropy_regime_3_app}, so long as $p$ is large enough such that $(1-p)^2\lambda_2 < p^2 \lambda_1$. 
When $p$ is smaller than that, Expansion 2 cannot be used. 

The alternative small $R$ expansion, Expansion 3, is valid for all values of $p$ in this regime and  gives \eqref{eq:entropy_regime_3_app}.

\begin{proof}
   Consider Expansion 1. 
   If $\lambda_1 e^{(A_2 - A_1)/4G} \ll 1$, the resolvent is always real and so does not contribute any eigenvalues.\footnote{In fact, the solution given by Expansion 1 for large values of $R$ does not actually appear as the resolvent for any value of $\lambda$. It gives a different sheet of the solution to the one given by the resolvent.}
   If $\lambda_1 e^{(A_2 - A_1)/4G} \gg 1$, then Expansion 1 works exactly as it did in Regime 2. 
   The smallest kept term is $e^{A_1/4G}/\lambda_1$.
   The largest dropped term is $\mathcal{O}((1-p)\sqrt{\lambda_2} e^{A_1/4G} / p \lambda_1^2 \sqrt{e^{(A_2 - A_1)/4G}})$. 
   The ratio dropped/kept is $\mathcal{O}((1-p)\sqrt{\lambda_2} / p \lambda_1 \sqrt{e^{(A_2 - A_1)/4G}})$,
   which is small so long as $ p^2 \lambda_1^2 e^{(A_2 - A_1)/4G} \gg (1-p)^2\lambda_2$.
   
    Consider Expansion 2. 
    If $\lambda_1 e^{(A_2 - A_1)/4G} \gg 1$, Expansion 2 works exactly as it did in Regimes 1 and 2. It is therefore valid (when using the adapted version) so long as $ p^2 \lambda_1^2 e^{(A_2 - A_1)/4G} \gg (1-p)^2\lambda_2$.
    
    If $\lambda_1 e^{(A_2 - A_1)/4G} \ll 1$,
    the largest dropped term in Expansion 2 is $\mathcal{O}((1-p)^2 \lambda_2 e^{A_2/4G}/p^2 \lambda_1)$, while
    the smallest kept term is either $e^{A_2/4G}$ or $(1-p)R/e^{A_2/4G}=(1-p)\sqrt{e^{(A_2 + A_1)/4G}}/p\sqrt{\lambda_1}$. 
    In the former case, the dropped/kept ratio is $\mathcal{O}((1-p)^2 \lambda_2 / p^2 \lambda_1)$, which is small unless $ p^2\lambda_1 \lesssim (1-p)^2\lambda_2$. 
    In the latter case, we find that the dropped/kept ratio is $\mathcal{O}((1-p)\lambda_2 \sqrt{\lambda_1 e^{(A_2 - A_1)/4G}} / p \sqrt{\lambda_1})$. 
    This is always small. 
    
    What about Expansion 3? The smallest term that we keep is $\mathcal{O}(e^{A_2/4G})$, while the largest term that we drop is $\mathcal{O}(\sqrt{e^{(3 A_2 - A_1)/4G}(p^3 \lambda_1^2 + (1-p)^3 \lambda_2^2)^2/(p^2 \lambda_1 + (1-p)^2 \lambda_2)^3})$. The ratio is small so long as $\lambda_1 \ll e^{(A_1 - A_2)/4G}$ or $p^2 \lambda_1 \ll (1-p)^2 \lambda_2$. Between the expansions, we can therefore cover all the possible regimes. 
    
    Note that for $p^2 \lambda_1 \ll (1-p)^2 \lambda_2 \ll p^2 \lambda_1^2 e^{(A_2 - A_1)/4G}$, both Expansions 1 and 2, and Expansion 3 are valid. However, Expansion 3 misses the existence of the second eigenvalue peak that appears in Expansion 2, even though it is a small $R$ expansion and this occurs at smaller $R$ than the main eigenvalue peak. This is because, for these intermediate values of $p$, the Taylor expansion of the $\lambda_1$ term in Expansion 3 was not under control, since $p \lambda_1 R \gg e^{(A_1 + A_2)/4G}$. We were only able to get away with the expansion anyway because both the true $\lambda_1$ term and our approximation of it were only small correction anyway (because $p$ was so small). Near the second eigenvalue peak itself, this isn't true because the true $\lambda_1$ term breaks down, and Expansion 3 breaks down. So we do need to use Expansion 2 here.

    We compute the entropy as follows.
    If $\lambda_1 e^{(A_2 - A_1)/4G} \gg 1$, then for $ p^2 \lambda_1^2 e^{(A_2 - A_1)/4G} \gg (1-p)^2\lambda_2$ we can use Expansions 1 and 2 to compute the entropy.
    For small $p$, we instead use Expansion 3. 
    If $\lambda_1 e^{(A_2 - A_1)/4G} \ll 1$, then we can always just use Expansion 3, although we can also use Expansion 2 if $p$ is not too small.
    In all cases, the entropy is given by \eqref{eq:entropy_regime_3_app}.
    
\end{proof}

\section{Solving the quadratic resolvent}\label{app:resfacts}

This appendix studies the quadratic resolvent equation,
\begin{align}
	(\lambda - W) R = X + \frac{R}{Y - ZR}~,
\end{align}
where $W,X,Y,Z$ are some fixed real numbers.
This equation has the solutions
\begin{align}\label{eq:quad_res_sol}
	R(\lambda) =& \frac{X Z + Y (\lambda - W) - 1}{2 Z (\lambda - W)} - \frac{Y}{2 Z (\lambda - W)}\sqrt{\lambda - \lambda_{+}}\sqrt{\lambda - \lambda_{-}}  ~,
\end{align}
where
\begin{align}
    \lambda_{\pm} = \frac{(1 \pm \sqrt{XZ})^2}{Y} + W ~.
\end{align}
The minus in front of the square root in \eqref{eq:quad_res_sol} is required by $R(\lambda \to \infty) = 0.$

\subsection*{Eigenvalues}

The density of eigenvalues $D(\lambda)$ is related by the formula
\begin{align}
	D(\lambda) = -\frac{1}{\pi} \lim_{\epsilon \to 0^+} \mathrm{Im}R(x + i\epsilon)~.
\end{align}
So, we need the imaginary part of the resolvent.
We can ignore everything not under the square root, because it will not contribute to $D(\lambda)$. 
Use the handy fact that the square root (with positive real part) of a complex number $a + ib$ can be written as $\sqrt{a + i b} = p + i q$ with 
\begin{align}
	p = \frac{1}{\sqrt{2}}\sqrt{\sqrt{a^2 + b^2} + a}~,~~~~~~q=\frac{\mathrm{sign}(b)}{\sqrt{2}}\sqrt{\sqrt{a^2 + b^2} - a}~.
\end{align}
The relevant piece of the imaginary part of $R$ gives
\begin{align}
    D(\lambda) = \frac{Y}{2 \pi Z (\lambda - W)}\sqrt{\lambda_{+} - \lambda}\sqrt{\lambda - \lambda_{-}}~,
\end{align}
for $\lambda \in [\lambda_{-},\lambda_{+}]$, and $D(\lambda) = 0$ otherwise.
To find the number of eigenvalues, we can integrate this using
\begin{align}
    \int_a^b dx \frac{\sqrt{(x-a)(b - x)}}{2 \pi x} = \frac{1}{4}\left( -2 \sqrt{ab} + a + b \right)~.
\end{align}
This gives a total number of eigenvalues
\begin{align}
    \mathrm{Number} =& \int_{\lambda_{-}}^{\lambda_{+}} d\lambda D(\lambda)
    =
    \begin{cases}
    X ,& XZ < 1 \\
    \frac{1}{Z} ,& XZ > 1~.
    \end{cases}\label{eq:eignum}
\end{align}
The average value of $\lambda$ is
\begin{align}\label{eq:eigavg}
    \lambda_{\mathrm{avg}} =& \frac{1}{\mathrm{Number}}\int_{\lambda_{-}}^{\lambda_{+}} d\lambda~D(\lambda)\lambda 
    =
    \begin{cases}
        \frac{1}{Y} + W ,& XZ < 1\\
        \frac{X Z}{Y} + W ,& XZ > 1~.
    \end{cases}
\end{align}

\subsection*{Entropy}
In principle we can compute the von Neumann entropy given a density of eigenvalues $D(\lambda)$ with
\begin{align}
   S = - \int_{0}^{1} \lambda \ln(\lambda) D(\lambda) d\lambda~.
\end{align}
The $\ln$ makes this difficult to evaluate in practice.
Fortunately, we can obtain a rather good approximation by expanding $\lambda$ around the average of the eigenvalue distribution.
Use
\begin{alignat}{2}
    &\int_{\lambda_{-}}^{\lambda_{+}} d\lambda D(\lambda) &&= 
    \begin{cases}
        X ,& XZ<1 \\
        \frac{1}{Z} ,& XZ>1 ~,
    \end{cases}\\
    &\int_{\lambda_{-}}^{\lambda_{+}} d\lambda D(\lambda) (\lambda - \lambda_\mathrm{avg}) &&= 0~,\\
    &\int_{\lambda_{-}}^{\lambda_{+}} d\lambda D(\lambda) (\lambda - \lambda_\mathrm{avg})^2 &&= 
    \begin{cases}
        \frac{X^2 Z}{Y^2} ,& XZ<1 \\
        \frac{X}{Y^2} ,& XZ>1~. 
    \end{cases}
\end{alignat}
The entropy is
\begin{equation}
\begin{split}
    S \approx& \int_{\lambda_{-}}^{\lambda_{+}} d\lambda D(\lambda)\left( \lambda_\mathrm{avg}\ln\left( \frac{1}{\lambda_\mathrm{avg}}\right) + (\lambda - \lambda_\mathrm{avg})\left( \ln\left(\frac{1}{\lambda_\mathrm{avg}}\right) - 1 \right) - \frac{1}{2\lambda_\mathrm{avg}}(\lambda - \lambda_\mathrm{avg})^2 +...\right) \\
    =& 
    \begin{cases}
        -X\left( \frac{1}{Y} + W \right)\ln\left( \frac{1}{Y} + W \right) - \frac{1}{2}\left( \frac{1}{\frac{1}{Y}+W}\right)\left( \frac{X^2 Z}{Y^2} \right) ,& X Z < 1 \\
         -\left( \frac{X}{Y} + \frac{W}{Z} \right)\ln\left( \frac{XZ}{Y} + W \right) - \frac{1}{2}\left( \frac{1}{\frac{XZ}{Y}+W}\right)\left( \frac{X}{Y^2} \right) ,& X Z > 1~.
    \end{cases}
\end{split}
\end{equation}

\subsection*{Resolvent values}
We are sometimes interested in evaluating the resolvent at the location of the average eigenvalue.
This is important in determining that the expansions used in Section \ref{sec:examples} and Appendix \ref{app:examples_full} provide accurate estimates of the eigenvalues associated to the cubic resolvent \eqref{eq:res}.

Plug \eqref{eq:eigavg} into \eqref{eq:quad_res_sol} to get
\begin{equation}
\begin{split}
    R(\lambda=\lambda_\mathrm{avg}) =& 
    \begin{cases}
    \frac{Y}{2Z}\left( X Z - \sqrt{XZ (XZ - 4)} \right),& X Z < 1 \\
    -\frac{Y}{2 X Z^2}\left( 1 - 2 X Z + \sqrt{1 - 4 X Z} \right) ,& X Z > 1 
    \end{cases}\\
    \approx&
    \begin{cases}
    -i\frac{Y}{Z}\sqrt{XZ} + \frac{XY}{2} + ... & X Z \ll 1 \\
    \frac{Y}{Z} - i\frac{Y}{Z}\frac{1}{\sqrt{XZ}} + ... & X Z \gg 1~.
    \end{cases}
\end{split}
\end{equation}
This implies
\begin{align}
    \frac{R(\lambda_\mathrm{avg})}{Y - Z R(\lambda_\mathrm{avg})} \approx& - i \sqrt{\frac{X}{Z}}~,
\end{align}
when either $XZ \ll 1$ or $XZ \gg 1$.
Also note
\begin{equation}
\begin{split}
    \lambda_\mathrm{avg} R(\lambda_\mathrm{avg}) =
    \begin{cases}
    -i\sqrt{\frac{X}{Z}}\left( 1 + W Y  \right) + ... & X Z \ll 1 \\
    X + \frac{WY}{Z} + ... & X Z \gg 1~.
    \end{cases}
\end{split}
\end{equation}
These are useful when comparing dropped terms to kept ones.

\subsection*{Example: bipartite tensor}
Apply this to a simple example. 
Consider a bipartite random tensor, with legs $A$ and $B$. 
The resolvent associated to $\rho_A$ satisfies
\begin{align}
	\lambda R = D_A + \frac{R}{D_A - \frac{R}{D_B}}~.
\end{align}
So, $W=0,~X = D_A,~Y = D_A$, and $Z = 1 / D_B$.
There are $\min( D_A, D_B )$ eigenvalues with average value $\lambda_\mathrm{avg} = \max( \frac{1}{D_A},\frac{1}{D_B})$. 
The ratio of the width of the peak to $\lambda_\mathrm{avg}$ is approximately ${\min(D_A,D_B)}/{\sqrt{D_A D_B}}$, and so the peak is very narrow when the dimensions are quite different (width relative to mean like $1/{\sqrt{D_\mathrm{larger}}}$) and widest when the dimensions are equal.
The entropy is
\begin{align}
    S_\mathrm{bip} = 
    \begin{cases}
        \ln\left( D_A\right) - \frac{1}{2}\frac{D_A}{D_B} ,& \frac{D_A}{D_B} < 1 \\
         \ln\left( D_B \right) - \frac{1}{2}\frac{D_B}{D_A} ,& \frac{D_A}{D_B} > 1~.
    \end{cases}
\end{align}
This is all just as expected.

\section{Numerics}\label{app:numerics}

\begin{figure}	
	\centering
	\includegraphics[width = \columnwidth]{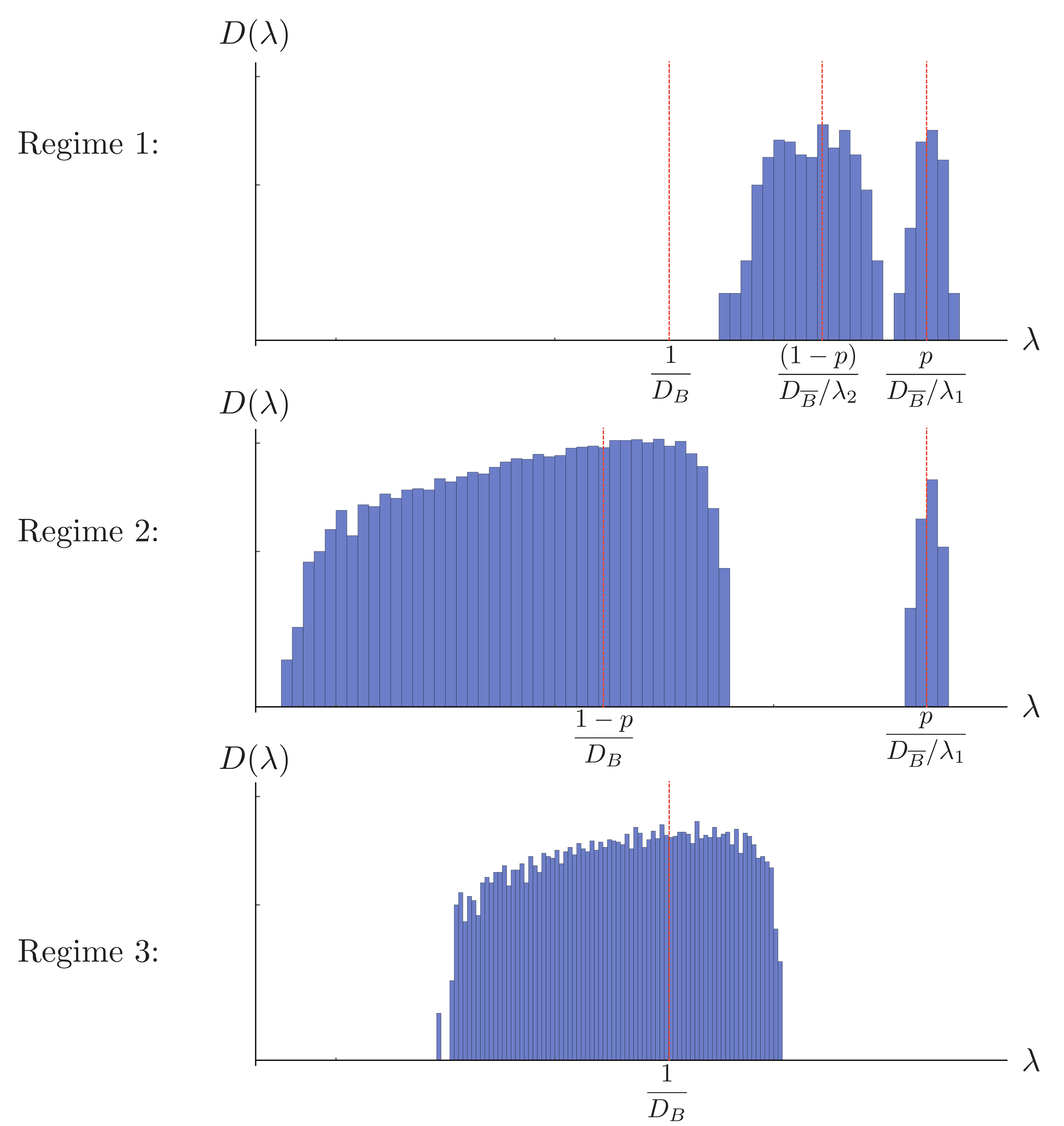}
	\caption{Log-log histogram of the eigenvalue density $D(\lambda)$ of $\rho_B$, for a random tensor with two ``boundary'' legs $B, \overline{B}$ and a ``bulk'' leg $b'$, for $80$ trials. All plots are made with boundary leg dimensions $D_B = 30,~ D_{\overline{B}} = 1$ and bulk leg state \eqref{eq:app_bulk_leg_state}, with $p = 1/2$. Regime 1 plot made with $\lambda_1 = 1,~\lambda_2 = 1/3$; Regime 2 plot $\lambda_1 = 1,~\lambda_2 = 1/45$; Regime 3 plot $\lambda_1 = 1/20,~\lambda_2 = 1/45$.
	Note the agreement with Figure \ref{fig:regimes}.}
	\label{fig:regimes_numerics}
\end{figure}

Here we present numerical evidence supporting the results of Section \ref{sec:examples}.
These numerics are of a single tripartite random tensor, with legs $B,~\overline{B},~b'$. 
As pointed out in \cite{Penington:2019kki}, computing the entropy of e.g. $B$ is equivalent non-perturbatively to the computation of $S(B)$ in a fixed-area state, like Figure \ref{fig:black_hole} with $\gamma_1$ fixed to area $\ln D_B$ and $\gamma_2$ fixed to area $\ln D_{\overline{B}}$. 
$D_B$ and $D_{\overline{B}}$ are the dimensions of legs $B$ and $\overline{B}$ respectively.
The leg $b'$ is the ``bulk'' leg, analogous to the state of the bulk fields between $\gamma_1$ and $\gamma_2$, and is projected into the state
\begin{equation}\label{eq:app_bulk_leg_state}
    \rho_{b'} = 
    \begin{pmatrix}
        p \lambda_1 \mathbb{1}_{\frac{1}{\lambda_1}} & 0 \\ 
        0 & (1-p) \lambda_2 \mathbb{1}_{\frac{1}{\lambda_2}}
    \end{pmatrix}~.
\end{equation}
Figure \ref{fig:regimes_numerics} displays the resulting eigenvalue density $D(\lambda)$ of density matrix $\rho_B$, in Regimes 1, 2, and 3 from Section \ref{sec:examples} and also Appendix \ref{app:examples_full}.

\section{One-shot decoupling}\label{app:one-shot_decoupling}
A great many facts in quantum information theory follow from the same basic principle: the decoupling theorem. 

While this powerful theorem was originally proven and used in the independent identically distributed (i.i.d.) setting \cite{hayden2008decoupling}, in which a large number of independent copies of the state are available, more recently a one-shot version has been proven \cite{Berta_2011}, effectively generalizing many key results to the one-shot setting. 
The chief difference between the two decoupling theorems is the replacement of the von Neumann entropy with the one-shot entropies, the min- and max-entropy. 

The setup is as follows. 
Consider a system $A = A_1 A_2$ (with dimensions $|A_1|$ and $|A_2|$) entangled with a system $R$.
The one-shot decoupling theorem provides a sufficient condition for the average unitary $U$ acting on $A$ to ``decouple'' $A_1$ and $R$. 
This is often used to provide a sufficient condition for something weaker, the existence of a unitary operator $U$ that decouples $A_1$ and $R$. 

For our purposes, the theorem says if
\begin{align}
    \ln |A_1| \le \ln |A_2| + H_\mathrm{min}(A | R) - 2 \ln \frac{1}{\varepsilon}~, 
\end{align}
then 
\begin{equation}
    \int dU \left\lVert \tr_{A_2}(U \rho_{AR} U^\dagger) - \frac{\mathbb{1}_{A_1}}{|A_1|} \otimes \rho_{R}  \right\rVert_1 \leq \varepsilon~,
\end{equation}
where $dU$ is the Haar measure on the group of unitaries acting on $\mathcal{H}_A$, normalized to $\int dU = 1$. 

We present the proof as Theorem 7 below, after some useful definitions and lemmas.

\subsubsection*{Definition 1.}
Let $X$ be an operator on Hilbert space $\mathcal{H}$. 
The $L^2$ norm, or \emph{Hilbert-Schmidt norm}, is defined as
\begin{equation}
    \left\lVert X \right\rVert_2 = \sqrt{\tr(X^\dagger X)}~.
\end{equation}
This upperbounds the $L^1$ norm ($ \left\lVert X \right\rVert_1 = \tr{\sqrt{X^\dagger X}})$, as $\left\lVert X \right\rVert_1 \le \sqrt{d} \left\lVert X \right\rVert_2$, where $d$ is the dimension of $\mathcal{H}$. 
This bound is involved in the i.i.d. proof of decoupling \cite{hayden2008decoupling}, but the one-shot version we are interested in requires a stronger bound.

\subsubsection*{Lemma 2.} 
(Lemma 5.1.3 of \cite{renner2005security})
Let $S$ be a Hermitian operator on Hilbert space $\mathcal{H}$, and $\sigma$ be a nonnegative operator on $\mathcal{H}$. 
Then
\begin{equation} \label{eq:lemma2}
    \left\lVert S \right\rVert_1 \le \sqrt{\tr{\sigma}} \left\lVert \sigma^{-1/4} S \sigma^{-1/4} \right\rVert_2~.
\end{equation}
\begin{proof}
    We first note that \eqref{eq:lemma2} can be rewritten as 
    \begin{equation}
        \tr \left| \sqrt{\sigma'} S' \sqrt{\sigma'} \right| \le \sqrt{\tr(S'^2)\tr(\sigma'^2)}~,
    \end{equation}
    where $\sigma' = \sqrt{\sigma}$ and $S' = \sigma^{-1/4} S \sigma^{-1/4}$~.
    Let $\ket{v}$ denote an orthonormal eigenbasis of  $S = \sigma'^{1/2}S'\sigma'^{1/2}$, and let $S' = \sum_y \alpha_y \ket{y}\bra{y}$ be a spectral decomposition of $S'$.
    Then
    \begin{align}
        \tr \left| \sqrt{\sigma'} S' \sqrt{\sigma'} \right| &= \sum_v \left| \bra{v} \sqrt{\sigma'} S' \sqrt{\sigma'} \ket{v} \right| \\
        &= \sum_v \left|\sum_y  \alpha_y \bra{v} \sqrt{\sigma'} \ket{y}\bra{y} \sqrt{\sigma'} \ket{v} \right| \\
        &\le \sum_v \sum_y  \left| \alpha_y \right| \bra{v} \sqrt{\sigma'} \ket{y}\bra{y} \sqrt{\sigma'} \ket{v} \\
        &= \sum_v   \bra{v} \sqrt{\sigma'} \left|S'\right| \sqrt{\sigma'} \ket{v}\\
        &= \tr( \sqrt{\sigma'} \left|S'\right| \sqrt{\sigma'} ) =\tr( \left|S'\right| \sigma' )\\
        &\le \sqrt{\tr(|S'|^2) \tr(\sigma'^2)} = \sqrt{\tr(S'^2) \tr(\sigma'^2)} ~.
    \end{align}
    The second inequality is the Cauchy-Schwarz inequality, applied to the  Hilbert space $\mathrm{End}(\mathcal{H})$ of operators on $\mathcal{H}$, with the inner product
        $\braket{A|B} = \tr(A^\dagger B)$.
\end{proof}

Having bounded the $L^1$ norm by this particular $L^2$ norm, we will later bound the relevant $L^2$ norm by something else. 
First, we need to define the conditional collision entropy $H_C(A|B)$, which, as we'll prove, bounds the conditional min-entropy $H_\mathrm{min}(A|B)$. 

\subsubsection*{Definition 3.}
The \emph{quantum conditional collision entropy} for density matrix $\rho_{AB}$ on Hilbert space $\mathcal{H}_{AB} = \mathcal{H}_A \otimes \mathcal{H}_B$ is defined as
\begin{equation}
    H_C(A|B) = - \inf_{\sigma_B} \ln \tr\left[ \left(  (\mathbb{1}_A \otimes \sigma_B^{-1/4} ) \rho_{AB} (\mathbb{1}_A \otimes \sigma_B^{-1/4} )   \right)^2 \right]
\end{equation}
where the infimum is taken over all density matrices $\sigma_B$ on Hilbert space $\mathcal{H_B}$. 
Note that $\sigma_B^{-1}$ is the `generalized inverse' of $\sigma_B$, defined as the inverse on its support. That is, $\sigma_B \sigma_{B}^{-1} = \sigma_{B}^{-1} \sigma_B = \sigma_B^0 = (\sigma_B^{-1})^0$.

\subsubsection*{Lemma 4.}
(Lemma B.3 of \cite{Berta_2011}) Let $\rho_{AB}$ be a nonnegative Hermitian matrix on Hilbert space $\mathcal{H}_{AB}$, with trace less than or equal to 1. 
Then
\begin{equation}
    H_\mathrm{min}(A|B) \le H_C(A|B)~.
\end{equation}
\begin{proof}
    By the definition of $H_\mathrm{min}(A|B)$ there exists some $\sigma_B$ such that $H_\mathrm{min}(A|B) = -D_\infty(\rho_{AB} || \rho_A \otimes \sigma_B)$.
    By the definition of $D_\infty(\rho_{AB} || \rho_A \otimes \sigma_B)$, we have
    \begin{align}
        e^\lambda \,\mathbb{1}_A \otimes \sigma_B \geq \rho_{AB}~.
    \end{align}
    if and only if $\lambda \geq -H_\mathrm{min}(A|B)$. Hence the largest eigenvalue of $\sigma_B^{-1/2} \rho_{AB} \sigma_B^{-1/2}$ is $e^{-H_\mathrm{min}(A|B)}$. We can rewrite this as
    \begin{equation}
        H_\mathrm{min}(A|B) = - \ln \max_{\omega_{AB}} \tr \left[ \omega_{AB} (\mathbb{1}_A \otimes \sigma_B^{-1/2}) \rho_{AB} (\mathbb{1}_A \otimes \sigma_B^{-1/2}) \right]~,
    \end{equation}
    where the maximization is over density matrices $\omega_{AB}$ and is achieved when $\omega_{AB}$ is a projector onto the largest eigenvalue of $\sigma_B^{-1/2} \rho_{AB} \sigma_B^{-1/2}$. 
    For $\kappa_B$ and $\omega_{AB}$ arbitrary density matrices on $\mathcal{H}_B$ and $\mathcal{H}_{AB}$ respectively,
    \begin{equation}
    \begin{split}
        H_C(A|B) =& - \ln \min_{\kappa_B} \tr \left[ \rho_{AB} (\mathbb{1}_A \otimes \kappa_B^{-1/2}) \rho_{AB} (\mathbb{1}_A \otimes \kappa_B^{-1/2}) \right] \\
        \ge& - \ln \tr \left[ \rho_{AB} (\mathbb{1}_A \otimes \sigma_B^{-1/2}) \rho_{AB} (\mathbb{1}_A \otimes \sigma_B^{-1/2}) \right] \\
        \ge& - \ln \max_{\omega_{AB}} \tr \left[ \omega_{AB} (\mathbb{1}_A \otimes \sigma_B^{-1/2}) \rho_{AB} (\mathbb{1}_A \otimes \sigma_B^{-1/2}) \right] \\
        =& H_\mathrm{min}(A|B)~.
    \end{split}
    \end{equation}
\end{proof}

\subsubsection*{Lemma 5.}
(Lemma C.1 of \cite{Berta_2011})
Let $F_{AB}$ denote the swap operator of $\mathcal{H}_A \otimes \mathcal{H}_B$. 
Let $A = A_1 A_2$. Then
\begin{equation}
    \int dU (U \otimes U)^\dagger (\mathbb{1}_{A_2 A_2'} \otimes F_{A_1 A_1'})(U \otimes U) \le \frac{1}{|A_1|} \mathbb{1}_{AA'} + \frac{1}{|A_2|}F_{AA'}~,
\end{equation}
where $dU$ is the Haar measure on the space of unitaries acting on $\mathcal{H}_A$, normalized to $\int dU = 1$.

\begin{proof}
    For any Hermitian $X$, it follows from Schur's lemma (see e.g. \cite{harrow2013church}) that
    \begin{equation}\label{eq:unitary_integral}
        \int dU (U \otimes U)^\dagger X (U \otimes U) = a_+(X) \Pi^+_A + a_-(X) \Pi^-_A~,
    \end{equation}
    where we have defined
    \begin{equation}
    \begin{split}
        \Pi^{\pm}_A &\equiv \frac{1}{2}( \mathbb{1}_{AA'} \pm F_{AA'} )~, \\
        a_\pm(X) &\equiv \frac{1}{\mathrm{rank}(\Pi_A^\pm)}\tr(X \Pi_A^\pm)~.
    \end{split}
    \end{equation}
    Plug in $X = (\mathbb{1}_{A_2 A_2'} \otimes F_{A_1 A_1'})$ and find
    \begin{equation}
        \tr\big(\Pi_A^\pm (\mathbb{1}_{A_2 A_2'} \otimes F_{A_1 A_1'})\big) = \frac{1}{2} |A_1|\cdot |A_2|^2 \pm |A_1|^2 \cdot |A_2|~.
    \end{equation}
    Using $\mathrm{rank}(\Pi_A^\pm) = \frac{1}{2}|A|(|A| \pm 1)$, we get
    \begin{equation}
        a_\pm\big(\mathbb{1}_{A_2 A_2'} \otimes F_{A_1A_1'}\big) = \frac{|A_2| \pm |A_1|}{|A_1|\cdot|A_2| \pm 1}~.
    \end{equation}
    Plugging all of this into \eqref{eq:unitary_integral} gives
    \begin{equation}
    \begin{split}
        \int dU (U \otimes U)^\dagger (\mathbb{1}_{A_2 A_2'} \otimes F_{A_1 A_1'})(U \otimes U) &= \frac{a_+ + a_-}{2}\mathbb{1}_{AA'} + \frac{a_+ - a_-}{2}F_{AA'} \\
        &\le \frac{1}{|A_1|} \mathbb{1}_{AA'} + \frac{1}{|A_2|}F_{AA'}~,
    \end{split}
    \end{equation}
    which is what we wanted to show. 
\end{proof}

\subsubsection*{Lemma 6.}
(Lemma C.2 of \cite{Berta_2011})
Let $\rho_{AR}$ be a density matrix on $\mathcal{H}_{AR}$, $A = A_1 A_2$, and $\sigma_{A_1 R}(U) = \tr_{A_2}\big( (U \otimes \mathbb{1}_R) \rho_{AR} (U \otimes \mathbb{1}_R)^\dagger \big)$.
Then
\begin{equation}
    \int dU \tr \big( \sigma_{A_1 R}(U)^2 \big) \le \frac{1}{|A_1|}\tr(\rho_R^2) + \frac{1}{|A_2|}\tr(\rho_{AR}^2)~,
\end{equation}
where $dU$ is the Haar measure on the space of unitaries acting on $\mathcal{H}_A$, normalized to $\int dU = 1$.

\begin{proof}
    Use Lemma 5 to get
    \begin{equation}
    \begin{split}
        \int dU & \tr \big( \sigma_{A_1 R}(U)^2 \big) \\
        &= \int dU \tr \big( (U_A \otimes U_{A'} \otimes \mathbb{1}_{RR'}) \rho_{A_1 A_2 R} \otimes \rho_{A'_1 A'_2 R'} (U^\dagger_A \otimes U^\dagger_{A'} \otimes \mathbb{1}_{RR'}) (F_{A_1 A'_1} \otimes \mathbb{1}_{A_2 A'_2} \otimes F_{RR'}) \big) \\
        &= \tr\bigg( (\rho_{A_1 A_2 R} \otimes \rho_{A'_1 A'_2 R'}) \int dU (U\otimes U)^\dagger (\mathbb{1}_{A_2 A_2'} \otimes F_{A_1 A_1'}) (U\otimes U) \otimes F_{RR'}\bigg) \\
        &\le \tr\bigg( (\rho_{A_1 A_2 R} \otimes \rho_{A'_1 A'_2 R'}) \big( \frac{1}{|A_1|}\mathbb{1}_{AA'} + \frac{1}{|A_2|}F_{AA'} \big) \otimes F_{RR'}\bigg) \\
        &= \frac{1}{|A_1|}\tr(\rho_R^2) + \frac{1}{|A_2|}\tr(\rho_{AR}^2)~,
    \end{split}
    \end{equation}
    which is what we wanted to show.
\end{proof}

We can finally combine these to prove the one-shot decoupling theorem.

\subsubsection*{Theorem 7.}
(Theorem III.1 of \cite{Berta_2011})
Consider a state $\rho_{AR}$ on Hilbert space $\mathcal{H}_A = \mathcal{H}_{A_1} \otimes \mathcal{H}_{A_2}$, with factors of dimensions $|A_1|$ and $|A_2|$ respectively, and Hilbert space $\mathcal{H}_R$. 
Then
\begin{align}
    \ln |A_1| \le \ln |A_2| + H_\mathrm{min}(A | R) - 2 \ln \frac{1}{\varepsilon} 
\end{align}
implies
\begin{equation}
    \int dU \left\lVert \tr_{A_2}(U \rho_{AR} U) - \frac{\mathbb{1}_{A_1}}{|A_1|} \otimes \rho_{R}  \right\rVert_1 \leq \varepsilon~,
\end{equation}
where $dU$ is the Haar measure on the group of unitaries acting on $\mathcal{H}_A$, normalized to $\int dU = 1$.
\begin{proof}
    Let $\sigma_{A_1 R}(U) = \tr_{A_2}\big( (U \otimes \mathbb{1}_R) \rho_{AR} (U \otimes \mathbb{1}_R)^\dagger \big)$.
    Because $H_C(A|R) \ge H_\mathrm{min}(A|R)$ by Lemma 4, and $\left\lVert S \right\rVert_1 \le \sqrt{\tr{\sigma}} \left\lVert \sigma^{-1/4} S \sigma^{-1/4} \right\rVert_2$ by Lemma 2, it suffices to show that
    \begin{equation}\label{eq:thm7_assumption}
        \ln |A_1| \le \ln|A_2| + H_C(A|R) - 2 \ln \frac{1}{\varepsilon} 
    \end{equation}
    implies
    \begin{equation}\label{eq:thm7_inequality}
        \int dU \left\lVert (\mathbb{1}_{A_1} \otimes \omega_R^{-1/4}) (\sigma_{A_1 R}(U) - \frac{\mathbb{1}_{A_1}}{|A_1|} \otimes \rho_{R}) (\mathbb{1}_{A_1} \otimes \omega_R^{-1/4}) \right\rVert_2^2 \le \frac{\varepsilon^2}{|A_1|}~,
    \end{equation}
    where $\omega_R$ is some density matrix on $\mathcal{H}_R$. 
    Define
    \begin{equation}
    \begin{split}
        \widetilde{\rho}_{AR} &\equiv \left( \mathbb{1}_A \otimes \omega_R^{-1/4} \right) \rho_{AR} \left( \mathbb{1}_A \otimes \omega_R^{-1/4} \right)~,\\
        \widetilde{\sigma}_{A_1 R}(U) &\equiv \tr_{A_2}\big( (U \otimes \mathbb{1}_R) \widetilde{\rho}_{AR} (U \otimes \mathbb{1}_R)^\dagger \big)~.
    \end{split}
    \end{equation}
    The left hand side of \eqref{eq:thm7_inequality} then equals
    \begin{equation}
    \begin{split}
        \int dU & \left\lVert \widetilde{\sigma}_{A_1 R}(U) - \frac{\mathbb{1}_{A_1}}{|A_1|} \otimes \widetilde{\rho}_R \right \rVert_2^2 = \int dU \tr\big( (\widetilde{\sigma}_{A_1 R}(U) - \frac{\mathbb{1}_{A_1}}{|A_1|} \otimes \widetilde{\rho}_R )^2 \big)\\
        &= \int dU \left[ \tr\big( \widetilde{\sigma}^2_{A_1 R}(U)\big) - 2 \tr\big( \widetilde{\sigma}_{A_1 R}(U) \frac{\mathbb{1}_{A_1}}{|A_1|} \otimes \widetilde{\rho}_R  \big) + \tr \big( \frac{\mathbb{1}_{A_1}}{|A_1|^2} \otimes \widetilde{\rho}^2_R  \big)\right] \\
        &= \int dU \left[\tr\big( \widetilde{\sigma}^2_{A_1 R}(U)\big) - \tr \big( \frac{\mathbb{1}_{A_1}}{|A_1|^2} \otimes \widetilde{\rho}^2_R  \big)\right] \\
        &= \int dU \tr\big( \widetilde{\sigma}^2_{A_1 R}(U)\big) - \frac{1}{|A_1|}\tr \big( \widetilde{\rho}_R^2 \big) \\
        &\le \frac{1}{|A_2|} \tr\big( \widetilde{\rho}_{AR}^2 \big) \le \frac{\varepsilon^2}{|A_1|}~, 
    \end{split}
    \end{equation}
    where in the third line we have used $\frac{\mathbb{1}_{A_1}}{|A_1|} \otimes \widetilde{\rho}_R = \int dU \widetilde{\sigma}_{A_1 R}(U)$, in the first inequality we have used Lemma 6, and in the final inequality we have used \eqref{eq:thm7_assumption}. 
\end{proof}

\bibliographystyle{jhep}
\bibliography{mybib2019}

\end{document}